\documentclass[11pt,fleqn]{article}
\PassOptionsToPackage{hyphens}{url}
\usepackage[utf8]{inputenc}
\usepackage[T1]{fontenc}
\usepackage[margin=1in]{geometry}
\usepackage{amsmath,amssymb}
\usepackage{graphicx}
\usepackage{float}
\usepackage{array}
\usepackage{longtable}
\usepackage{newunicodechar}
\usepackage{textcomp}
\usepackage[hidelinks]{hyperref}
\usepackage{parskip}
\newunicodechar{’}{'}
\newunicodechar{‘}{`}
\newunicodechar{“}{``}
\newunicodechar{”}{''}
\newunicodechar{–}{--}
\newunicodechar{©}{\textcopyright{}}
\newunicodechar{β}{\ensuremath{\beta}}
\newunicodechar{ρ}{\ensuremath{\rho}}
\newunicodechar{α}{\ensuremath{\alpha}}
\newunicodechar{θ}{\ensuremath{\theta}}
\newunicodechar{Δ}{\ensuremath{\Delta}}
\newunicodechar{σ}{\ensuremath{\sigma}}
\newunicodechar{τ}{\ensuremath{\tau}}
\newunicodechar{κ}{\ensuremath{\kappa}}
\newunicodechar{ε}{\ensuremath{\varepsilon}}
\newunicodechar{Σ}{\ensuremath{\Sigma}}
\newunicodechar{δ}{\ensuremath{\delta}}
\newunicodechar{ψ}{\ensuremath{\psi}}
\newunicodechar{π}{\ensuremath{\pi}}
\newunicodechar{μ}{\ensuremath{\mu}}
\newunicodechar{Λ}{\ensuremath{\Lambda}}
\newunicodechar{·}{\ensuremath{\cdot}}
\newunicodechar{−}{\ensuremath{-}}
\newunicodechar{≈}{\ensuremath{\approx}}
\newunicodechar{≤}{\ensuremath{\le}}
\newunicodechar{≥}{\ensuremath{\ge}}
\newunicodechar{±}{\ensuremath{\pm}}
\newunicodechar{×}{\ensuremath{\times}}
\newunicodechar{∈}{\ensuremath{\in}}
\newunicodechar{→}{\ensuremath{\rightarrow}}
\newunicodechar{√}{\ensuremath{\surd}}
\newunicodechar{∂}{\ensuremath{\partial}}
\newunicodechar{≳}{\ensuremath{\gtrsim}}
\newunicodechar{≲}{\ensuremath{\lesssim}}
\newunicodechar{⟨}{\ensuremath{\langle}}
\newunicodechar{⟩}{\ensuremath{\rangle}}
\newunicodechar{⟹}{\ensuremath{\Longrightarrow}}
\newunicodechar{Õ}{\ensuremath{\tilde{O}}}
\newunicodechar{ᵢ}{\textsubscript{i}}
\newunicodechar{ₖ}{\textsubscript{k}}
\newunicodechar{₀}{\textsubscript{0}}
\newunicodechar{₁}{\textsubscript{1}}
\newunicodechar{₂}{\textsubscript{2}}
\newunicodechar{₃}{\textsubscript{3}}
\newunicodechar{₄}{\textsubscript{4}}
\newunicodechar{²}{\textsuperscript{2}}
\newunicodechar{³}{\textsuperscript{3}}
\newunicodechar{¹}{\textsuperscript{1}}
\newunicodechar{⁻}{\textsuperscript{-}}
\newunicodechar{⁴}{\textsuperscript{4}}
\newunicodechar{⁵}{\textsuperscript{5}}
\newunicodechar{⁶}{\textsuperscript{6}}
\newunicodechar{ã}{\~{a}}
\newunicodechar{É}{\'{E}}
\newunicodechar{ó}{\'{o}}
\newunicodechar{í}{\'{i}}
\newunicodechar{á}{\'{a}}
\newunicodechar{ő}{\H{o}}
\newunicodechar{å}{\aa{}}
\newunicodechar{é}{\'{e}}
\newunicodechar{è}{\`{e}}
\newunicodechar{ü}{\"{u}}
\newunicodechar{ö}{\"{o}}
\newunicodechar{ä}{\"{a}}
\newunicodechar{ñ}{\~{n}}
\title{\textbf{A Scenario-Based Evaluation of CRQC+AI Vulnerability Spectrum for TLS 1.3 Cryptographic Dependencies}}
\author{Noel Grover 1,*, Mussie Haile 2, Brad Pedersen 1, Eric Uner 3 and Bradley J Erickson 1}
\date{July 23rd, 2026; revised September 7th, 2026; revised November 3rd, 2026}
\begin{document}
\maketitle
\begin{center}\small
1 EnQuanta, USA; ngrover@enquanta.com (N.G.); bpedersen@enquanta.com (B.P.); berickson@enquanta.com (B.J.E.) \\
2 MOYA Technologies, Inc., USA; mhaile@moyatech.com (M.H.) \\
3 Lab33, USA; eric.uner@lab33.us (E.U.) \\
* Correspondence: ngrover@enquanta.com
\end{center}
\begin{center}\footnotesize © 2026 The Authors. Published under a Creative Commons Attribution 4.0 International License (CC BY 4.0). arXiv:2608.23785 [cs.CR].\end{center}
\begin{abstract}

This paper evaluates quantum and AI-accelerated risks to TLS 1.3 cryptographic dependencies under an evidence-tiered model, distinguishing mechanism-backed threats (Shor’s algorithm against RSA and ECC) from contingency-backed risks to lattice-based post-quantum cryptography (PQC) and hypothesis-only risks to hash-based and symmetric primitives. We do not identify any known breaks of ML-KEM, ML-DSA, SLH-DSA, or AES-256. Instead, we use explicit scenario assumptions, organized as a four-scenario capability model with parameters and pseudocode for reproducibility, to stress-test migration timelines accompanied by parameter sensitivity analysis and explicit falsification analysis. The primary methodological contribution is a reproducible scenario-estimation instrument together with its explicit update mechanics: every parameter is a named, anchored quantity that can be varied and the model rerun; a stated protocol maps observed conformance to, or deviation from, the modeled curves onto revisions of specific parameters, so progressive refinements can be tested against accumulating historical data. The paper is a methodological companion to quantum resource-estimation studies and to expert-elicitation timeline surveys such as the Global Risk Institute quantum threat reports, with its revision rules stated explicitly. As of late 2026, the model does not show any NIST-approved algorithms as broken. Instead, the vulnerability spectrum under different scenarios shows RSA risk crossing the 50\% threshold between 2029 and 2032 and the PQC risk becoming a non-zero risk after 2031 to 2034 under contingency scenarios conditional on the unproven dimension-collapse. We urge that PQC migration proceed as mandated by the 2030 and 2031 federal deadlines and by Mosca’s HNDL reasoning, and that crypto-agility and hybrid cryptographic deployment be treated as necessary complements to any PQC migration effort.


\end{abstract}
\section*{1. Introduction}
\addcontentsline{toc}{section}{1. Introduction}
In May 2021, the \textit{Journal of Cryptology} published a special issue devoted to security of the Transport Layer Security (TLS) 1.3 protocol. [1] Since then, progress in the fields of quantum computing, quantum algorithms, and AI has exploded and heightened the need for NIST-standardized post-quantum cryptography (PQC) to address the threats posed to classical public key cryptography. [2] Quantum-enhanced artificial intelligence (QAI), quantum machine learning (QML) and quantum deep learning (QDL) are the three classes of quantum-accelerated AI whose bearing on the PQC migration timeline this paper takes up. Figure 1 sets out the high-level interfaces the paper assumes: a quantum processor, the classical control stack that drives and reads it, and the AI systems on the far side of that stack, with machine learning a subset of AI and deep learning a subset of machine learning. The arrows carry the contributions that Section 3.3 later prices: learned decoding, calibration and code search flowing toward the processor, and syndrome and calibration data flowing back.

\begin{figure}[H]\centering
\includegraphics[width=0.82\linewidth]{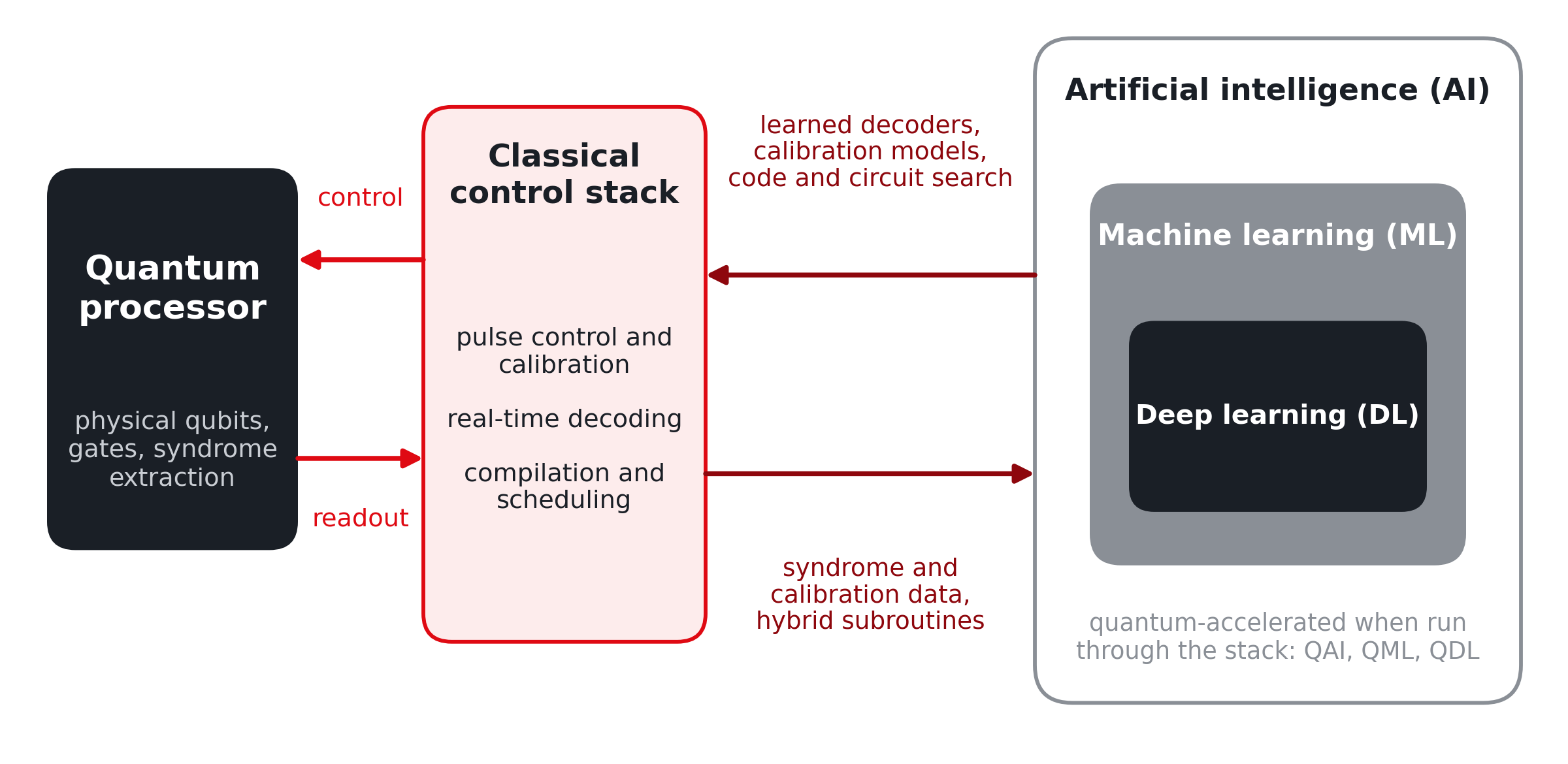}
\par\vspace{3pt}{\small\textbf{Figure 1: Simplified Quantum-Classical Artificial Intelligence Interface \\[2pt](the authors' schematic; conceptual framing only).}}
\end{figure}
The National Institute of Standards and Technology (NIST) finalized its first PQC standards, for key encapsulation (FIPS 203) and digital signatures (FIPS 204 and 205), in August 2024. [4] But less than six months later, the federal timelines for implementing these PQC algorithms for certain classes of infrastructure systems were accelerated by executive order; and the most recent executive order as of June 2026 mandates that all federal agencies migrate all high value assets and high-impact systems (excluding National Security Systems) to PQC for key establishment by December 31, 2030, and for digital signatures by December 31, 2031 (Figure 2). [5] Comparable updates to accelerate the PQC transition timelines have happened in the European Union and Australia. [6]

\begin{figure}[H]\centering
\includegraphics[width=0.82\linewidth]{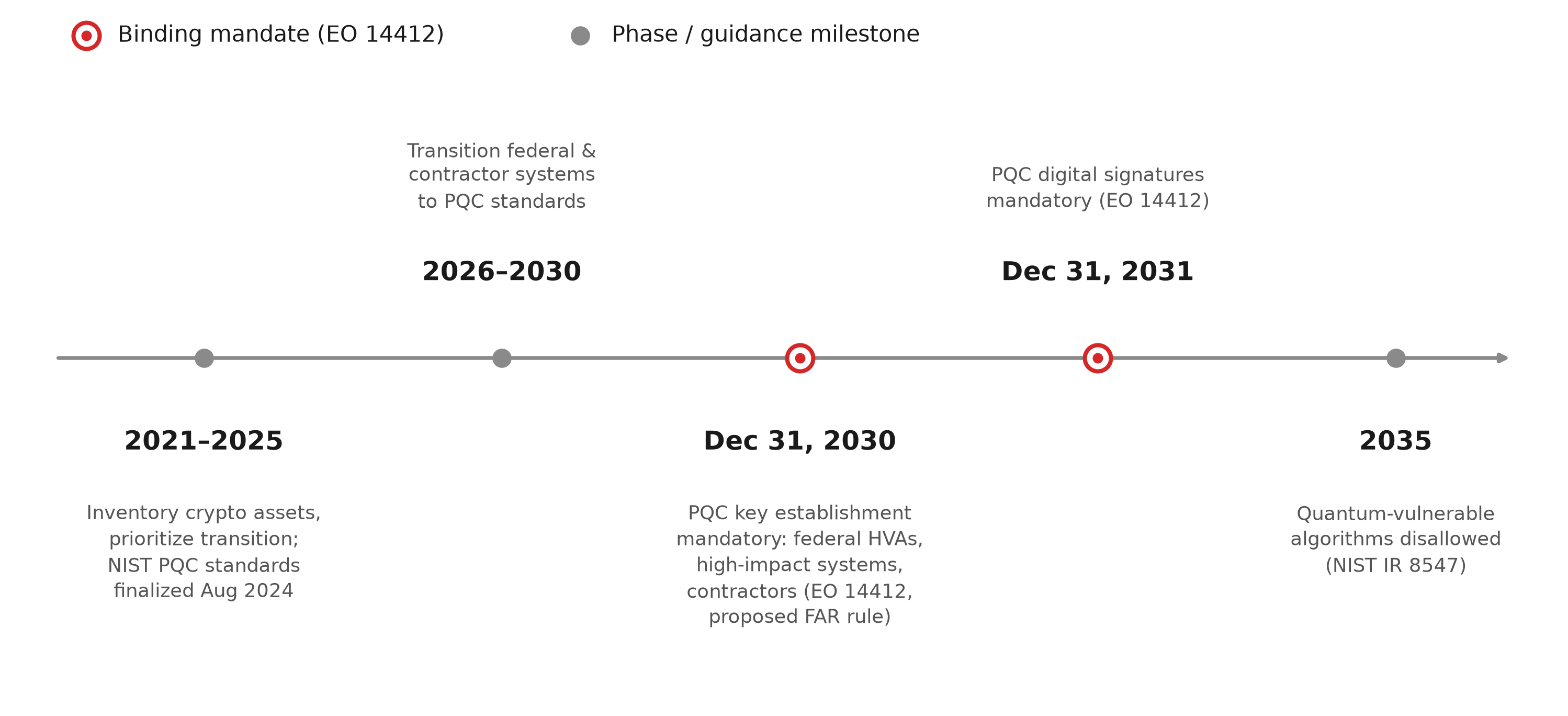}
\par\vspace{3pt}{\small\textbf{Figure 2: Federal PQC Transition Compliance Deadlines}}
\end{figure}
The global efforts to implement PQC in various communication and identification protocols are underway. [7] Regardless of the actual transition deadlines for PQC, there is an open question as to whether implementation of PQC alone will be enough to protect TLS 1.3 from potential cryptographically relevant quantum computer (CRQC) and AI threats. These threats include not only quantum-accelerated AI, ML and DL attacks on the primitives, but also the potential use of AI to enhance both quantum algorithms and quantum computers. [8] Some of the different ways in which this synergistic relationship between AI and quantum-acceleration can impact different aspects of both quantum hardware and software are shown in Figure 3.

\begin{figure}[H]\centering
\includegraphics[width=0.82\linewidth]{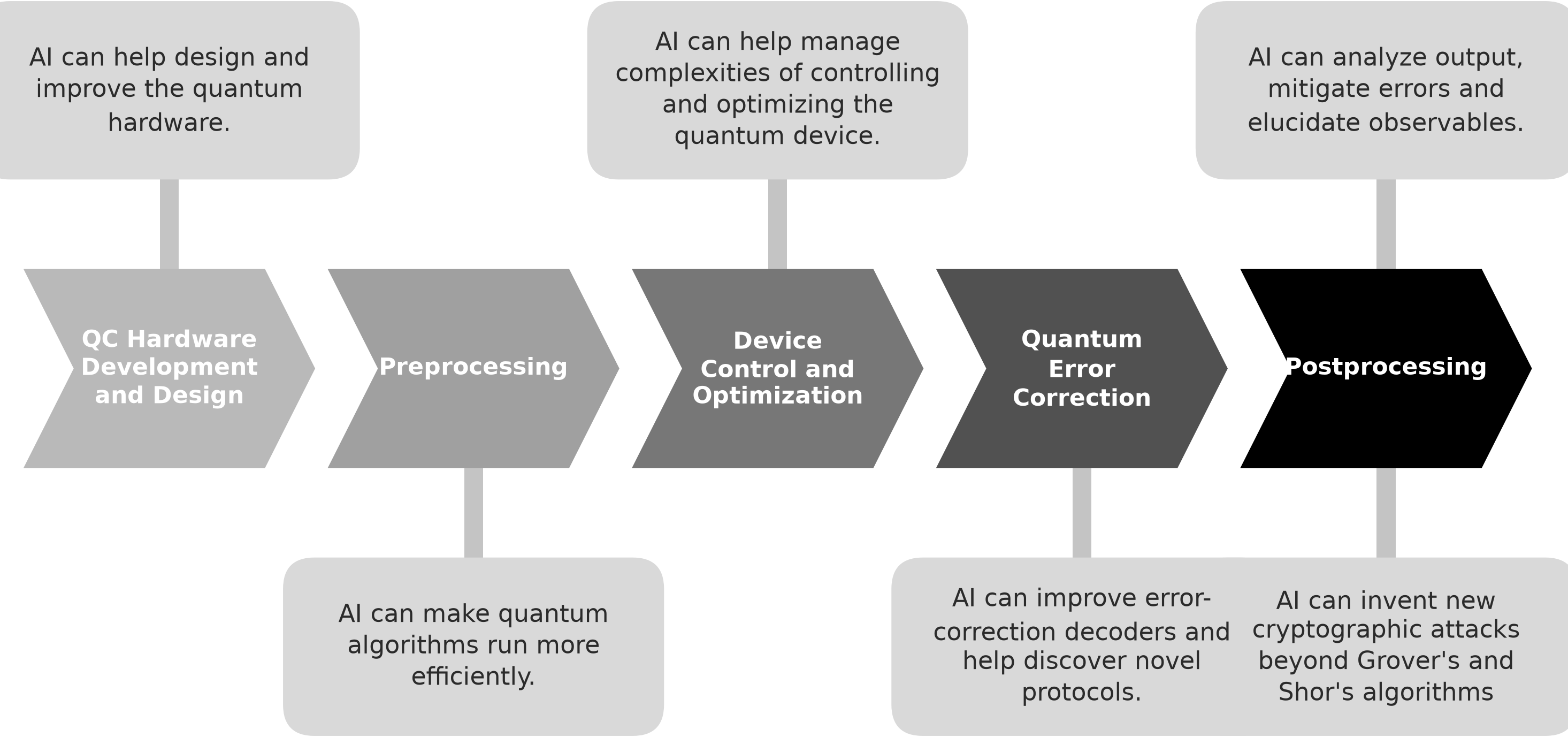}
\par\vspace{3pt}{\small\textbf{Figure 3: Bi-Directional Impact of Quantum and AI Exponential Development}}
\end{figure}
This paper will collectively refer to such vulnerabilities as Cryptographically Relevant Quantum-Accelerated AI and AI-Enhanced Quantum (CRQC+AI) threats. This paper investigates the most recent issues impacting the CRQC+AI vulnerability timelines of PQC adoption as part of the TLS 1.3 protocol and advocates for more research and development of crypto-agility for quantum-resilient hybrid cryptographic frameworks capable of withstanding CRQC+AI attacks and gracefully migrating away from compromised ciphers. [9] [167] [168] [169] [173]

Starting with quantum algorithms, this paper explores cryptographically relevant algorithms that extend beyond the well-known Grover’s algorithm [10] and Shor’s algorithms. [11] At the forefront of these is the Harrow-Hassidim-Lloyd (HHL) algorithm, a quantum algorithm designed to efficiently solve certain systems of linear equations, particularly those involving sparse matrices with low condition numbers. [12] Initially perceived as theoretical, recent advancements in AI suggest that practical implementations of HHL could become reality sooner than previously expected. [13] The PQC encryption algorithms first approved by NIST use lattice-based cryptography. [14] This frequently involves solving linear equations and can be reliant on the Learning with Errors (LWE) problem, [15] which raises the question, examined and substantially qualified in Section 2.1, of whether AI-enhanced quantum linear-algebra methods like HHL could erode the margins of lattice-based cryptography and, with it, PQC as a defense against Harvest-Now, Decrypt-Later (HNDL) attacks. [16]

Other issues this paper investigates in Section 2.2 include a combined quantum-geometric approach to cryptanalysis that integrates CRQC+AI with low-dimensional representations in geometric algebra. Low-dimensional representations in geometric algebra simplify complex cryptographic problems into geometrically accessible forms, enabling CRQC+AI optimization techniques such as quantum annealing or variational quantum eigensolvers as examined in Sections 2.3 and 2.4 to more effectively identify and exploit vulnerabilities within cryptographic schemes, including lattice-based PQC. Numerous additional recent CRQC+AI approaches, including side-channel attacks, are also surveyed in Section 3.

Many quantum hardware limitations continue to be resolved; as Section 2.1 details, however, some of the hurdles facing HHL (notably the density and conditioning of the matrices arising in LWE) are intrinsic to the problem rather than temporary engineering limits. The pairing of CRQC linear-algebra methods such as HHL with the AI geometric simplifications provided by low-dimensional representations in geometric algebra is investigated here as a potential cryptanalytic attack vector against PQC. As Sections 2.1 and 2.2 establish, no mechanism for such a break is presently demonstrated. The HHL approach is considered as a polynomial acceleration effort, and the geometric algebra approach is treated as an exploratory, falsifiable hypothesis rather than an established threat. The paper concludes with an exploration of various author-parameterized, script-generated CRQC+AI vulnerability timelines for both RSA/ECC and lattice-based PQC encryption algorithms in Sections 3 and 4. These sections demonstrate that the CRQC+AI risk profiles are more severe and accelerate sooner than originally projected, underscoring the critical need for immediate action toward crypto-agile hybrid cryptography as a defense.

The contribution of this paper is therefore twofold: (i) an evidence-graded evaluation of the CRQC+AI attack surface against TLS 1.3 that separates mechanism-backed threats (Shor against RSA and ECC) from contingent ones (the lattice schemes, exposed only under a falsifiable structural collapse), and also from hypothesis-only ones (hash-based and symmetric primitives); and (ii) an AI-accelerated timeline model projecting runtime attack estimates and when a 50\% likelihood crossover plausibly occurs for each of these attack surfaces against TLS 1.3 under different scenarios. Of these, the timeline model is the paper’s primary methodological contribution: a reproducible, parameterized scenario-estimation instrument in which the parameters can be varied and the model rerun, and whose explicit update mechanics (Sections 3.3.5 and 5) specify how future observations that follow, or deviate from, the modeled curves are mapped onto revisions of specific parameters, so that successive runs of the model track progressively closer to the accumulating historical record. In this regard, the paper contributes an instrument built to be recalibrated, rather than a one-time forecast, to the quantum resource-estimation work, [79] [82] expert-elicitation timeline studies such as the Global Risk Institute quantum threat reports, [80] [83] and quantitative risk-assessment methodology. The combined findings demonstrate that even the earliest and best-grounded attack surfaces can arrive sooner than the 2030 to 2035 PQC migration window originally assumed. These findings should motivate crypto-agility as a PQC migration strategy, rather than implementation of single or even static-hybrid encryption algorithms as a PQC migration response. To support external scrutiny, the model is reproducible from explicit parameters and pseudocode (Appendix A), accompanied by a parameter sensitivity analysis (Appendix A.4), and paired with an explicit falsification section (Section 5).

Adopting hybrid cryptography and crypto-agile frameworks that can withstand CRQC+AI threats and gracefully migrate away from compromised ciphers is now more essential than ever. Equally critical is intensifying research into resilient cryptography solutions to mitigate against HNDL attacks and extend the CRQC+AI vulnerability timeline for TLS 1.3 well beyond the next decade. All of the authors have some affiliation with EnQuanta, which develops commercial quantum-resilient and crypto-agile cryptographic products; this interest is detailed in the Conflicts of Interest statement.

To keep the evidence base of this analysis explicit, the CRQC+AI threats analyzed in this paper fall into three tiers that are distinguished by how firmly each is grounded in a known mechanism (Table 1A).

\par\vspace{4pt}\noindent\begin{minipage}{\linewidth}
\begin{center}\footnotesize\setlength{\tabcolsep}{4pt}\begin{tabular}{|p{0.22\linewidth}|p{0.22\linewidth}|p{0.22\linewidth}|p{0.22\linewidth}|}\hline
\textbf{\textbf{Tier}} & \textbf{\textbf{Algorithms / primitives}} & \textbf{\textbf{Mechanism status}} & \textbf{\textbf{This paper’s claim}} \\ \hline
Mechanism-backed & RSA, ECC & Shor’s algorithm gives a known polynomial-time quantum attack once sufficient fault-tolerant quantum resources exist & Migration away from RSA and ECC is mandatory \\ \hline
Contingency-backed & ML-KEM (Kyber) ML-DSA (CRYSTALS-Dilithium) & Requires an undiscovered reduction in effective lattice hardness, modeled here as the collapse parameter ρ & Conditional stress scenario, not a known break \\ \hline
Hypothesis-only & SLH-DSA, AES-256 & No current structural attacks: Grover-only analysis is already accounted for & Included only as unknown-unknown risk, not a forecast \\ \hline
\end{tabular}\end{center}
\vspace{11.25pt}\begin{center}\small\textbf{Table 1A: Evidence tiers for the quantum and CRQC+AI threats to TLS 1.3 cryptographic dependencies.}\end{center}
\end{minipage}\par\vspace{4pt}
This paper does not claim a known break of current standardized PQC algorithms in TLS 1.3. It presents a scenario-based stress test of how much quantum hardware progress, AI-assisted engineering, and hypothetical cryptanalytic structure would be required for earlier-than-expected vulnerability compromise. The paper supports a conclusion that the rational response to the CRQC+AI vulnerability spectrum for TLS 1.3 is to encourage crypto-agile replaceable deployment of PQC in a hybrid cryptographic framework.

\par\vspace{4pt}\noindent\begin{minipage}{\linewidth}
\begin{center}\footnotesize\setlength{\tabcolsep}{4pt}\begin{tabular}{|p{0.22\linewidth}|p{0.22\linewidth}|p{0.22\linewidth}|p{0.22\linewidth}|}\hline
\textbf{\textbf{Tier}} & \textbf{\textbf{Confidence a reader may place}} & \textbf{\textbf{Promotes when}} & \textbf{\textbf{Demotes when}} \\ \hline
Mechanism-backed (RSA, ECC via Shor) & High; a standard result & (not applicable) & (not applicable) \\ \hline
Contingency-backed (lattice under ρ-collapse) & Conditional; holds only if the collapse is discovered & a geometric embedding with ρ > 0 is demonstrated, or a sub-exponential or polynomial-time algorithm for the dihedral coset problem survives six months of open verification and composes through a published Module-LWE reduction (Appendix D), which is a change of functional form rather than a value of ρ & the conjecture is refuted, a lower bound closes it, or a no-go theorem narrows the algorithm class, as the 2026 Fourier-label barrier did for one class of dihedral-coset algorithms [230] \\ \hline
Hypothesis-only (SLH-DSA, AES-256) & None as a dated claim & a concrete attack is published & remains undated \\ \hline
\end{tabular}\end{center}
\vspace{11.25pt}\begin{center}\small\textbf{Table 1B: Assurance and promotion/demotion by evidence tier (complements Table 1A).}\end{center}
\end{minipage}\par\vspace{4pt}
The promotion and demotion triggers above follow the structural-claim indicator and its verification logging in Section 3.3.5; they should be read as the conditions under which a reader may raise or lower the confidence attached to each tier, not as predictions that those conditions will be met.

\subsection*{1.1 Scope and Exclusions}
\addcontentsline{toc}{subsection}{1.1 Scope and Exclusions}
In scope. This paper's instrument estimates when five cryptographic primitives, RSA-2048, ML-KEM-768/1024, ML-DSA-44, SLH-DSA and AES-256, as deployed in TLS 1.3, become feasibly attackable under combined quantum and AI acceleration. It measures primitive erosion.

Out of scope. The instrument does not model information technology (IT) organizational estate architecture, inventory, rotation authority or change control, nor does it transfer to other protocols. TLS 1.3 is deliberately the most tractable migration surface in the IT estate: keys are ephemeral and rotatable, algorithms are negotiated, and certificates are short-lived. SSH, IPsec, DNSSEC, and code and firmware signing carry materially worse migration profiles (long-lived or unrotatable keys, trust anchors with decade-plus lifetimes, and size constraints), and the paper's Cloudflare citation, showing over half of human-initiated traffic already post-quantum encrypted, indicates how far ahead TLS already is. Conclusions about primitive erosion in TLS 1.3 therefore represent a best case and do not transfer to those surfaces, or to estate-level recommendations, without the additional analysis in the sections below. Literature cut-off. This version tiers results posted through November 3, 2026; claims posted after that date are not tiered here and wait for the next versioned rerun.

\subsection*{1.2 Approach and Instrument Map}
\addcontentsline{toc}{subsection}{1.2 Approach and Instrument Map}
Approach. Three instruments are used, in this order. First, the channel model (Section 3.3) decomposes engineering progress toward a CRQC into physical-qubit count, gate fidelity, code efficiency and decoder latency, and yields the per-channel times τᵢ and the binding crossover T, calibrated to agree with the hardware budget B(t) of the cost model. Second, the capability model (Appendix A) generates the runtime and feasibility curves from two exponential drivers, one for hardware and one for AGI-gated software. Third, the cost model (Section 4.5) restates the lattice and hash contingencies in explicit cryptanalytic units.

Instrument map. The capability model generates the curves; the cost model and the channel model are checks on it. The cost model's collapse parameter ρ and the capability model's software couplings are two parameterizations of one conjecture, an undiscovered effective-dimension collapse, so agreement among them is consistency, not independent corroboration (see Section 3.3.4). What was done, in what order, on what evidence, and what was excluded is stated in the Scope section above.

\section*{2. Background}
\addcontentsline{toc}{section}{2. Background}
This section reviews the CRQC+AI techniques that bear on the attack pathways analyzed later in the paper. It includes methods such as quantum annealing and the variational quantum eigensolver not because they break standardized cryptography on their own (the analysis below concludes they do not), but because establishing what does not contribute to a near-term break is as important to the argument as identifying what might.

\subsection*{2.1 Harrow-Hassidim-Lloyd Background}
\addcontentsline{toc}{subsection}{2.1 Harrow-Hassidim-Lloyd Background}
The Harrow-Hassidim-Lloyd (HHL) algorithm is a quantum algorithm designed to solve systems of linear equations exponentially faster than classical methods under certain conditions. This efficiency arises when dealing with sparse matrices that have low condition numbers, meaning the ratio of the largest to smallest eigenvalues is small.

Lattice-based encryption schemes, such as those based on the LWE problem, rely on the computational hardness of certain lattice problems, which can often be reduced to solving linear equations. The rapid advancement of quantum algorithms such as HHL and the linear combination of Hamiltonian simulation (LCHS) technique introduces a growing area of concern for these cryptographic systems. [17] The most prominent recent attempt in this direction, a 2024 claim of a polynomial-time quantum algorithm for LWE, was withdrawn by its own author within days after a bug was identified, which underscores that no quantum attack has yet overturned lattice hardness. [18] If quantum hardware evolves to efficiently implement the HHL algorithm, it could accelerate the linear-algebra steps underlying lattice problems; whether that amounts to a threat to lattice-based encryption is the question examined below.  

Quantum hardware barriers such as error rates and limited scalability are easing steadily, with advances in error correction and hybrid quantum-classical techniques narrowing the gap between theory and practice. [19] [20] Not every obstacle facing HHL is of this kind, however. The requirement for sparse, well-conditioned matrices is a property of the problem instance rather than of the hardware, and as the formal analysis below shows, it is precisely where lattice problems fail to cooperate.

Formally, for an N × N linear system Ax = b with sparsity s and condition number κ, solved to precision ε, the HHL algorithm and its near-optimal refinement by Childs, Kothari, and Somma [21] run in time:

\begin{center}T\textsubscript{HHL} = Õ(s · κ · polylog(N/ε))     [near-optimal in ε]\end{center}
\begin{center}T\textsubscript{HHL} = Õ((log N) · s² · κ² / ε)     [original HHL]\end{center}
The decisive structural features are that the runtime is polylogarithmic in the dimension N but polynomial in the sparsity s and the condition number κ, and that the output is a quantum state |x⟩ proportional to the solution rather than the classical solution vector. Useful information is obtained by estimating an observable ⟨x|M|x⟩; recovering the individual solution coordinates removes the exponential advantage. For LWE the relevant system matrices are effectively dense (s on the order of N) and not well-conditioned, and a key-recovery attack must read the secret out coordinate by coordinate. In addition, LWE is not an exact linear solve but a bounded-distance decoding problem (recovering the secret from As + e = b with small noise e), so the clean Ax = b formulation that HHL assumes does not directly apply. An unmodified HHL therefore does not by itself break LWE; its role, together with the LCHS technique for non-unitary dynamics, is confined to accelerating polynomial factors of an inner linear-algebra step. The cost model of Section 4.5 makes this boundary explicit. As with the annealing and variational methods discussed below, HHL is treated here as a candidate attack capability to be evaluated on its merits rather than assumed. The finding that it contributes at most polynomial acceleration to an inner step, rather than a standalone break, is reported as deliberately as any positive result and forms part of the due-diligence basis for the timeline analysis that follows.

\subsection*{2.2 The Geometric-Algebra Hypothesis (Contingency Tier)}
\addcontentsline{toc}{subsection}{2.2 The Geometric-Algebra Hypothesis (Contingency Tier)}
The intersection of CRQC+AI and low-dimensional representations in geometric algebra presents intriguing theoretical pathways to breaking both current cryptographic systems and PQC encryption. [22] Unlike the methods in Sections 2.1, 2.3, and 2.4, which rest on established algorithms with known performance bounds, the geometric-algebra approach is presented as an exploratory hypothesis rather than a demonstrated capability: no mechanism is presently known by which it yields a concrete cryptanalytic advantage. It is included, and evaluated, as a candidate direction precisely so that its speculative status is explicit and falsifiable rather than assumed, on the same due-diligence basis applied to the other methods in this section.

\subsubsection*{2.2.1 Quantum AI-Driven Optimization of Cryptanalysis}
\addcontentsline{toc}{subsubsection}{2.2.1 Quantum AI-Driven Optimization of Cryptanalysis}
The first idea is to apply quantum optimizers such as quantum annealing and the variational quantum eigensolver, treated in Sections 2.3 and 2.4 below, to a geometric encoding of a cryptographic problem, in the hope that a low-dimensional geometric form exposes structure that an unstructured search would miss. The limitation carries over from those sections: such optimizers carry no demonstrated scaling advantage on unstructured problems, so this step helps only if the geometric encoding itself supplies exploitable structure, which is the open question rather than a settled result.

\subsubsection*{2.2.2 Geometric Algebraic Representation of Cryptographic Spaces}
\addcontentsline{toc}{subsubsection}{2.2.2 Geometric Algebraic Representation of Cryptographic Spaces}
The load-bearing idea is representational. If a lattice or its underlying linear system could be embedded in a low-dimensional geometric algebra, operations in that space might expose linear dependencies or simplifications that are invisible in the standard basis. [23] Everything else in this section depends on whether such an embedding exists and is efficiently computable, and no construction achieving this against standardized lattices is presently known.

\subsubsection*{2.2.3 A Conjectured Quantum-Geometric Pathway Against PQC}
\addcontentsline{toc}{subsubsection}{2.2.3 A Conjectured Quantum-Geometric Pathway Against PQC}
Our precise conjecture is that such an embedding would shrink the lattice’s effective block size β, the parameter that sets the cost of sieving, an effect captured by the collapse fraction ρ defined at the close of this section and quantified in Section 4.5. This is the single quantity on which any geometric quantum break of lattice PQC would rest; this paper states how large ρ would need to be rather than asserting that value is attainable.

\subsubsection*{2.2.4 Geometrically Inspired Quantum AI Algorithms}
\addcontentsline{toc}{subsubsection}{2.2.4 Geometrically Inspired Quantum AI Algorithms}
A more speculative variant uses quantum machine learning over the geometric representation to learn structural weaknesses from data rather than derive them analytically. It shares the trainability and classical simulation limits discussed for the variational methods in Section 2.4 and has no demonstrated application to standardized PQC. It is noted for completeness, not advanced as a near-term threat.

\subsubsection*{2.2.5 Parameterized Geometry as an Attack Vector}
\addcontentsline{toc}{subsubsection}{2.2.5 Parameterized Geometry as an Attack Vector}
Finally, large AI models could in principle search systematically over geometric representations and their parameters, an automated hunt for an embedding that yields a nonzero ρ. This describes how such an attack might be discovered, not evidence that it exists. Absent a known mechanism, the parameterized geometry search has no established target.

In summary, the geometric-algebra direction is a single hypothesis presented in several forms: namely, that some low-dimensional embedding could expose structure in a lattice that classical algebra does not. Its entire cryptanalytic force reduces to whether such an embedding can shrink the effective hardness of the lattice, and no construction achieving that against standardized schemes is currently known. The section is retained not as evidence of a break, but rather as an explicit, falsifiable hypothesis, stated precisely enough in the paragraph that follows to be tested or discarded.

To make this hypothesis precise enough to test, this paper represents the conjectured effect of a low-dimensional geometric embedding as a reduction of the operative lattice block size, written β → β(1 − ρ) for a collapse fraction ρ ∈ [0, 1). No mechanism is presently known that achieves ρ > 0 against the random q-ary lattices underlying standardized LWE; ρ is introduced here as an explicit, falsifiable parameter rather than an established result. Section 4.5 quantifies the value of ρ that the modeled 2031 to 2034 lattice crossover vulnerabilities would require. A second structural pathway, the dihedral-coset route through which a surviving polynomial-time quantum algorithm for the dihedral coset problem would reach Module-LWE, is set out in Appendix D. It is not a value of ρ but a change of functional form, and the promotion trigger of Table 1B is written to catch either.

\subsection*{2.3 Quantum Annealing Background }
\addcontentsline{toc}{subsection}{2.3 Quantum Annealing Background }
By way of framing, quantum annealing is presented here as background: the analysis in Section 3.3.5 concludes that it does not, on its own, yield a standalone cryptanalytic break of the schemes considered in this paper. It is surveyed as part of the adjacent quantum-optimization landscape, alongside the variational methods of Section 2.4, rather than as a demonstrated attack path.

Quantum annealing is a specialized quantum computing technique used to solve complex optimization and combinatorial problems by leveraging quantum superposition, entanglement, and tunneling. [24] It is based on the principles of adiabatic quantum computation (AQC). [25] A quantum system implementing AQC is initialized in the ground state of an easy-to-prepare Hamiltonian [26] and then slowly evolved into a Hamiltonian that encodes the solution to a given optimization problem.

\subsubsection*{2.3.1 Key Aspects of Quantum Annealing}
\addcontentsline{toc}{subsubsection}{2.3.1 Key Aspects of Quantum Annealing}
\paragraph*{2.3.1.1 Energy Minimization Approach}
\addcontentsline{toc}{subparagraph}{2.3.1.1 Energy Minimization Approach}
Quantum annealing works by finding the lowest energy state (global minimum) of an objective function mapped onto a quantum system, commonly represented as a mathematical optimization problem.

Concretely, the objective is encoded in a problem Hamiltonian H\textsubscript{P} whose ground state is the optimal solution. In Ising form,

\begin{center}H\textsubscript{P} = Σ\textsubscript{i} h\textsubscript{i}σ\textsubscript{i}\textsuperscript{z} + Σ\textsubscript{i < j} J\textsubscript{ij}σ\textsubscript{i}\textsuperscript{z}σ\textsubscript{j}\textsuperscript{z}\end{center}
or equivalently as a quadratic unconstrained binary optimization (QUBO) problem, the minimization over x ∈ \{0,1\}\textsuperscript{n} of x\textsuperscript{T} Q x. Cryptanalytic search problems are attacked by mapping them into this form.

\paragraph*{2.3.1.2 Quantum Tunneling}
\addcontentsline{toc}{subparagraph}{2.3.1.2 Quantum Tunneling}
Unlike classical annealing methods (such as simulated annealing) that rely on thermal fluctuations to escape local minima, quantum annealing can leverage quantum tunneling to explore multiple solutions simultaneously, potentially leading to a faster convergence.

The advantage is sharpest for tall, narrow barriers: the quantum tunneling rate through a barrier of height h and width w scales as exp(−w√h), whereas the thermal hopping used by classical simulated annealing scales as exp(−h / k\textsubscript{B}T). Where the energy landscape has thin barriers, tunneling can traverse them at a rate the thermal process cannot match.

\paragraph*{2.3.1.3 Hamiltonian Evolution}
\addcontentsline{toc}{subparagraph}{2.3.1.3 Hamiltonian Evolution}
The system starts with a simple Hamiltonian whose ground state is known. Over time, this Hamiltonian is gradually transformed into a more complex problem-specific Hamiltonian while ideally maintaining the ground state.

Formally the system follows a time-dependent interpolation between an easy initial Hamiltonian H\textsubscript{0} (with H\textsubscript{0} = −Σ\textsubscript{i}σ\textsubscript{i}\textsuperscript{x}, a transverse field whose ground state is an equal superposition) and the problem Hamiltonian H\textsubscript{P}:

\begin{center}H(s) = (1 − s)·H\textsubscript{0} + s·H\textsubscript{P},     s = t/T,     t ∈ [0, T]\end{center}
By the adiabatic theorem the system stays in the ground state only if the evolution is slow relative to the inverse square of the minimum spectral gap:

\begin{center}T ≳ max\textsubscript{0 ≤ s ≤ 1} |⟨E\textsubscript{1}(s)|dH/ds|E\textsubscript{0}(s)⟩| / g\textsubscript{min}\textsuperscript{2},     g\textsubscript{min} = min\textsubscript{s} [E\textsubscript{1}(s) − E\textsubscript{0}(s)]\end{center}
The runtime is therefore governed by g\textsubscript{min}. For benign landscapes the gap stays open and annealing is fast, but for the hardest combinatorial and cryptanalytic instances the gap can close exponentially in the problem size n, in which case the required annealing time grows exponentially, and no quantum speedup is obtained. This gap-dependence is the precise reason quantum annealing offers advantage on some structured problems while providing no general guarantee against well-designed cryptography.

\paragraph*{2.3.1.4 Quantum Superposition \& Entanglement}
\addcontentsline{toc}{subparagraph}{2.3.1.4 Quantum Superposition \& Entanglement}
The qubits in a quantum annealer exist in a superposition of states and can be entangled, allowing for parallel exploration of multiple solutions.

\subsubsection*{2.3.2 Specialized Hardware}
\addcontentsline{toc}{subsubsection}{2.3.2 Specialized Hardware}
Companies like D-Wave Systems have developed quantum annealers that use superconducting qubits for optimization problems. [27] Unlike universal quantum computers, which perform arbitrary quantum algorithms, quantum annealers are primarily tailored for solving specific optimization and machine learning tasks.

\subsubsection*{2.3.3 Applications of Quantum Annealing}
\addcontentsline{toc}{subsubsection}{2.3.3 Applications of Quantum Annealing}
Quantum annealing is applied wherever a problem can be cast as energy minimization over a discrete configuration space, typically by encoding it as a quadratic unconstrained binary optimization (QUBO) or Ising instance and letting the annealer relax toward a low-energy state. [28] The three domains below are the most developed. Practical, if still modest, value has appeared in the first two; the third is the domain most relevant to this paper and, at present, also the most speculative. We treat annealing, like the variational eigensolver methods in Section 2.4, as a candidate attack capability to be evaluated on its merits rather than assumed. We report the negative findings as deliberately as the positive ones, because ruling a method out is itself part of the due-diligence basis for the timeline analysis that follows.

\paragraph*{2.3.3.1 Optimization Problems}
\addcontentsline{toc}{subparagraph}{2.3.3.1 Optimization Problems}
Logistics, scheduling, portfolio optimization, traffic flow management.

These are the workloads quantum annealing was built for, and they share a common shape: a combinatorial objective with many competing constraints that maps naturally onto a QUBO or Ising cost function. Vehicle routing and traffic flow have been the most visible pilots, including Volkswagen’s traffic-routing demonstrations, [29] while portfolio optimization recasts the Markowitz mean-variance objective as a binary selection problem, [30] and scheduling problems (job-shop, crew, and gate assignment) translate their constraints into penalty terms in the Hamiltonian. [31] The practical caveat is that a clear advantage over mature classical heuristics (simulated annealing, tabu search, and modern mixed-integer solvers) has not been established, even in these pilots. Problems where quantum annealing helps tend to be on landscapes with tall, narrow energy barriers that favor tunneling over thermal hopping, and the benefit is instance-specific rather than universal. Embedding the logical problem onto the fixed hardware graph (e.g., the minor-embedding step on the Chimera and Pegasus topologies) also consumes physical qubits rapidly, which currently bounds the size of problems that can be run natively.

\paragraph*{2.3.3.2 Machine Learning}
\addcontentsline{toc}{subparagraph}{2.3.3.2 Machine Learning}
feature selection, clustering, and deep learning acceleration.

In machine learning the annealer is used in two ways: as an optimizer and as a sampler. As an optimizer it handles discrete subproblems such as feature selection, balanced clustering, and the training of structured models, each expressed as a QUBO. [32] These are proof-of-concept demonstrations rather than evidence of advantage; the early deep-network training results, for example, were small and coarse-grained and were subsequently matched by classical methods. As a sampler it draws low-energy configurations that approximate a Boltzmann distribution, which has been used to train restricted and quantum Boltzmann machines and to seed generative models. [33] This sampling role carries an important qualification: the hardware does not produce a clean quantum Boltzmann distribution but something closer to a classical distribution at an unknown, instance-dependent effective temperature, distorted by freeze-out during the anneal, and the quantum Boltzmann machine itself remains largely a theoretical and simulation construction rather than a demonstrated hardware advantage. The appeal is that sampling from complex distributions is expensive classically, so a hardware sampler could in principle accelerate part of a learning pipeline. The same quantum hardware limits apply, however: restricted qubit counts, sparse connectivity, embedding overhead, and analog control noise keep demonstrations small. On standard benchmarks, classical machine learning remains dominant over currently known annealer implementations. For the threat model of this paper the relevance is indirect: annealing-assisted learning could speed components of a cryptanalytic pipeline, for example extracting structure from side-channel traces, but it acts as a contributing tool rather than a standalone capability.

\paragraph*{2.3.3.3 Cryptography \& Cybersecurity}
\addcontentsline{toc}{subparagraph}{2.3.3.3 Cryptography \& Cybersecurity}
Enhancing cryptographic algorithms and security models.

On the offensive side, cryptanalysis can be recast as optimization and handed to an annealer. Integer factorization is the canonical example: writing N = p · q and minimizing (N − p · q)² over the binary digits of the factors turns factoring into a QUBO, and small semiprimes (numbers ranging from the hundreds up to the hundreds of thousands) have been factored this way on annealing hardware, usually with substantial classical preprocessing. [34] Lattice problems have received similar treatment, with the shortest-vector and closest-vector problems, and by extension LWE, formulated as Ising or QUBO instances for both annealers [35] and quantum enumeration variational solvers [36]. These are existence results rather than evidence of scaling; the gate-model lattice work, for instance, solves only a minority of instances even in small dimensions. In every case the same two limits recur: the qubit count and connectivity required grow rapidly with problem size. There is no scaling advantage over the best classical attacks (the general number field sieve for factoring, lattice reduction for LWE) that has been demonstrated at or near cryptographic parameters. Current analyses of RSA-scale factoring on a D-Wave annealer document the wall rather than a breach. [37] This is consistent with the gap-dependence established in Section 2.3.1.3: because the minimum spectral gap can close exponentially for hard, well-structured instances, annealing carries no general guarantee against properly parameterized cryptography. On the defensive side, the same optimization machinery is applied constructively, to network-security resource and sensor placement, intrusion-detection feature selection, anomaly detection, and attack-graph and vulnerability analysis. The honest summary, and the one this paper adopts, is that quantum annealing is a component of the broader CRQC+AI optimization toolkit rather than a standalone route to breaking standardized cryptography.

In summary, while quantum annealing is promising for solving certain classes of problems, it does not provide exponential speedups for all computational tasks. It has shown instance-specific promise in real-world optimization and as a contributing tool in AI-related tasks, and no advantage against standardized cryptography.

\subsection*{2.4 Variational Quantum Eigensolver Background}
\addcontentsline{toc}{subsection}{2.4 Variational Quantum Eigensolver Background}
The Variational Quantum Eigensolver (VQE) is a hybrid quantum-classical algorithm designed to find the ground state energy of a quantum system, particularly useful for solving problems in quantum chemistry and materials science. [38] It combines quantum computing with classical optimization to efficiently approximate the lowest eigenvalue (ground state energy) of a given Hamiltonian.

\subsubsection*{2.4.1 How VQE Works}
\addcontentsline{toc}{subsubsection}{2.4.1 How VQE Works}
VQE parameterizes a quantum circuit, the ansatz U(θ), and optimizes it with a classical loop. The target Hamiltonian is written as a weighted sum of Pauli operators, H = Σₖ cₖ Pₖ, a form the quantum device can measure directly. The ansatz prepares a parameterized state |ψ(θ)⟩ = U(θ)|0⟩, and the variational principle guarantees that its energy expectation E(θ) = ⟨ψ(θ)|H|ψ(θ)⟩ is bounded below by the true ground-state energy E₀, with equality when the ansatz reaches the ground state. The device assembles ⟨H⟩ from the separately measured Pauli terms, a classical optimizer minimizes E(θ) over θ, obtaining gradients directly on hardware through the parameter-shift rule, and the loop repeats until the energy converges.

Two limitations bound this approach in the cryptanalytic setting. The optimization landscape is subject to barren plateaus: for many ansatz families the gradient ∂E/∂θ vanishes exponentially in the qubit count, so the variational search becomes effectively untrainable at the scale a cryptographic problem would demand. [39] More fundamentally, VQE is a ground-state estimation method for physical and chemical Hamiltonians and provides no established reduction from integer factoring, discrete logarithm, or lattice problems. It is surveyed here as part of the quantum-optimization landscape, not as a demonstrated route to breaking standardized cryptography.

\subsubsection*{2.4.2 Key Advantages of VQE}
\addcontentsline{toc}{subsubsection}{2.4.2 Key Advantages of VQE}
VQE’s appeal on near-term hardware rests on three properties [40]. First, it is hybrid-efficient: the quantum processor runs only short, shallow circuits to prepare the trial state and measure Pauli expectations, while the classical optimizer carries the search, so the hardware does only what is classically hard. [38] [41] Second, it is comparatively noise-resilient, because the variational principle tolerates the small state-preparation and readout errors that would corrupt quantum phase estimation, and dedicated error-mitigation techniques can extend its usable reach on noisy devices [42] [43]. Third, the ansatz is a free design choice, ranging from hardware-efficient circuits matched to a device’s native gates, [44] to adaptive, problem-tailored constructions that grow the circuit only as needed. [45] These same properties bound its cryptanalytic power: the shallow circuits and heuristic optimization that make VQE practical, together with the trainability limits noted above [46], are also what keep it from performing the deep, structured computation that an attack on factoring, discrete-logarithm, or lattice problems would require.

In summary, VQE is one of the most promising algorithms for Noisy Intermediate-Scale Quantum (NISQ) devices, offering a practical approach to leveraging quantum computation before fully error-corrected quantum computers become viable. [47]

\section*{3. CRQC+AI Vulnerability Timelines}
\addcontentsline{toc}{section}{3. CRQC+AI Vulnerability Timelines}
\subsection*{3.1 Current Timeline Projections}
\addcontentsline{toc}{subsection}{3.1 Current Timeline Projections}
The discussion of timelines in this section should be understood as the authors’ best current estimates based on the factors discussed. In presenting these timelines, we acknowledge our inherent limitations to accurately estimate timelines of future technologies. Humans have always been challenged in making predictions as our abilities to predict are rooted in linear projections, not the exponential growth of accelerating change, as Kurzweil argued in The Singularity Is Near (2005).

Current timeline projections for when RSA, ECC, and similar factoring- and/or elliptic-curve-based public key encryption algorithms need to be replaced by lattice-based PQC encryption algorithms have focused primarily on the growth of quantum computing processing power evaluated in terms of numbers of logical gates in conjunction with the use of Shor’s and Grover’s quantum algorithms. [49] The approach used by NIST with respect to timelines for when new generations of encryption solutions should be adopted in the face of threats from quantum computers is based on the principles of Mosca’s Theorem. [50] This approach correlates the intended duration of data protection and the length of time needed for an organization to implement a new encryption solution in view of the potential for HNDL attacks. [51]

Mosca expresses this balance as a simple inequality. Let X be the length of time the data must remain confidential, Y the time an organization needs to migrate to new cryptography, and Z the time until a cryptographically relevant quantum computer exists. Then whenever

\begin{center}X + Y > Z\end{center}
the organization is already exposed, because data harvested today under HNDL attacks may still be sensitive by the time the capability to decrypt it arrives. The practical force of the inequality is that it does not require a known value of Z: any statistically significant probability of a break within the secrecy horizon X is sufficient to make migration and crypto-agility rational today.

NIST's general expectation, though never stated as a formal prediction, was that PQC encryption would serve as a replacement for RSA and similar algorithms for multiple decades after RSA encryption is compromised. [52] Unfortunately, even as an informal expectation this durability now appears more wishful than analytical. Importantly, none of these expectations, nor the formal projections for RSA encryption itself, accounted for the accelerating changes of CRQC+AI seen in the last few years and the potential for AGI in the next few years. [54] [55] [56] [57]

The current approach to timeline projections for these CRQC+AI vulnerabilities is sometimes referred to as Q-Day, which analogizes to Y2K and assumes the single-point deadline based on a single-point failure. Google recently bumped its PQC transition time frame goals from 2030 to 2029. [59] As noted by IBM in a recent article, however, the better approach is not to assume Q-Day is a single day for a single event, but rather to model a spectrum of risks that will have a range of dates for ranges of risks across a range of breaches across a spectrum of different types of cryptographic algorithms. [60] More recent IBM articles characterize the challenge presented by the PQC transition as the most significant computing revolution in 60 years and as the digital equivalent of needing to change the locks and keys on every single door in the world. [61] [173]

Accordingly, the CRQC+AI spectrum projections below are model-generated scenario estimates, not empirical measurements. The projections are based on a range of factors that form a CRQC+AI risk band plotted over time between a lower bound based on linear extrapolation of potential CRQC improvement based on known quantum and AI developments and an upper bound based on an exponential extrapolation of potential CRQC+AI improvements based on an assumption that there is a meaningful probability that AGI can be achieved by the end of 2028. [62]

\subsubsection*{3.1.1 Historical Analogues for the PQC Transition: the 2008 Financial Crisis and Y2K}
\addcontentsline{toc}{subsubsection}{3.1.1 Historical Analogues for the PQC Transition: the 2008 Financial Crisis and Y2K}
The appropriate historical analogue for a CRQC+AI cryptographic break is not Y2K alone. Y2K was a contained engineering problem: a single, calendar-fixed deadline whose cost was remediation, largely completed before the date passed. A cryptographic break has a different character. It is systemic, cross-institutional and confidence-driven: certificate trust, key establishment and signed software underpin essentially every digital relationship at once, and traffic harvested today can be decrypted retroactively long after the fact. The closer analogue for this kind of systemic failure mode is the 2008 financial crisis, in which the damage was dominated not by the direct cost of the failing instruments but by contagion, the collapse of counterparty confidence, and cross-institutional propagation. The systemic estimate this paper itself adopts, the Citi and Hudson Institute figure for a single-day attack on one major bank's payment-rail access, is of that order rather than of Y2K's remediation history. This section therefore leads with the 2008 analogue for the scale and character of the exposure and retains the Y2K comparison for the narrower and still-useful lesson it teaches, namely how a fixed, unpostponable deadline concentrates remediation, a property the PQC transition largely lacks.

Systemic and economic stakes. The scale implied by this analogue is not hypothetical. Independent modeling by the Hudson Institute (Butler, Prosperity at Risk, April 2023) [172], adopted in the Citi Institute's 2026 analysis [70], estimates the direct cost of a single-day quantum-enabled attack on one top-five U.S. bank's Fedwire access at roughly \$730 billion to \$1.95 trillion, and the indirect impact, the contagion cost measured as GDP-at-risk over a resulting recession, at \$2.0 to \$3.3 trillion, about 10 to 17 percent of GDP. The gap between the two is the point: as in 2008, the dominant cost is not the direct failure but the systemic contagion that follows a loss of confidence in the financial plumbing. This is a single modeled scenario on a single payment rail, offered as an order-of-magnitude indication rather than a comprehensive economic estimate; the sector-by-sector variation in exposure is taken up in Section 6.2.

Unlike the Y2K threat at the end of 1999 that was a single weakness based on a single critical date, the CRQC+AI threats underlying the PQC transition represent a broad spectrum of threats and risks. Even though these risks are orders of magnitude greater than risks tolerated for other kinds of high-impact fault events like nuclear power plant meltdowns, many of those who will be most impacted by these risks have been slow to realize, plan for, and implement the recommended strategies NIST has promulgated over the last decade. [63] The contrast in tolerated risk is stark: modern nuclear reactor designs are held to a core damage frequency approaching one in ten million per reactor year, [64] while the chief executive of a leading AI laboratory has publicly put the probability that advanced AI development goes catastrophically wrong at roughly one in four. [65] The two different kinds of threat risks are not measured on the same basis, yet the tolerated risk gap between a one in ten million engineering target for nuclear risks and a one in four expert estimate for catastrophic AI outcomes underscores how unevenly these risks are treated. This also explains why awareness and response for cyber risks follow an exponential pattern, with slow leadups to a dramatically increasing push as a deadline looms, or, even worse, after a crisis occurs.

\begin{figure}[H]\centering
\includegraphics[width=0.82\linewidth]{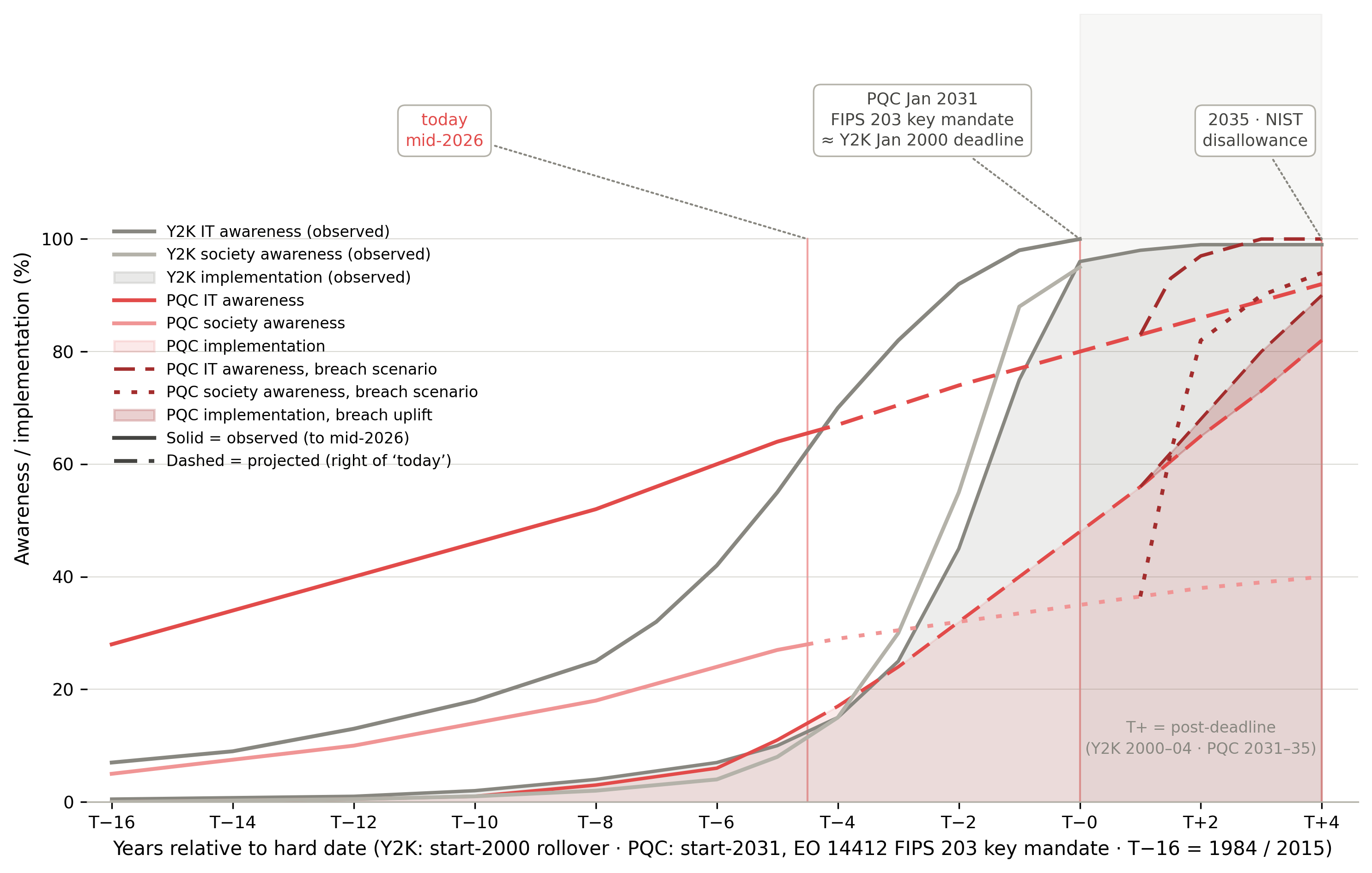}
\par\vspace{3pt}{\small\textbf{Figure 4A: Comparison of Awareness and Response for Y2K vs. PQC Transition}}
\end{figure}
Figure 4A compares awareness and response for Y2K versus PQC transition on the relative scale of corresponding 20-year timelines. The starting dates at T-16 correspond to the equivalent historical origins for Y2K (1984) [66] and PQC (2015). [67] The similarity in both the public awareness and implementation curves as of today (mid-2026 at T-4.5) provides historical credibility for this comparison, even though both PQC awareness curves have led their Y2K equivalents up to this point. In comparing these two threat response timelines, it is important to note that the Y2K portions are entirely historical, whereas the PQC portions after today (mid-2026) are projections whose milestone calibration was drafted with the Anthropic Claude Fable 5.1 AI model in July 2026 from various government reports and expert projections, and which are rendered by the deterministic script of Appendix A.3. [5] [51] [68] [69] [70] [71] [72] [73] The figure is rendered from a deterministic generation script so that the curves can be regenerated and audited; curve values are illustrative reconstructions calibrated to documented milestones and point-in-time survey data, not measured time series. The Y2K awareness curves are truncated at T-0 for legibility. It should also be noted that public attention to Y2K threats collapsed to roughly 20\% after the uneventful rollover. It is also interesting to note that the PQC projections without a publicly disclosed breach scenario match our human inclination toward linear projections, whereas historical Y2K responses and the publicly disclosed breach scenario (est. in 2032) match the technological curve of increasing exponential growth. Although the two threats share a similar 20-year arc that spans early insider warnings, mid-course disclosure and standardization of responses, and endgame sprints, the PQC transition differs from Y2K in four decisive respects.

\paragraph*{3.1.1.1 Deadlines.}
\addcontentsline{toc}{subparagraph}{3.1.1.1 Deadlines.}
The Y2K deadline of T-0 was an entirely calendar-driven deadline (Jan. 1, 2000) that could not be postponed, which drove remediation to effective completion. The PQC 2031 transition deadline is entirely regulatory and is chosen by fiat as a single date (Jan. 1, 2031), which is the day after the effective date of the Dec. 31, 2030, EO 14412 mandate for FIPS 203 (ML-KEM) key establishment. [5] This single date is chosen as the best reasonable approximation of the single Y2K date across what is a broad spectrum of both regulatory deadlines from different governmental organizations and risk profile dates for RSA-2048 as presented in Section 4. The harder PQC cutoff is 2035, when NIST disallows classical RSA/ECC algorithms, which puts the estimated publicly disclosed breach scenario (est. in 2032) inside and in the middle of the expert-estimated CRQC+AI threat emergence window of 2029 to 2035.

\paragraph*{3.1.1.2 Awareness.}
\addcontentsline{toc}{subparagraph}{3.1.1.2 Awareness.}
Y2K produced genuine mass concern (>85\% of the public in 1999), which created bottom-up pressure on institutions. [74] PQC may not attract a similar mass concern. [75] Public recognition of the PQC cryptographic threat is approximately 25 to 30\% today, with the threat effectively invisible to roughly 3 out of 4 ordinary people [75]. This means that pressure for remediation must be driven primarily by regulation and will not be driven by public recognition until after there is an actual publicly disclosed breach.

\paragraph*{3.1.1.3 Breach.}
\addcontentsline{toc}{subparagraph}{3.1.1.3 Breach.}
The breach scenario for RSA-2048 is shown in Figure 4A as a reasonable example with two important caveats: (i) it is an estimated projection, in this case based on an approximately 75\% modeled feasibility estimate from Figure 6B at a projected attack runtime of between a day and a month per Figure 6A; and (ii) as noted in the Citi report [70] it assumes that a breach of RSA-2048 would become publicly disclosed as of or shortly after the breach occurred, which is an unlikely assumption if the breach was the result of actions by a state-actor. It must also be noted that even a Y2K-1999-scale emergency response to a publicly disclosed breach does not reach 100\% PQC implementation by 2035. Unlike Y2K, the scope and pervasiveness of systems requiring PQC upgrades (protocols, PKI roots, HSMs, embedded devices, and long-lived OT hardware) make it unlikely without significant new innovations that all these systems could be breach-migrated in just three years. In addition, the uplift in implementation assumes that it is publicly disclosed. A covert-breach variant from the Citi report would leave all three dark-red curves collapsed onto their baselines, which is the report's most severe case. The stakes are material: the Citi Institute analysis, drawing on Hudson Institute modeling from 2023, places the GDP at risk from a single-day quantum-enabled attack on the Fedwire access of one of the top five U.S. banks at \$2.0 to \$3.3 trillion. [70]

\paragraph*{3.1.1.4 Cost.}
\addcontentsline{toc}{subparagraph}{3.1.1.4 Cost.}
The cost of remediation efforts for Y2K was in the hundreds of billions-dollar scale: \textasciitilde{}\$100B in the U.S. alone and an estimated \$300B to \$500B+ globally. [74] [76] Direct spending forecasts for the commercial PQC market (software, hardware, and services) are substantially smaller than the likely total migration cost. Future Market Insights projects the PQC migration market to grow from approximately \$1.9B in 2025 to \$12.4B by 2035. [77] Large-scale migration, however, requires inventory discovery, software redevelopment, certificate replacement, testing, compliance, hardware refresh, and operational deployment across virtually every digitally connected enterprise: Moody's estimates that PQC migration may consume roughly 2.5\% of enterprise IT budgets, [78] while Citi characterizes the transition as a trillion-dollar security race for its systemic economic implications rather than as a quantified migration budget. [70] These figures measure different quantities (market size, migration spending, and economic exposure) and should not be read as a single global cost of replacing vulnerable cryptography. The response to Y2K is unfortunately not a reassuring precedent for our ability to respond to this kind of cyber-threat level event; but rather this comparison is a warning about the scale, time, and effort of what a full-scale remediation effort for the PQC transition will most likely require from a future historical perspective (Figure 4B).

\begin{figure}[H]\centering
\includegraphics[width=0.82\linewidth]{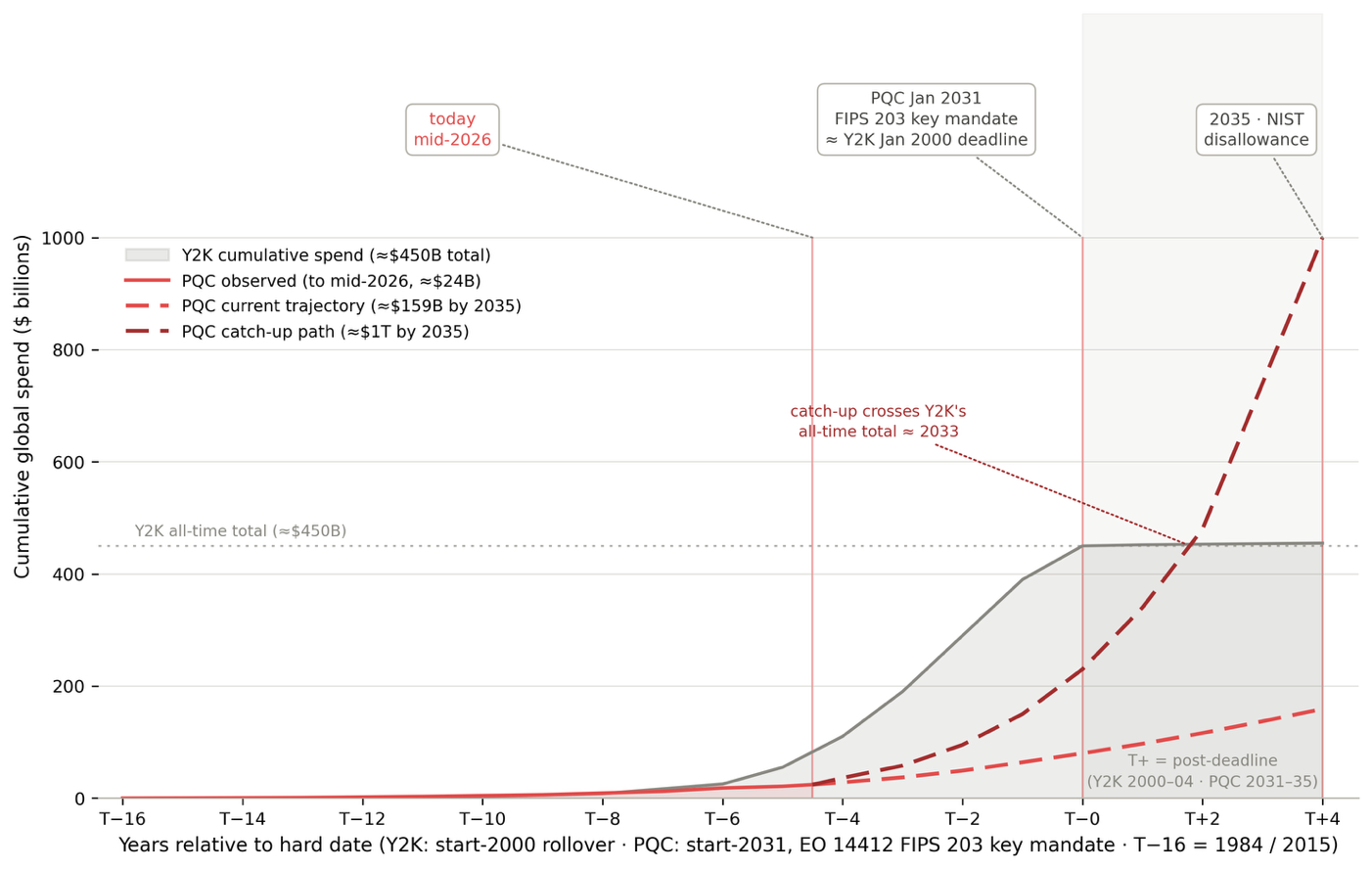}
\par\vspace{3pt}{\small\textbf{Figure 4B: Comparison of Cumulative Global Spend for Y2K vs. PQC Transition}}
\end{figure}
\subsubsection*{3.1.2 RSA/ECC Vulnerability Spectrum}
\addcontentsline{toc}{subsubsection}{3.1.2 RSA/ECC Vulnerability Spectrum}
An academic analysis first posted in 2019 and published in 2021 estimated that 2048-bit RSA encryption could be broken in about 8 hours using Shor’s algorithm and a fault-tolerant quantum computer built from roughly 20 million noisy physical qubits with surface code error correction. [79] In 2022, the White House directed NIST and the National Security Agency (NSA) to prioritize “the goal of mitigating as much of the quantum risk as is feasible by 2035.” [63] Based on surveys of expert opinions of the growth in quantum computing processing power into the 2030s, as of 2023 it was projected that the so-called Q-Day for RSA/ECC encryption was most likely to happen around the mid-to-late 2030s. [80]

But less than six months after NIST approved the first PQC algorithms in 2024, it became necessary to advance the federal timeline for migrating high value assets and high-impact systems off RSA/ECC by five years, to 2030. [5] At least one organization has an online countdown timer to the so-called Y2Q date for RSA/ECC that currently predicts this date to be less than four (4) years away. [81] In May 2025, the same researcher at Google who predicted it would take 20 million qubits to break RSA-2048 in 2021 presented an approach for achieving a 20X reduction in this estimate. [82] Most recently in March 2026, the same research team at Google published a further reduction in their estimates [20], and days earlier moved its internal company deadline for PQC transition from 2030 to 2029. [59] In July 2026, an independent research team was purportedly able to exceed Google’s results using a zero-knowledge proof approach. [170]

Two scaling relationships explain both why the estimates for potentially breaking these algorithms continue to fall and why the qubit estimates quoted below differ so widely. Shor’s algorithm factors an n-bit modulus (n = log\textsubscript{2} N) in a polynomial number of gates, on the order of Õ(n²) to Õ(n³), using roughly 2n + O(1) logical qubits, whereas the best classical method, the general number field sieve, runs in sub-exponential time:

\begin{center}L\textsubscript{N} = exp[ (1.92 + o(1)) · (ln N)\textsuperscript{1/3} · (ln ln N)\textsuperscript{2/3} ]\end{center}
It is this collapse from sub-exponential to polynomial that dooms RSA and ECC. The second relationship governs the gap between logical and physical qubits. Under the surface code each logical qubit consumes on the order of 2d² physical qubits, where the code distance d is fixed by the target logical error rate:

\begin{center}p\textsubscript{L} ≈ A · (p\textsubscript{phys} / p\textsubscript{th})\textsuperscript{(d+1)/2}\end{center}
The widely differing figures cited below are therefore not contradictory: counts near 100,000 are physical-qubit estimates inclusive of error-correction overhead, while counts in the hundreds to low thousands (1,193 and about 640 for ECC-256, Section 3.1.2.6) are logical or idealized-circuit estimates taken before that overhead is applied. Comparing them requires fixing whether the code distance d, and hence the physical multiplier 2d², has been included.

The compression of the various deadlines for PQC transition is reflected in the ever-increasing power of CRQC developments. The most recent briefing by Global Risk Institute reflects the implications of the continual reductions in estimated number of qubits needed for CRQC relevant to different NIST algorithms. [83] This report includes a figure whose caption reads: “Figure 1 (Cain et al.) [19] Estimated number of physical qubits needed to run Shor’s algorithm compared over the years of publication for previous resource estimates and for Cain et al. Please note that the vertical scale is logarithmic: published resource estimates have dropped by orders of magnitude over time.” Figure 5 below is based on these various published estimates of physical qubits required for Shor's algorithm by year of publication, with a fitted exponential decay dashed line per homogeneous series representing the trendline. The data are from Cain et al. [19] as compiled in the Global Risk Institute briefing [83]; the figure is redrawn by the authors, with the exponential decay fits added.

\begin{figure}[H]\centering
\includegraphics[width=0.82\linewidth]{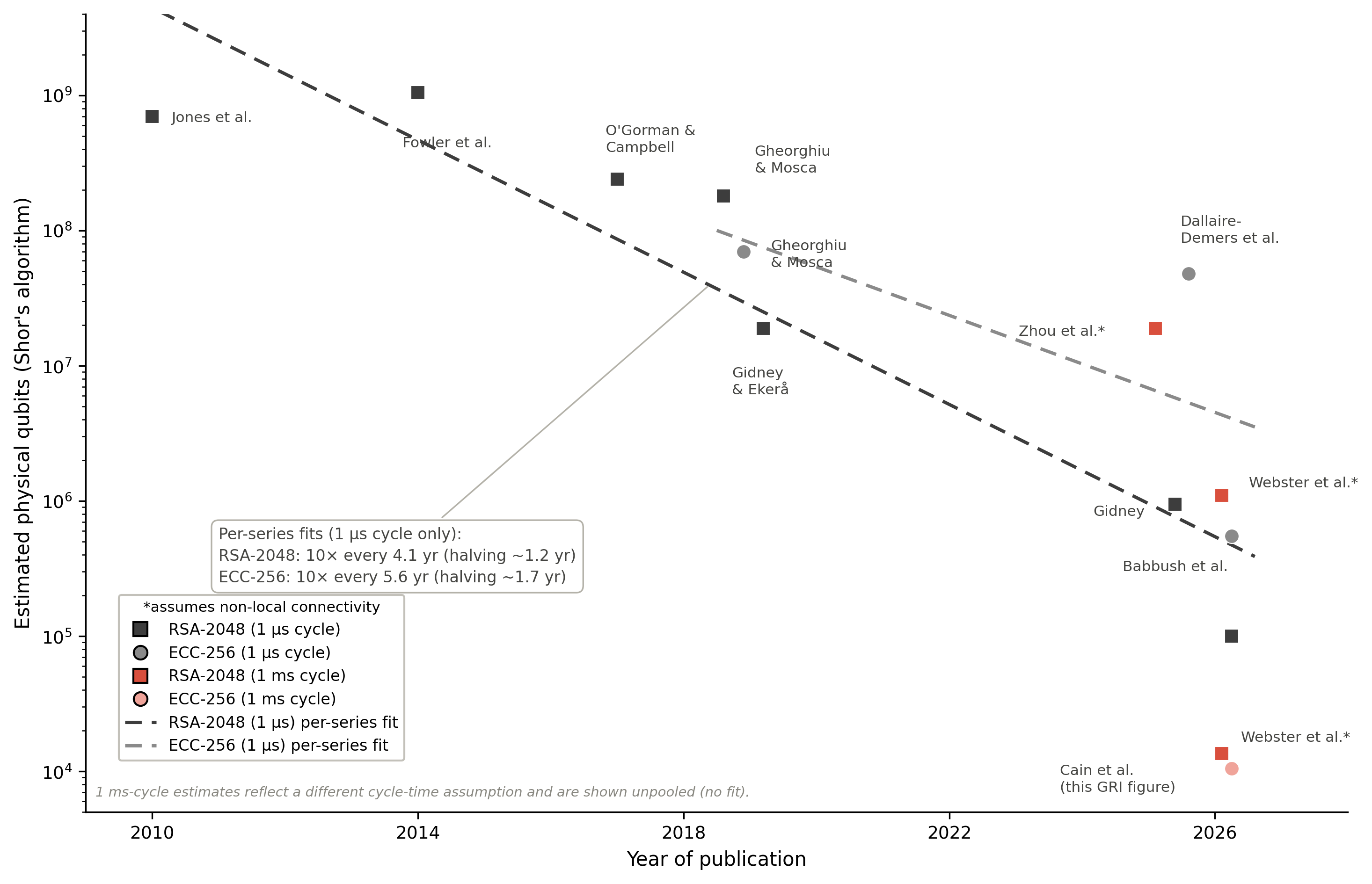}
\par\vspace{3pt}{\small\textbf{Figure 5: Exponential Decay Fit of Global Risk Institute Estimated Number of Qubits for Shor’s Algorithm. Plotted qubit estimates are approximate reads of the published figure rather than values from a source table; the fit is reported per homogeneous series (RSA-2048 and ECC-256), not as a single pooled decay.}}
\end{figure}
This graph includes representations of some of these significant advancements in the first half of 2026 in reducing the estimated number of qubits needed for CRQC: 

\paragraph*{3.1.2.1 Pinnacle.}
\addcontentsline{toc}{subparagraph}{3.1.2.1 Pinnacle.}
The Iceberg Quantum Pinnacle Architecture estimates a reduction in qubit count against RSA-2048 (2048-bit RSA integers factored with fewer than 100,000 physical qubits). [84]

\paragraph*{3.1.2.2 Caltech/IQIM.}
\addcontentsline{toc}{subparagraph}{3.1.2.2 Caltech/IQIM.}
Caltech/IQIM applies QLDPC codes to reconfigurable atom qubits running Shor’s algorithm against ECC-256 and RSA-2048 (Shor’s algorithm possible with as few as 10,000 reconfigurable atomic qubits). [19]

\paragraph*{3.1.2.3 Inria/CNRS.}
\addcontentsline{toc}{subparagraph}{3.1.2.3 Inria/CNRS.}
Inria/CNRS reduces the qubits needed against ECC (only 1,193 logical qubits required for ECC-256, half the 2,124 of the previous best estimate and roughly 42\% below Gidney's 2,043 logical qubits for RSA-3072). [85] 

\paragraph*{3.1.2.4 Hansung/Nanyang.}
\addcontentsline{toc}{subparagraph}{3.1.2.4 Hansung/Nanyang.}
Hansung/Nanyang takes an alternative approach to the computation of point multiplication in Shor’s algorithm for improved quantum circuits against ECC (up to a 40\% improvement in the product of qubit count and circuit depth). [86]

\paragraph*{3.1.2.5 QuEra.}
\addcontentsline{toc}{subparagraph}{3.1.2.5 QuEra.}
QuEra, a startup quantum computing company, promises thousands of logical qubits with a logical error rate at the 7-sigma level by 2028 to 2029 by cutting the overhead to 20 hardware qubits per logical qubit. [87]

\paragraph*{3.1.2.6 ECC-256 Estimates and the RSA-2048 Proxy.}
\addcontentsline{toc}{subparagraph}{3.1.2.6 ECC-256 Estimates and the RSA-2048 Proxy.}
The mechanism-backed curve of Section 4.1 is computed for RSA-2048 and is used as a conservative proxy for ECC-256, the primitive TLS 1.3 negotiates (X25519, P-256) and signs with (ECDSA). Three estimates already cited place ECDLP-256 at or below the RSA-2048 cost: the Google Quantum AI whitepaper of March 2026 [20], at 1,200 to 1,450 logical qubits and under 500,000 physical qubits with minutes of runtime on a superconducting surface-code architecture; Chevignard, Fouque and Schrottenloher [85]; and Kim et al. [86]. Two 2026 estimates sharpen the picture: IonQ's fully compiled blueprint [180] puts a 256-bit discrete logarithm on secp256k1 at 1,457 logical and 19,397 physical trapped-ion qubits for 25.7 days per attempt, at 40.7 percent success on Mosca's rigorous bound and 63.3 percent on Ekerå's heuristic estimate, both after a 26.45 percent logical-error allowance (five devices in parallel raise the latter to 99.3 percent at no wall-clock cost), with two-qubit gate noise assumed at 10⁻⁴; and Jo and Lee [181] give a 5n/2 + o(n) logical-qubit construction, about 640 logical qubits for a 256-bit curve, a width result which, like the 1,193-qubit Chevignard, Fouque and Schrottenloher circuit [85], buys width with orders of magnitude more non-Clifford work and does not lower the practical physical range. The Pauli Group's challenge ladder [182] places secp256k1 in a 2027 to 2033 window by its common-ruler method, a window its third version of 29 October 2026 leaves unchanged, which brackets the RSA Track 0 to Track 3 crossovers of Section 4.6 from the ECC side; the authors present that window as a calibrated overlay of algorithmic cost on published roadmaps rather than a prediction, its early end rests on cat-code architectures (their surface-code rows give about 813,000 physical qubits and 22 hours, or 4.9 million and 670 hours), and they note that no group has yet run Shor's algorithm for a prime-field discrete logarithm end to end, which is the natural first rung of indicator 7 in Section 3.3.5. None of this dates an ECC attack on its own: secp256k1 is not X25519 or P-256, and a lower qubit count does not date an attack without the runtime and success-probability assumptions behind it. What the estimates establish is the direction of the proxy: the ECC-256 exposure begins no later than the RSA-2048 crossover shown, so the RSA curves of Section 4.1 are to be read as the mechanism-backed tier for every TLS 1.3 key exchange and signature that rests on Shor's algorithm.

A sixth estimate arrived in October 2026. Alice \& Bob's end-to-end compilation of Shor's algorithm on cat qubits [232] puts RSA-2048 at 22,593 cat qubits and 3.14 days, or 1.25 hours on about 611,000, and the 256-bit discrete logarithm on secp256k1 at most 29,588 cat qubits and 61.8 minutes (Table I of [232]), on a 1 μs error-correction cycle, a 10 μs decoder reaction time, circuit-level phase-flip noise of 5 × 10⁻⁴ and no bit flips. That last assumption corresponds to a bit-flip time of about 100 years per qubit (101 years in their Table XXXI), more than eight orders of magnitude beyond the 22 seconds measured on a single cat qubit; an additional distance-3 bit-flip code relaxes it to about an hour and raises the RSA-2048 estimate to 44,850 cat qubits, a figure scaled from the 98,560 computed for the repetition-cat architecture. The count is more than forty times below Gidney's 2025 figure of under one million physical qubits [82], on a different qubit type and noise model, and it is an architecture estimate pending demonstration at any scale rather than a dated forecast. For the proxy of this section the comparison is instructive: on every published architecture the elliptic-curve problem finishes first, and on Alice \& Bob's it needs more qubits than RSA-2048 yet runs roughly seventy times sooner, because 1.03 expected shots suffice where RSA-2048 needs 9.6 (Table I of [232]). Alice \& Bob also report each design's space-time volume in qubit-days [232]: 7.09 × 10⁴ for their RSA-2048 compilation (22,593 qubits for 3.14 days) against the several million of Gidney's 2025 estimate of under a million qubits for under a week [82], a single figure that joins qubit count and runtime; by the authors' own comparison the cut is forty-fold in qubits at a shorter runtime, and five orders of magnitude against the first surface-code estimates of 2010. The metric is the field's rather than this paper's; the tracking indicators of Section 3.3.5 can adopt it once more compilations report it, and Table 4 does not carry a column for it because most of its rows state no runtime.

\subsubsection*{3.1.3 PQC Vulnerability Spectrum for Lattice-Based PQC}
\addcontentsline{toc}{subsubsection}{3.1.3 PQC Vulnerability Spectrum for Lattice-Based PQC}
Given the challenges of the current estimates of a timeline of at least a decade for migration to new encryption solutions, it is understandable why NIST has not made any formal predictions for CRQC threat risk for lattice-based PQC or for a migration timeline to the next generation of cryptography beyond PQC. At various points in the initial proposal for and review of proposed PQC algorithms, the impressions from the NIST proposals, presentations and reports can be interpreted as a general desire that the proposed PQC encryption algorithms would be quantum-resilient for at least a couple of decades after approval. [88] 

The focus of much of the external analysis of the proposed PQC encryption algorithms was centered around use of Shor’s and Grover’s algorithms implemented by quantum computers or hybrid quantum/classical computer attacks, and on the strength of lattice-based cryptography against these kinds of attacks. NIST confirmed this focus on the presumed strength of lattice-based cryptography in its recent Fourth Status Report which selected for evaluation a set of PQC algorithms that do not use lattice-based cryptography to continue in the approval process as a potential PQC backup algorithm. (“The security of ML-KEM (FIPS-203) is based primarily on the presumed hardness of certain computational problems in lattices.”) [89] 

Unfortunately, what is missing from such a “presumed hardness” assumption for projecting the cryptographic strength of lattice-based cryptography based on the LWE hard problem is any evaluation of the impact of new or improved quantum algorithms or quantum computing technologies on the computational problems underlying lattice-based schemes, particularly where such improvements are undertaken with the assistance of AI and AGI. Recent papers suggest the use of modular arithmetic to boost the performance of ML models in attacks on LWE. [90] [91] The presumed hardness assumption is also suspect based on two recent surveys of proofs for PQC schemes which found that PQC security proofs suffer a larger reduction loss as compared to classical ECC/RSA security proofs, which negatively affects overall security. [92] [93]

To anchor these qualitative concerns in the quantitative security model on which the standardized parameters rest, the best-known attacks on LWE proceed by lattice reduction (the BKZ algorithm) with a block size β, whose cost is captured by the core-SVP model:

\begin{center}cost ≈ 2\textsuperscript{c·β},    c = 0.292 (classical sieving),     c = 0.265 (quantum sieving)\end{center}
The required block size β is fixed by the scheme parameters through the root-Hermite factor δ(β):

\begin{center}δ(β) = [ (πβ)\textsuperscript{1/β} · β / (2πe) ]\textsuperscript{1/(2(β−1))}\end{center}
with the primal attack succeeding approximately when

\begin{center}√β · σ ≲ δ(β)\textsuperscript{2β−d−1} · q\textsuperscript{m/d}\end{center}
where d is the lattice dimension, q the modulus, and σ the noise width. This is a simplified primal-uSVP heuristic (with d the full embedded lattice dimension); an exact estimator additionally carries lower-order dimension and error-norm factors that do not materially change the block sizes reported here (the −1 in the exponent scales the right-hand side by δ⁻¹ ≈ 0.998 at β ≈ 877, shifting the required block size by less than one unit). For the Kyber-1024 parameter set this yields β ≈ 877, giving an estimated classical cost near 2\textsuperscript{256} and a quantum cost near 2\textsuperscript{232} operations. These figures are the published basis for the scheme’s claimed security, and any argument that the scheme falls earlier must show concretely how one of these two quantities, the exponent c or the block size β, is reduced. Section 4.5 develops exactly that accounting.

Two caveats temper these figures, and both cut against an earlier break rather than toward one. First, the quantum exponent c = 0.265 assumes Grover-amplified sieving with idealized, low-cost quantum random access memory (QRAM); once realistic memory-access and circuit-depth costs are charged, recent analyses indicate that the quantum advantage over classical sieving narrows substantially and may largely vanish, so 0.265 is optimistic for the attacker and the classical 0.292 may be closer to the operative exponent. Second, the core-SVP model counts only a single shortest-vector call and omits the additional reduction calls and overhead that a full attack incurs, so the values near 2\textsuperscript{256} and 2\textsuperscript{232} are conservative lower bounds and the true work factor is higher. Both points widen the margin that any earlier-break argument must overcome.

The current best quantum sieving time exponent, 2\textsuperscript{0.2563·d} (Bonnetain, Chailloux, Schrottenloher and Shen, EUROCRYPT 2023), was achieved by human cryptanalysts and published before the assumed 2028 AGI onset. Erosion from 0.265 to 0.2563 has therefore already occurred without AGI. This narrows the room left for AGI-driven erosion, supporting the contention that the collapse parameter must carry any break, and it rebuts the objection that the model gives human cryptanalysis too little credit. The same literature that fixes the list size (Cho, Hhan, Kim, Lee and Shen, IACR ePrint 2024/1700) also achieves 2\textsuperscript{0.2653·d} time with QRAM reduced to 2\textsuperscript{0.05778·d}, which cuts against the QRAM caveat; the same source proves a lower bound of 2\textsuperscript{0.2925·d − 2s} at QRAM size 2\textsuperscript{s}, with no quantum speedup available without QRAM under its assumptions, which cuts back in its favor. Both readings are recorded here so that the caveat is not selective.

\subsection*{3.2 Survey of Potential Improvements to Quantum and AI Technologies}
\addcontentsline{toc}{subsection}{3.2 Survey of Potential Improvements to Quantum and AI Technologies}
Given the rapid expansion in both the capability and reach of AI over the last few years, and the even greater potential of AGI in the next few years, the biggest opportunities for potential improvements in quantum algorithms or quantum computing technologies are likely to be driven by the application of AI and then AGI to the design, optimization and innovation of quantum algorithms and computing technologies. 

This section will briefly highlight some of the potential improvements related to CRQC+AI, as well as other quantum and AI related technologies that support the high-level presentation of these in Figure 3. The AI-capability results and quantum-hardware advances surveyed here are cited as evidence of the pace of AI-assisted progress and as enabling technologies; none of them constitutes a demonstrated cryptanalytic advance against standardized post-quantum cryptography.

\subsubsection*{3.2.1 AI Improvements for CRQC+AI}
\addcontentsline{toc}{subsubsection}{3.2.1 AI Improvements for CRQC+AI}
Some very recent examples of the unexpected advancements and innovations achieved by these new AI models include:

\paragraph*{3.2.1.1  Using AI to Improve AI Models }
\addcontentsline{toc}{subparagraph}{3.2.1.1  Using AI to Improve AI Models }
Developments in AI models have reached an inflection point as of 2025, where AI models and constructs are now being improved upon by AI. [94]

\paragraph*{3.2.1.2  Claude Mythos AI Models }
\addcontentsline{toc}{subparagraph}{3.2.1.2  Claude Mythos AI Models }
Anthropic has introduced a new class of frontier models known as Mythos. These models are notably capable at discovering software vulnerabilities: independent reporting describes the Mythos-class models as particularly adept at detecting software flaws, including some that had gone undiscovered for years. [95] [96] Anthropic initially limited the availability of these models to a small set of organizations as part of its Project Glasswing. [97] A more broadly available variant, Fable 5.1, was released with specific safeguards intended to reduce the likelihood of misuse for cybersecurity tasks, and became available again, after a temporary suspension, as of July 1, 2026. [98] A systems-engineering analysis has further examined the implications of such frontier-class models for post-quantum cryptographic migration. [99]

On 28 July 2026 two results from the same model class were published that bear directly on the research-acceleration factor R of Section 4.5. [183] [184] The first improved the best known attack on 7-round AES-128, unchanged since 2013, from 2\textasciicircum{}99 to between 2\textasciicircum{}89.3 and 2\textasciicircum{}91.4 operations at 2\textasciicircum{}105 data, a gain of roughly 8 to 10 bits found by an autonomous search whose published, model-edited chain of thought records a brief to invent a sixth attack family outside the five known ones [185]; the attack is impractical and the full ten-round cipher is untouched. A human-authored improvement on the same Derbez et al. attack, through a cross-ratio property of the multiplicative inverse, was posted a day later with no figure in its abstract [241]; the two are independent, and the 2013 figure fell twice in two days, once by autonomous search and once by human cryptanalysis. The second found a previously unused symmetry in the HAWK lattice that reduces HAWK-n key recovery to a shortest-vector problem in dimension n/2 + 1, reported by Straznickas and Weis and announced on the NIST pqc-forum [186]. Keys were recovered end to end for the challenge-size HAWK-256, from about 2\textasciicircum{}64 to 2\textasciicircum{}38 operations; for the level-1 set HAWK-512 the published estimate falls from about 2\textasciicircum{}150 to at most 2\textasciicircum{}108 gates in the specification's own model; and the HAWK team, confirming that the block size needed for an equivalent key roughly halves, withdrew the scheme from the third round of NIST's additional-signature process on 29 July 2026 [187]. Neither result is a lattice-exponent improvement against a standardized scheme, and HAWK carries structure that ML-KEM's random module lattices are designed not to have. What the two results establish is that R was above zero in 2026, two years before this paper's AGI gate. That observation, and no calibration, is the anchor behind the frontier-model slope band of Appendix A.2 (0.08 to 0.12 per year): a judgment to be replaced by a measured time series once a benchmark such as CryptanalysisBench [188] carries one. A preliminary result on 13 of LEA's 24 rounds reported alongside, key recovery from under 2\textasciicircum{}30 plaintexts in under an hour against a prior best of 2\textasciicircum{}86 work at 2\textasciicircum{}98 pairs, is left out of R pending its full publication.

\paragraph*{3.2.1.3 OpenAI Math Proof AI Models }
\addcontentsline{toc}{subparagraph}{3.2.1.3 OpenAI Math Proof AI Models }
OpenAI recently announced that one of its general-purpose reasoning models, not a system trained specifically for mathematics, disproved an 80-year-old conjecture in discrete geometry concerning the planar unit distance problem first posed by Erdős in 1946. [100]

Two 2026 results outside cryptography bear on the same capability class and are cited here as capability evidence only; neither touches a cryptographic primitive, and no link to RSA, ECC or lattice cost is drawn or implied. On 6 October 2026 OpenAI released 722 machine-written manuscripts in 372 families [189]. One of them states a zero-free half-plane Re s > 7/8 for the Riemann zeta function and Dirichlet L-functions, the quasi-Riemann hypothesis. The 199-page statement is formalized and machine-checked in Lean, and an independent re-run of the single zeta statement on two kernels under the standard axioms succeeded [190]; the 49-page 11/12 variant that human experts are studying is not formally verified and is the one part of the release a person edited. The experts quoted largely agree the result is correct, while Sarnak will not be satisfied that it is a theorem until humans check and understand all the details [191]; Visser's remark, in a number-theory note that builds on the 7/8 result, that most of the 722 preprints are unreadable and would be desk-rejected concerns readability, not correctness [192], and by 7 October three manuscripts had been withdrawn for a sign error and fourteen revised, none in the zeta family [189]. Separately, a Lean-verified proof that there is no percolation at the critical probability in every dimension, θ(p\_c) = 0, obtained through Kozma and Nitzan's Conjecture 3, was posted on 28 August 2026 [193], two days before Duminil-Copin, a non-interested party, wrote in a public essay, without reference to it, that AI would probably prove θ(p\_c) = 0 before humans do [194]. The Lean development checks under the standard axioms with no unproven placeholder in the solution, and it has not been refereed by anyone independent of its author. A partial claim on Navier-Stokes blow-up remains contested and is not relied on. What these results share with the cryptanalytic results of Section 3.2.1.2 is the search profile: no new concepts, a known family others had passed over, pursued against the field's own expectation.

\paragraph*{3.2.1.4 AI-Engineered Classical Pipelines}
\addcontentsline{toc}{subparagraph}{3.2.1.4 AI-Engineered Classical Pipelines}
Two factoring records of September 2026 show the boundary of Section 3.3.1 in the wild. RSA-260 was factored on 3 September 2026 by a team at Cognition using a GPU-accelerated general number field sieve built and operated with its Devin coding system, at about 4,900 GPU-days (13.5 GPU-years, roughly \$400,000 at market prices) and about three weeks from first prompt to factors [195]; RSA-896 followed on 19 September 2026 with a frontier model doing the GPU porting and orchestration [196]. Both runs used the unchanged number field sieve and sit on its complexity curve, and RSA-2048 remains on the order of 10\textasciicircum{}11 times beyond classical reach. The AI contribution was engineering, porting, orchestration and tuning, with a calendar compression of roughly four to eight times on a software-heavy task. That is the rule of Section 3.3.1 observed directly, and it is the first anchor of the right type for the engineering acceleration factors of Section 3.3.3, recorded there without raising them. It also makes RSA-1024 a present classical exposure at an estimated \$30 million of market compute, independent of any quantum timeline (Section 6.4).

\subsubsection*{3.2.2 Quantum Computing Improvements for CRQC+AI}
\addcontentsline{toc}{subsubsection}{3.2.2 Quantum Computing Improvements for CRQC+AI}
IBM and Quantinuum have updates on the current state of quantum computing technology and roadmaps for future quantum computing developments that currently project hundreds to thousands of logical qubits capable of running billions of quantum gates. [101] [102] A recent announcement by Amazon and QuEra projected useful, error-correcting quantum computing by 2028. [103] Numerous other related and complementary improvements could further enhance the capabilities of quantum computing. The advances surveyed below are enabling technologies whose bearing on practical cryptanalysis is mostly indirect, and several remain early-stage research rather than near-term threats, as discussed for Rose’s Law below. These include:

\paragraph*{3.2.2.1 Qudits/Qutrits.}
\addcontentsline{toc}{subparagraph}{3.2.2.1 Qudits/Qutrits.}
Beyond qubits, which resolve to just two states upon measurement, work is progressing on the development of qudits and qutrits, which can resolve to m-dimensional states upon measurement, with the possibility of n entangled qudits representing m\textasciicircum{}n possible combinations. [104]

\paragraph*{3.2.2.2 Noise/Heat Reductions.}
\addcontentsline{toc}{subparagraph}{3.2.2.2 Noise/Heat Reductions.}
Work on noise and thermal-excitation reduction is also advancing, including hyper-entanglement of neutral atoms using a novel form of cooling to detect and correct thermal motion excitations, [105] and the use of mirrored reflections to eliminate quantum noise. [106]

\paragraph*{3.2.2.3 Quantum Switch.}
\addcontentsline{toc}{subparagraph}{3.2.2.3 Quantum Switch.}
Cisco has recently announced a Universal Quantum Switch, a working research prototype that routes quantum information between quantum processors by entanglement at room temperature. [107] 

\paragraph*{3.2.2.4 Synthetic Dimensions for Photonic/Magnon/Quantum States.}
\addcontentsline{toc}{subparagraph}{3.2.2.4 Synthetic Dimensions for Photonic/Magnon/Quantum States.}
Various organizations are developing enhanced methods for manipulating photonic/magnon/quantum states to enhance coherence, entanglement and quantum information processing by using synthetic dimensions and quantum entanglement. [108] [109] [110] [111] [112] [113] [114] [171]

\paragraph*{3.2.2.5 Quantum Circuits for SHA-1.}
\addcontentsline{toc}{subparagraph}{3.2.2.5 Quantum Circuits for SHA-1.}
Hansung University has developed a SHA-1 quantum circuit with a minimized circuit depth. [115] IBM has demonstrated a Quantum Fourier Transform (QFT) on its Quantum Heron processor. [116]

\paragraph*{3.2.2.6 Quantum Hardware/Error Correction.}
\addcontentsline{toc}{subparagraph}{3.2.2.6 Quantum Hardware/Error Correction.}
Most of the recent efforts on exploring the relationship between AI and quantum computing have focused on using AI to enhance quantum computing hardware and error correction and quantum computing speedups. [117] [118] [119] [120] [121] [122]

\paragraph*{3.2.2.7 Rose’s Law.}
\addcontentsline{toc}{subparagraph}{3.2.2.7 Rose’s Law.}
Just as Moore’s Law projects the accelerating growth of classical computer technologies, [123] Rose’s Law makes a similar projection for quantum computing technologies. [124] The natural human assumption that the spectrum of threats posed by CRQC+AI can safely be assumed to be limited to merely linear growth is a fallacious assumption that cannot be the basis for ignoring PQC transition mandates and the need for crypto-agility to address these exponentially expanding threat profiles.

Stated quantitatively, Rose’s Law is an exponential, not linear, growth model for usable quantum capability:

\begin{center}Q(t) = Q\textsubscript{0} · 2\textsuperscript{(t − t₀)/τ}\end{center}
where Q\textsubscript{0} is the capability at a reference year t\textsubscript{0} and τ is the doubling time. This is the growth that feeds the hardware budget of Section 4.5: because that budget is the base-2 logarithm of the usable operations an attacker can afford, an exponential Q(t) yields a budget B(t) that rises linearly in t, which is exactly the constant-doubling form assumed there. Treating the threat as linear understates it by the difference between t and 2\textsuperscript{t}.

One qualification is essential here. Rose’s Law is an empirical observation about the qubit counts of quantum annealers (the D-Wave line), and annealer qubits are neither gate-model qubits nor fault-tolerant logical qubits; they do not by themselves bound the resources available for Shor, Grover, or lattice sieving. The hardware budget B(t) of Section 4.5 is therefore properly anchored to fault-tolerant logical-qubit and error-rate roadmaps for gate-model machines, with Rose’s-Law scaling retained only as a loose upper envelope on the pace of progress, not as a direct count of cryptanalytically usable qubits.

The hardware driver in this paper is deliberately platform-agnostic in the same sense: it keys on effective error-corrected capability rather than on the roadmap of any single modality. Superconducting programs (IBM, Google), trapped-ion systems (Quantinuum, IonQ), photonic architectures (PsiQuantum), and neutral-atom platforms differ in qubit counts, error rates, clock speeds, and scaling cadence, and none is singled out by the model; what enters as the driver is the aggregate pace at which fault-tolerant logical capacity becomes available, from whichever platform or combination delivers it.

\subsubsection*{3.2.3 Quantum Algorithm Improvements for CRQC+AI}
\addcontentsline{toc}{subsubsection}{3.2.3 Quantum Algorithm Improvements for CRQC+AI}
Some of the work of CRQC+AI relative to CRQCs has focused on optimizing and/or hacking both classical and PQC cryptography, including the use of CRQC+AI to enhance lattice reduction algorithms. [125] Current estimates based on topology classification are in the range of 100-200 different kinds of quantum algorithms. [126] Given the complexity of both quantum algorithms and programming quantum computers, it is likely that the various kinds of AI/ML/DL learning models and tools may be substantially faster at exploring variations and even completely different approaches to design and optimization of quantum programming of quantum algorithms in ways that humans might not even consider. 

\paragraph*{3.2.3.1 Language Models}
\addcontentsline{toc}{subparagraph}{3.2.3.1 Language Models}
Examples of various AI models potentially useful for improving lattice reduction algorithms and other CRQC relevant quantum algorithms include language model programs and dynamic word embedding. [127] [128]

\paragraph*{3.2.3.2 Quantum Algorithms}
\addcontentsline{toc}{subparagraph}{3.2.3.2 Quantum Algorithms}
Reinforcement learning has already discovered faster classical algorithms, for example improved sorting routines found by deep reinforcement learning, [129] which is direct evidence that AI can autonomously surpass human-designed algorithms. The same approach is expected to yield new kinds of quantum algorithms beyond the HHL and low-dimensional geometric algebra approaches described in this paper, including better approaches to reducing optimization problems via decoded quantum interferometry [130] and traveling salesman solutions [131], and solutions to quantum permutation puzzles. [132] Contests for using quantum algorithms to break Bitcoin encryption are examples of how these new kinds of quantum algorithms are being encouraged. [133]

\subsubsection*{3.2.4 Side-Channel Attacks}
\addcontentsline{toc}{subsubsection}{3.2.4 Side-Channel Attacks}
Numerous types of new side-channel attacks are being developed for both classical and PQC algorithms using CRQC+AI. [134] These include attacks such as practical fault attacks on randomness, [135] single-trace passive side-channel attacks on HQC, [136] first-order masked ML-KEM attack using multiple distinct chosen ciphertexts, [137] and exploitation of noisy single-bit leakage. [138] The 2026 record extends the list: horizontal fusion attacks recovering ML-DSA-87 keys from four traces [197], improved ILWE solving that needs 62 times fewer signatures against ML-DSA [198], Falcon key recovery from floating-point conversion leakage in 20 to 100 signatures [199], a first-order masked Kyber-768 key recovered from about 2,200 decapsulation traces by belief propagation alone [200], and key recovery from deployed ML-KEM implementations whose Fujisaki-Okamoto re-encryption check compared a truncated ciphertext [201]. The four-trace attack is non-profiling, needing no cloned device with known keys, and the leakage it exploits is the module lattice's repeated matrix-vector product rather than an implementation slip [197].

The model does not separately price implementation or side-channel compromise, that is, trace-acquisition cost, success probability, or countermeasure state. On any reasonable assessment, implementation compromise of a deployed PQC stack before 2035 is more probable than a structural lattice break and requires no quantum computer. An independent review of the first version proposed pricing that compromise as a third driver, independent of the AGI onset [234]; a quantified term is deferred to a later rerun.

\subsubsection*{3.2.5 Key Recovery Attacks}
\addcontentsline{toc}{subsubsection}{3.2.5 Key Recovery Attacks}
Numerous types of new key attacks are being developed for both classical and PQC algorithms using CRQC+AI. These include attacks such as neighborhood search attacks on Oracle MLWE for ML-KEM, [139] data repetition and stepwise regression, [140] quantum truncated differential attacks, [141] multi-instance key degradation, [142] as well as hybrid-adaptive-LDPC attacks on ML-KEM. [143] On the algorithmic side, three 2026 provable dual-attack analyses [202] [203] [204] give ML-KEM-1024 a provable security of 410 to 426 bits against a heuristic estimate near 256; their 22- to 68-bit reductions are against the provable baseline and leave the heuristic figures, and the block size β used in Section 4.5, unchanged. AH-BKZ [205] sets the implementation frontier of lattice reduction at a shortest-vector record in dimension 163. Cid, Elkouss and Goulão [206] obtain full key recovery against the original Dilithium and its round-1 and round-2 versions in the superposition-query model and find the final ML-DSA closed to the technique; Kuninets and Bakharev [207] halve the dimension of a direct-sum Barnes-Wall lattice-isomorphism instance through its hull, a second 2026 structural result on a structured lattice of the HAWK family; and Lin and Vasudevan [208] give a provably subexponential algorithm for LWE with tiny errors and few samples whose constants sit far above concrete security. Structured lattices keep yielding; the standardized schemes have not, and the NIST process is part of why.

\subsubsection*{3.2.6 Hybrid Attacks}
\addcontentsline{toc}{subsubsection}{3.2.6 Hybrid Attacks}
Numerous types of hybrid attacks are being developed for both classical and PQC algorithms using CRQC+AI. For PQC algorithms, these include primal hybrid attacks against sparse LWE [144] [145], Polynomial LWE (PLWE) [93], Ring-LWE [146], and dual attacks [147].

\subsection*{3.3 An Acceleration-Factor Model for AI-Compressed Timelines}
\addcontentsline{toc}{subsection}{3.3 An Acceleration-Factor Model for AI-Compressed Timelines}
The upper-bound scenarios in Section 3.1 assume that AI compresses the calendar time to a CRQC, but that assumption does most of the work in the aggressive projections and deserves to be stated as an explicit, falsifiable model rather than a qualitative claim. This section sketches such a model. Its purpose is not to make the timeline more aggressive but to make whatever timeline it produces auditable: every input is a named quantity with a baseline anchored in published progress, an acceleration factor anchored in observed machine-learning results, and a stated condition under which the projection would be revised. Because every quantity is named and anchored, the model is built to be rerun: any reader can vary the parameters, regenerate the curves, and compare successive runs against the milestones that have since become historical data points, in the same way that published quantum resource estimates and expert-elicitation surveys are periodically revised. The model deliberately holds the quantum algorithm fixed. AI is allowed to accelerate engineering and design; it is not allowed to change the complexity of Shor’s algorithm or to lower the fault-tolerance threshold.

\subsubsection*{3.3.1 What the Acceleration Factor Scales, and What It Does Not}
\addcontentsline{toc}{subsubsection}{3.3.1 What the Acceleration Factor Scales, and What It Does Not}
The single most important boundary in this model is between what AI can plausibly accelerate and what it cannot. AI is permitted to compress the engineering and design loop: decoder design and real-time decoding, pulse-level gate calibration and optimal control, fabrication and materials search, error-correcting-code discovery, and circuit compilation and optimization. AI is not permitted to alter the underlying complexity theory or physics: it does not change the polynomial scaling of Shor’s algorithm, it does not lower the fault-tolerant error-correction threshold, and it does not repeal qubit coherence limits. This separation is what keeps the model honest. A projection in which AI accelerates everything uniformly is indistinguishable from assuming the conclusion; a projection in which AI accelerates only the identifiable engineering bottlenecks can be checked channel by channel against the published record.

\subsubsection*{3.3.2 Four Channels and Their Baselines}
\addcontentsline{toc}{subsubsection}{3.3.2 Four Channels and Their Baselines}
We decompose progress toward a CRQC into four engineering channels, each with its own baseline rate of improvement and its own acceleration factor. Treating them separately, rather than as a single global rate, is what lets each factor be argued and falsified independently. The channels, their published baselines, and the engineering tasks where AI plausibly contributes are summarized in Table 2 below. Each acceleration factor below is calibrated rather than assumed: its baseline rate is taken from a published progress trajectory and its value from a measured machine-learning speedup, with the supporting quantities given in the calibration Table 3 that follows. The factors are kept as ranges, low to central to high, to carry the uncertainty honestly, but they are not free parameters.

Magic-state distillation, a further cost of fault-tolerant Shor, is treated within the code-efficiency and logical-qubit channels rather than as a separate one. At the modeled binding point the raw logical-qubit count is the constraint (Table 3), and whether distillation would bind later than that count is not established; recent qLDPC codes and improved protocols have reduced its overhead. It is therefore not carried as an independent channel. Should a distillation demonstration change that assessment, the tracking indicators of Section 3.3.5 promote it to its own factor rather than revising the whole projection at once.

The channel model prices resources, not throughput: it says when a machine of sufficient size and quality can exist, not how long a run takes. The 2026 compilations of ECDLP-256 make the gap visible. The two lowest agree on qubit count within about 1.5 times, 19,397 trapped-ion qubits [180] against at most 29,588 cat qubits [232], and differ by roughly six hundred times in runtime, 25.7 days against 61.8 minutes, almost entirely through the assumed error-correction cycle, millisecond-scale for trapped ions (1 to 5 ms in [222]) against the 1 μs assumed for cat qubits, while the Google surface-code estimate sits at under 500,000 physical qubits and minutes [20]. Shot count is a second dimension the channels do not carry: 9.6 expected shots for RSA-2048 against 1.03 for ECDLP-256 on the cat-qubit compilation (Section 3.1.2.6). The non-binding status of channel 4 on trapped ions follows from that slow cycle rather than from decoder progress alone (Table 3). A clock-rate channel is deferred to the next versioned rerun: adding one changes the per-channel times τᵢ and the figures, and it should enter with an independent re-review rather than by amendment.

\par\vspace{4pt}\noindent\begin{minipage}{\linewidth}
\begin{center}\footnotesize\setlength{\tabcolsep}{4pt}\begin{tabular}{|p{0.22\linewidth}|p{0.22\linewidth}|p{0.22\linewidth}|p{0.22\linewidth}|}\hline
\textbf{\textbf{Channel}} & \textbf{\textbf{Pre-AI baseline trajectory (anchor)}} & \textbf{\textbf{What AI plausibly accelerates}} & \textbf{\textbf{Acceleration factor Aᵢ (low / central / high)}} \\ \hline
Physical qubit count and modular interconnect & Vendor roadmaps; for example, on the order of 200 logical qubits by 2029 and 1,000 or more by the early 2030s [148] & Fabrication and yield search, module layout and packaging, calibration at scale & A₁ = 1.0 / 1.3 / 2.0 \\ \hline
Two-qubit gate fidelity and physical error rate & Below-threshold physical error rates demonstrated on current processors [149] & Pulse-level optimal control, calibration, crosstalk mitigation & A₂ = 1.0 / 1.5 / 2.5 \\ \hline
Logical error rate per cycle and code efficiency Λ & Λ ≈ 2.14 per two units of code distance; 0.143\% per cycle at distance 7 [149]; Quantinuum's color code on Helios reaches Λ ≈ 12 per distance step in memory at distances 3, 5 and 7, with a memory logical error rate as low as 5 × 10⁻⁶ [239] & Decoder accuracy (learned decoders [122]) and error-correcting-code discovery & A₃ = 1.0 / 1.4 / 2.0 \\ \hline
Code-cycle time and real-time decoder latency & Real-time decoder latency on the order of 60 μs; cycle time ≈ 1 μs [149] & Faster learned decoders and real-time inference on FPGA or ASIC & A₄ = 1.0 / 1.3 / 1.8 \\ \hline
\end{tabular}\end{center}
\vspace{11.25pt}\begin{center}\small\textbf{Table 2: Engineering channels toward a CRQC, with their pre-AI baseline trajectories 
and the machine-learning-anchored acceleration factors.}\end{center}
\end{minipage}\par\vspace{4pt}
The required-improvement factor for the qubit-count channel is fixed by the resource estimate for the target attack, not chosen freely. Factoring RSA-2048 was estimated in 2019 at roughly twenty million physical qubits running for about eight hours, [79] and revised in 2025 to under one million qubits running for under a week. [82] The qubit-count target has therefore moved by more than an order of magnitude in six years through algorithmic and error-correction advances rather than hardware alone, and the model treats that moving target explicitly rather than assuming a fixed endpoint.

Setting t₀ = 2026, the required factors and baseline rates, with their anchors, give the following per-channel calibration in Table 3. The time each channel needs at no acceleration is τᵢ = ln(Fᵢ) / rᵢ; the latest of these is the binding constraint.

\par\vspace{4pt}\noindent\begin{minipage}{\linewidth}
\begin{center}\scriptsize\renewcommand{\arraystretch}{0.95}\setlength{\tabcolsep}{4pt}\begin{tabular}{|p{0.176\linewidth}|p{0.176\linewidth}|p{0.176\linewidth}|p{0.176\linewidth}|p{0.176\linewidth}|}\hline
\textbf{\textbf{Channel}} & \textbf{\textbf{Required factor Fᵢ (anchor)}} & \textbf{\textbf{Baseline rate rᵢ, e-folds per year (anchor)}} & \textbf{\textbf{τᵢ at Aᵢ = 1 (years)}} & \textbf{\textbf{Implied year at Aᵢ = 1}} \\ \hline
\textbf{Logical qubit count (binding)} & ≈ 10³ (2 to 48 error-corrected logical qubits on Helios at about 10⁻⁵ to 10⁻⁴ per round, the 48-qubit run a single round [220], and up to 96 distance-4 logical qubits in postselected neutral-atom circuits with no per-round rate reported [221], far from Shor grade, to ≈ 1,400 logical for RSA-2048 [82]) & 1.1 (IBM roadmap: order-1 to ≈ 2,000 logical by 2033 [148]) & 6.3 & ≈ 2032 \\ \hline
Gate fidelity and physical error rate & ≈ 3 (few × 10⁻³ today to 10⁻³ target [149], [82]) & 0.4 (≈ 10× error reduction over ≈ 6 years; 99\% to 99.9\% [149]) & 2.7 & ≈ 2029 \\ \hline
Code efficiency and overhead & ≈ 3 (qLDPC overhead headroom of order 10× [148]); the 2026 color-code result on Helios [239] and Google's magic-state cultivation on Willow, a 40-fold error reduction to 99.99\% T-state fidelity [240], are upside anchors on this channel and leave F₃ and r₃ unchanged & 0.4 (recent ≈ 90\% overhead reduction via qLDPC [148]) & 2.7 & ≈ 2029 \\ \hline
Code-cycle time and decoder latency & ≈ 60 (≈ 60 μs to ≈ 1 μs [149], a superconducting-specific baseline; a single-CPU real-time decoder for 408 logical qubits at 1 to 5 ms cycles is reported for trapped ions [222], on simulated syndromes rather than device data, and conditional on the 1 to 5 ms trapped-ion cycle, about 1,000 to 5,000 times a superconducting cycle, which lengthens the computation by the same factor) & 1.0 (Relay-BP 5 to 10 times faster [148]; learned decoders [122]) & 4.1 & ≈ 2030 \\ \hline
\end{tabular}\end{center}
\vspace{11.25pt}\begin{center}\small\textbf{Table 3: Per-channel required factors, baseline rates, and the implied CRQC \\[2pt]readiness year under baseline progress (acceleration factor of one).}\end{center}
\end{minipage}\par\vspace{4pt}
\subsubsection*{3.3.3 Empirical Anchors for the Acceleration Factors}
\addcontentsline{toc}{subsubsection}{3.3.3 Empirical Anchors for the Acceleration Factors}
Each acceleration factor should be calibrated to an observed machine-learning result rather than chosen to produce a target date. For the decoder and logical-error channel, a neural-network decoder has reported on the order of six percent fewer errors than the best slow decoder and roughly thirty percent fewer than the best real-time decoder on the same hardware, [122] a measured improvement that bounds the near-term contribution of learned decoding. For the logical-error and code-cycle channels, the demonstration of below-threshold surface-code operation, with a logical-error suppression factor of about 2.14 per two units of code distance and a real-time decoder latency on the order of sixty microseconds, [149] sets the baseline trajectory those factors accelerate. Vendor roadmaps targeting on the order of two hundred logical qubits by 2029 and a thousand or more in the early 2030s [148] anchor the qubit-count channel. Where no measured machine-learning speedup exists for a channel, its acceleration factor should default to one, that is, no acceleration, rather than to an optimistic guess. The RSA-260 and RSA-896 factorizations of September 2026 (Section 3.2.1.4) are the first calendar-compression anchors of the right type for the engineering channels, roughly four to eight times on a software-heavy task; they are recorded here and the factors are left unchanged, because one publication-selected event in the domain where coding systems are strongest cannot move the binding channel.

\subsubsection*{3.3.4 From Channels to a Distribution Over the Crossover Date}
\addcontentsline{toc}{subsubsection}{3.3.4 From Channels to a Distribution Over the Crossover Date}
For each channel let Fᵢ be the factor by which that channel must still improve to reach the CRQC target, rᵢ the baseline rate of improvement (in e-folds per year) drawn from the published trajectory, and Aᵢ the acceleration factor. The calendar time the channel requires is then

\begin{center}τᵢ = ln(Fᵢ) / (Aᵢ · rᵢ)\end{center}
and, because a CRQC requires every channel to reach its target, the crossover date is set by the slowest, or binding, channel:

\begin{center}T = t₀ + maxᵢ τᵢ\end{center}
Treating the Aᵢ, and where appropriate the Fᵢ and rᵢ, as ranges rather than point values turns this into a distribution over T rather than a single date. The output of interest is then not a single Q-Day but a probability that a cryptographically relevant capability arrives before a given year, P(T ≤ year), which is exactly the quantity Mosca’s inequality needs for Z: any nonzero probability mass inside the secrecy horizon X is sufficient to make migration rational today. This formulation also makes the bounds of Section 3.1 explicit: the linear lower bound corresponds to every Aᵢ set to one, and the exponential upper bound corresponds to the high end of the acceleration ranges combined with the most aggressive resource estimates.

As a worked example, take t₀ = 2026 and the figures in the calibration table. At no acceleration the four channels reach their targets in roughly 2029, 2029, 2030, and 2032, so the binding channel is the raw logical-qubit count, and the no-acceleration crossover is about 2032. Applying the acceleration range to that binding channel, T = 2026 + ln(10³) / (A₁ · 1.1), gives about 2031 at the central factor A₁ = 1.3 and about 2029 at the high factor A₁ = 2.0. These outputs, roughly 2032, 2031, and 2029, reproduce the RSA-2048 crossovers shown in Figure 9, where the no-AGI Track 0 crosses by about 2032 on hardware scaling alone and the aggressive Track 3 crosses in 2029 (2029.5 on the grid), and they bracket IBM’s own internal 2029 target. [148] The acceleration-factor model is therefore not an independent or more aggressive forecast; it is the mechanism that produces the same spread of dates the figures show, expressed in auditable parameters.

This decomposition is the granular form of the aggregate Rose’s-Law hardware growth of Section 3.2 and the budget B(t) of Section 4.5, and the two are calibrated to agree: the binding-channel crossover of about 2032 at no acceleration coincides with the Track 0 hardware-only RSA crossover the aggregate model produces. The engineering acceleration factors Aᵢ are also distinct from the research-acceleration factor R of Section 4.5 and are not interchangeable with it. The Aᵢ compress the time to build a fault-tolerant machine, the hardware track that drives the RSA and ECC crossovers, whereas R compresses cryptanalytic research effort, the software track on which the lattice and hash-based contingencies of Section 4.5 separately depend; one set of numbers governs hardware and the other governs cryptanalysis, and the timeline uses both rather than a single blended rate. Two simplifications should be stated plainly. The per-channel rates rᵢ are averages over the published roadmap horizon and are in practice front-loaded, with the steep early logical-qubit ramp expected to slow in later years. And channels one and three are physically coupled, since better codes lower the qubit-count target, so treating them as independent is conservative for the binding channel rather than optimistic.

Two features of this calibration are worth stating plainly. First, the binding channel is the one where AI has the least direct evidence of acceleration: raw qubit-count scaling. The strongest measured machine-learning contributions, in decoding, calibration, and code design, fall in the non-binding channels, which already reach their targets earlier and so cannot move the crossover on their own. Second, the route by which those evidenced contributions do move the date is indirect but real: better codes, higher fidelity, and faster decoding lower the qubit-count target F₁, which is exactly the mechanism behind the fall of the RSA-2048 estimate from twenty million to under one million qubits between 2019 and 2025. [79], [82] The aggressive end of the timeline therefore depends less on AI building qubits faster than on AI continuing to shrink the number of qubits required, a dependency this model makes explicit rather than absorbing into a single factor.

\subsubsection*{3.3.5 Falsification Conditions and Tracking Indicators}
\addcontentsline{toc}{subsubsection}{3.3.5 Falsification Conditions and Tracking Indicators}
Because the model is built from named quantities, it can be falsified or updated by observation rather than argued in the abstract. The following indicators, tracked over the next eighteen to thirty-six months, could move the projected crossover date earlier or later:

\begin{itemize}\item physical qubit counts and modular interconnect scaling meeting, beating, or missing the published vendor roadmap milestones;\item logical error rate per cycle continuing to fall by a factor of two or more per two units of code distance, or instead stalling above the level the target computation requires;\item real-time decoder latency reaching and holding the microsecond regime at increasing code distance, or failing to keep pace with the cycle time;\item magic-state production and injection moving from theory to a first hardware factory demonstration, or remaining undemonstrated;\item the published resource estimate for the target attack continuing to fall, as it did from twenty million to under one million qubits, or instead stabilizing;\item structural claims against lattice assumptions and their verification outcome, logged as an indicator in its own right: in 2026 the Luo claim against Module-LWE was not accepted, the HAWK automorphism attack held and ended a NIST candidate, and the Simon dihedral-coset claim was refuted within days (Appendix D);\item AI-assisted records on public challenges, including the RSA Factoring Challenge, the Darmstadt SVP and LWE challenges, the ECDLP ladder and the FFDLP-512 instance that Alice \& Bob propose as an intermediate benchmark [232], a full run of Shor's algorithm at a bit-flip requirement about 600 times less demanding than RSA-2048's (1.9 months against 101 years) and at a size whose 512-bit parameters the 2015 Logjam scan still found in production, as a direct read on the research-acceleration factor R.\end{itemize}
Each of the first five indicators maps to one of the four channels, and the sixth and seventh map to the software driver, so an observation that beats or misses a milestone updates a specific acceleration factor rather than the whole projection at once. This converts the timeline from a single prediction into a tracker, consistent with the spectrum-of-risk-dates framing adopted in Section 3.1.

These indicators define the model’s explicit update mechanics. The intended protocol includes the following: 

\begin{itemize}\item a quarterly recalibration cycle, with an event-triggered rerun whenever the structural-claim indicator records a verified result, and no shorter published cadence: \item each observed milestone is mapped to its channel; \item the corresponding required factor Fᵢ, baseline rate rᵢ, or acceleration factor Aᵢ is re-estimated from the new anchor; the distribution over T is recomputed; and \item the revised parameter set is published as a versioned rerun alongside its predecessors, so that conformance to, or deviation from, the previously modeled curves is itself preserved as calibration data. \end{itemize}
Over successive update cycles this procedure narrows the acceleration-factor ranges toward what the accumulating historical record supports, in the same spirit in which expert-elicitation surveys are re-fielded and published resource estimates are revised. This periodic update protocol is part of what qualifies the model as an estimation instrument that improves with observation rather than a static projection. An independent review of the first version places the paper's claim on a rung of evidential warrant it calls the living instrument, earned only if the versioned reruns appear [234]; the quarterly cadence and event triggers above are that commitment, and the reproduction archive of Appendix A lets outside teams run the same protocol and publish competing reruns.

Finally, the model makes the paper’s central claim explicit and bounded: it accelerates the arrival of fault-tolerant Shor’s algorithm, the one method this paper does not rule out, rather than creating a new near-term attack. The methods evaluated in Sections 2.3 and 2.4, quantum annealing and the variational quantum eigensolver, do not contribute a standalone break, and the linear-algebra and geometric-algebra approaches of Sections 2.1 and 2.2 contribute at most polynomial assistance or remain exploratory. The novel content of the timeline is therefore concentrated in the acceleration factors above, which is precisely why they are stated as named, anchored, and falsifiable quantities, rather than being folded into a single qualitative assumption.

Reproducibility versus validation. Appendix A permits exact regeneration of every curve from the stated equations and seeds; that is reproducibility, and it is distinct from validation. Three levels of validation should be kept apart. First, outcome validation, whether reality reaches a modeled crossover, is prospective: the earliest testable event is the RSA-2048 crossing of 2029 to 2032, so as of November 2026 no projection has yet been checked against its own predicted event. Second, concordance validation is available now and is reported in Table 4: the model's crossover ranges are consistent with five independent expert forecasts, standards timelines and vendor roadmaps (Mosca and the Global Risk Institute, NIST IR 8547, CNSA 2.0, IBM, and Google/Gidney), which is convergent support rather than proof; the Gershteyn and Alber blend added to Table 4 in v3 is the one discordant row, and Section 5 states why. Third, leading-indicator validation is also available now: the seven tracking indicators of this section, observable over the next eighteen to thirty-six months and each mapped to a channel or to the software driver, give early corroborating or falsifying signal well before 2030, and one of them, the fall in the published resource estimate from roughly twenty million to under one million qubits, is a retrodictive anchor the model's trend already matches. The quarterly recalibration protocol above turns these into a standing check. Accordingly, the auditability and reproducibility claims describe repeatability; outcome correctness remains open, to be established by the tracking protocol as the record accumulates. A second, like-for-like retrodiction is developed in Section 4.5: the observed decline of the quantum sieving exponent runs about seven times too slow to reach the sieving floor by 2033, independent historical support for the contingency on the collapse parameter.

\section*{4. Using AI to Project the Advancement of CRQC+AI Vulnerabilities}
\addcontentsline{toc}{section}{4. Using AI to Project the Advancement of CRQC+AI Vulnerabilities}
The rapid rise in the capability and reach of AI in recent years has opened previously unforeseeable opportunities for improvements in quantum algorithms and quantum computing technologies. Both AI-enhanced CRQCs and AGI-enhanced CRQCs are expected to advance the design, optimization and innovation of quantum algorithms and computing technologies. This section presents the CRQC+AI vulnerability timelines generated by the released scripts of Appendix A.3 from author-specified parameters and equations; a frontier model (Anthropic Claude Fable 5.1) [150] was used for drafting and review, as the provenance statement at the end of this section records. These author-parameterized charts (all modeling assumptions, parameters, and calculations author-controlled) evaluate four different scenarios with respect to the risk profile for cracking classical asymmetric RSA/ECC public key encryptions and PQC key encapsulation (FIPS 203) and digital signatures (FIPS 204), with two additional algorithms evaluated in Appendix C and not in the body of this paper, hash-based signatures (FIPS 205) and AES symmetric encryption (FIPS 197). [151] [152]

The end-of-2028 horizon used in these scenarios reflects publicly stated developer timelines rather than an assumption unique to this paper. Anthropic’s founder and chief executive has projected powerful AI on a 2026 to 2027 timeline, [153] [154] and its submission to the United States Office of Science and Technology Policy places such systems within the current administration term, that is, before the end of 2028. [155] These developer timelines are echoed across the field by forecasts that bracket the 2028 horizon rather than uniformly confirming it: Google DeepMind’s Chief AGI Scientist has for over a decade assigned roughly even odds to AGI by 2028, [156] DeepMind’s chief executive gives a somewhat longer five-to-ten-year horizon, [157] the chief executive of Microsoft AI expects AI to reach human-level performance on most professional, computer-based tasks within roughly 12 to 18 months, a professional-task horizon near 2027 rather than a forecast of full general intelligence, [158] and a former OpenAI researcher argues AGI is plausible by 2027. [159] A prominent dissent holds that current large language models are not on a path to human-level intelligence and that AGI would require different architectures on a longer horizon. [160] Such forecasts come from interested parties and remain contested, and the horizon is adopted here only as a scenario input rather than a settled date; Track 0 accordingly retains a fully no-AGI baseline for readers who do not accept the premise. As of 2 November 2026 Metaculus places the community median for weakly general AI at 28 August 2027, with the question still open while its resolution is reviewed, and for the stricter general-AI question at 18 September 2030 [209] [210]; the AI Futures Project's August 2026 update puts its three authors' medians for an automated coder, the point at which a leading laboratory would rather drop its human software engineers than forgo AI coding, between November 2027 and January 2030 [211]; and a non-interested party, Duminil-Copin, has written publicly that frontier models are far better than he is at many mathematical tasks and that AI will probably settle his field's central open problem before humans do [194]; on that record the lower tail of the onset distribution of Appendix A.5 is widened to 2026 while the 2028 mode is kept. The four different scenarios shown in each chart track the risk threats for breaking the relevant encryption algorithms under different assumptions based on AGI being available for CRQC by the end of 2028:

\newpage
\begin{itemize}\item Track 0: No-AGI baseline (zero AGI-gated software capability).\item Track 1: Low AGI-driven software capability.\item Track 2: Moderate AGI-driven software capability.\item Track 3: High AGI-driven software capability (most aggressive regime).\end{itemize}
These four scenario tracks are distinct from the four engineering channels of Section 3.3: the channels decompose engineering progress toward a CRQC, whereas the tracks grade the intensity of the AGI-gated software driver (its rate r\_k and ceiling B\_k in Appendix A). The four tracks share one functional form and differ only in that intensity, so the labels above denote capability level rather than distinct attack mechanisms. The runtime and feasibility curves in the figures that follow are produced by the combined hardware and software capability model documented in full, with parameters and pseudocode, in Appendix A; the cost model of Section 4.5 provides a separate analytical restatement in explicit cryptanalytic units; because it shares the same effective-dimension-collapse conjecture, it is a consistency check rather than independent corroboration.

The curves in Figures 6A/6B through 8A/8B, together with the Appendix C stress-case Figures C1A/C1B and C2A/C2B, are generated from a single parameterized capability model rather than from individual expert or analyst projections. Cryptanalytic capability is expressed as the accumulated orders-of-magnitude reduction in attack runtime over time and is modeled as the sum of two exponential drivers: a hardware driver representing the accelerating growth of effective quantum resources (Rose’s-Law scaling together with the qubit-count reductions surveyed in Section 3), that is present in every track because hardware progress is not contingent on AGI; and an AGI-gated software driver representing AI-assisted advances in cryptanalysis, which differs by track. The full specification, with parameters, calibration, and pseudocode, is given in Appendix A. One reading convention applies to all these figures: the vertical runtime axis is a normalized index, not a per-scheme physical estimate. Every scheme starts from the same baseline value R0 (Appendix A), so the curves support within-scheme trends and crossover timing, but differences between schemes on this axis do not represent their true differences in absolute attack cost, which span dozens of orders of magnitude and are accounted for in the cryptanalytic units of Section 4.5.

The Anthropic Claude Fable 5.1 AI model as of July 2026 was used as a drafting and visualization assistant for these charts and for an editorial review of the paper. All modeling assumptions, parameters, and calculations are author-controlled and reproducible from the equations and calibration tables given in Sections 3 and 4. Every curve in this paper is generated by the released scripts of Appendix A.3; the frontier model was used for drafting and review, not for curve generation, which is the deterministic path an independent review of the first version asked for [234].

\subsection*{4.1 RSA-2048 Estimates of CRQC+AI Vulnerability: Mechanism-Backed Tier}
\addcontentsline{toc}{subsection}{4.1 RSA-2048 Estimates of CRQC+AI Vulnerability: Mechanism-Backed Tier}
Figures 6A and 6B evaluate the runtime attack and modeled feasibility estimates for breaking RSA-2048 public key encryption under each of the different vulnerability tracks for the four different assumption scenarios. In TLS 1.3 specifically, RSA key transport was removed; RSA remains relevant for certificate authentication, where RSA signatures use RSASSA-PSS, and for legacy interoperability and downgrade exposure, while ephemeral elliptic-curve Diffie-Hellman (ECDHE) and ECDSA are the direct TLS 1.3 key-exchange and authentication mechanisms. The RSA-2048 curves are read as the mechanism-backed tier for the ECDHE and ECDSA primitives as well, as a conservative proxy for ECC-256 (Section 3.1.2.6).

\begin{figure}[H]\centering
\includegraphics[width=0.82\linewidth]{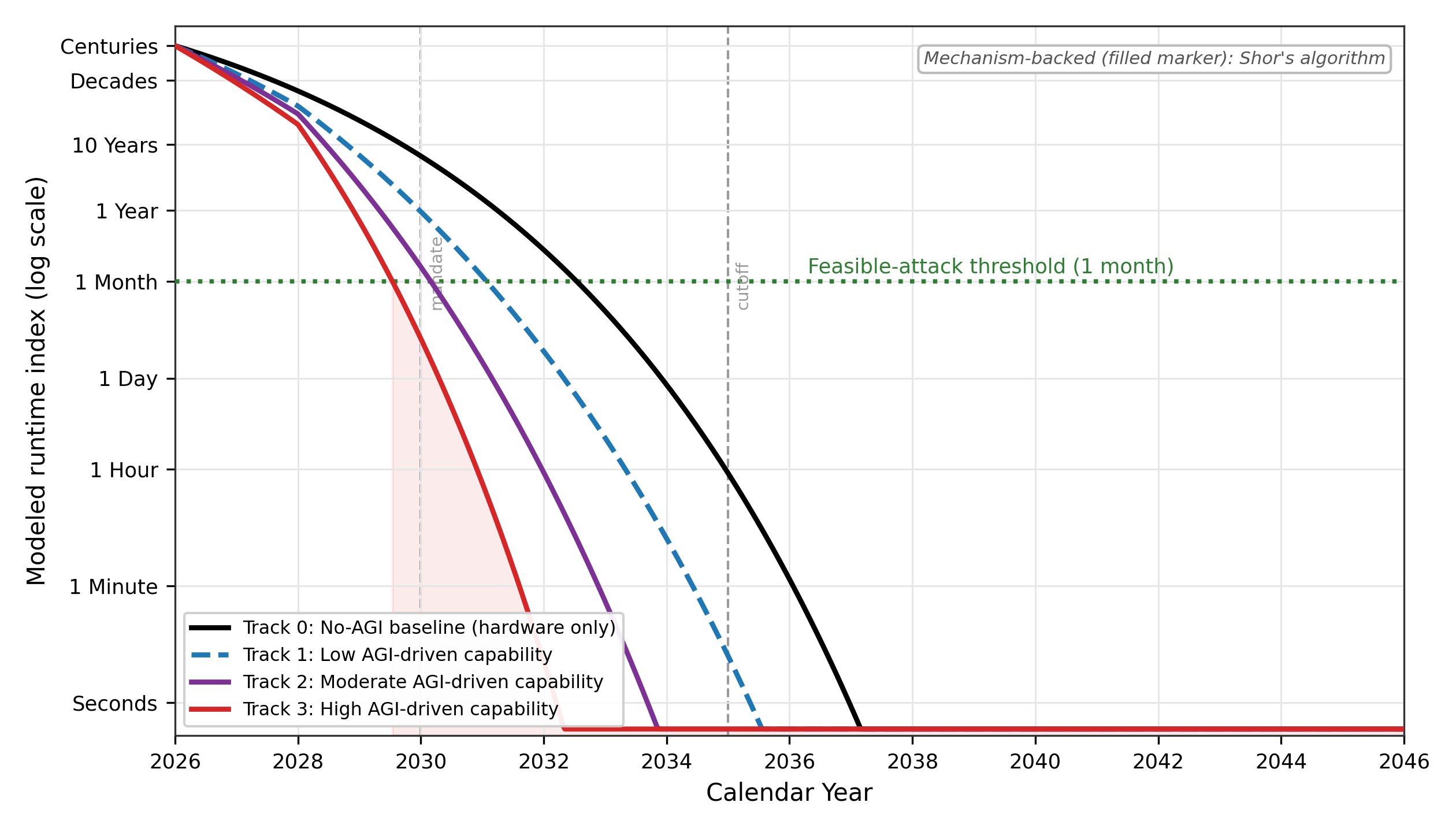}
\par\vspace{3pt}{\small\textbf{Figure 6A: Projected Attack Runtime Against RSA-2048 (classical RSA).  \\[2pt]Certificate Authentication / Legacy TLS Exposure}}
\end{figure}
\begin{center}\textit{The vertical runtime axis is a within-scheme index on a common baseline (R0); }\end{center}
\begin{center}\textit{values are not comparable across schemes (see Section 4).}\end{center}
\begin{figure}[H]\centering
\includegraphics[width=0.82\linewidth]{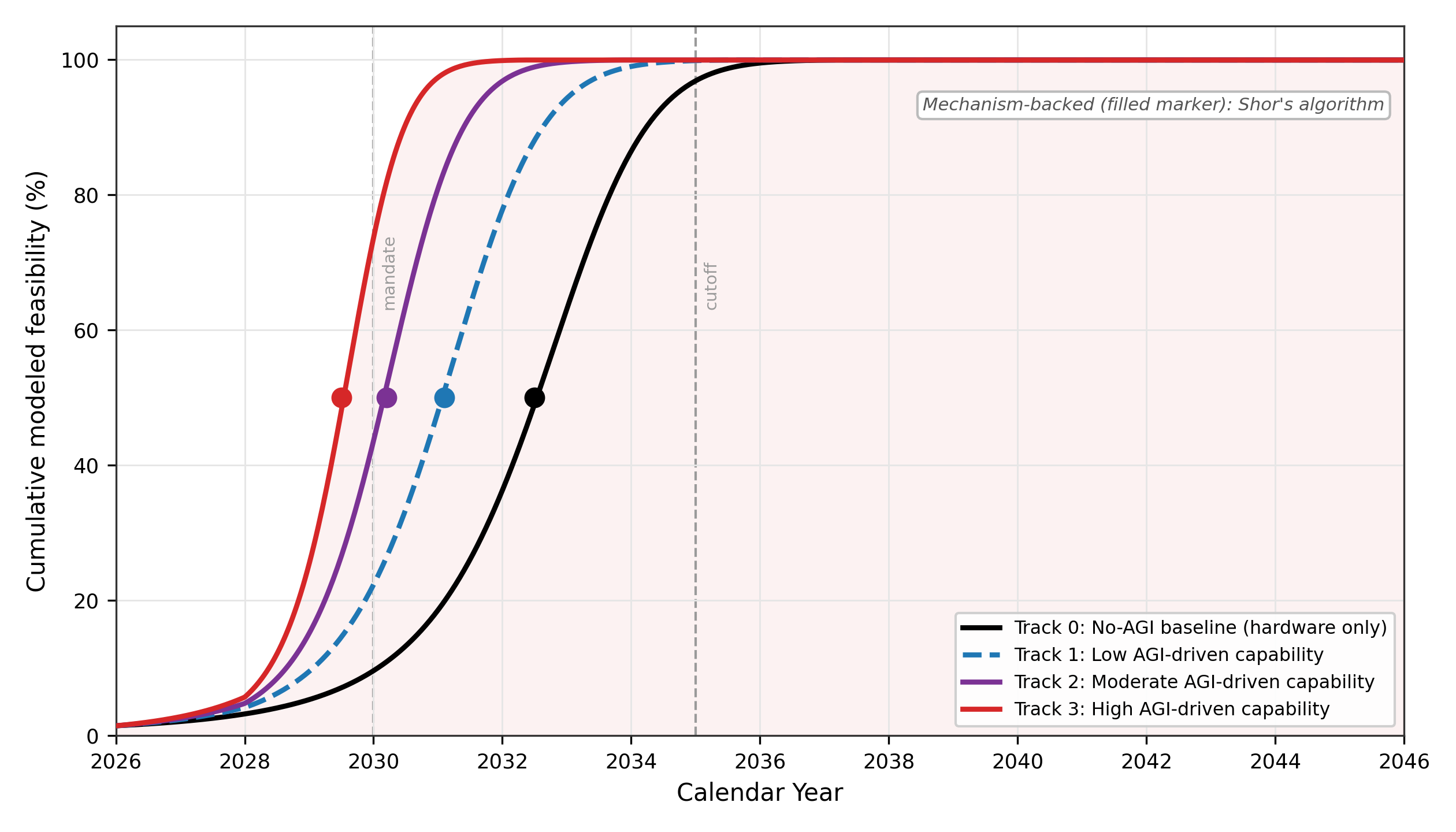}
\par\vspace{3pt}{\small\textbf{Figure 6B: Modeled Feasibility Estimates for RSA-2048 (classical RSA) \\[2pt]Certificate Authentication / Legacy TLS Exposure}}
\end{figure}
\subsection*{4.2 ML-KEM Kyber-768/1024 Estimates of CRQC+AI Vulnerability (FIPS 203): Contingency Tier }
\addcontentsline{toc}{subsection}{4.2 ML-KEM Kyber-768/1024 Estimates of CRQC+AI Vulnerability (FIPS 203): Contingency Tier }
Figures 7A/7B and 7C/7D evaluate the runtime attack and modeled feasibility estimates for breaking Kyber-768 and Kyber-1024 public key encryption (FIPS 203) under both the most common configuration in TLS 1.3, the hybrid X25519 + ML-KEM-768 group, which major browsers enable by default and which Cloudflare reports as its widely deployed post-quantum key agreement, with over half of human-initiated traffic to Cloudflare now post-quantum encrypted, [71] and the standalone ML-KEM-1024 configuration required by the National Security Agency (NSA) under Commercial National Security Algorithm Suite 2.0 (CNSA 2.0), [164] under each of the different vulnerability tracks for the four different assumption scenarios.

Because TLS 1.3 runs on ML-KEM-768 (NIST category 3) in hybrid with classical ECDH (X25519) [161] and combines the two public key exchange results, an attacker must beat both the lattice-based ML-KEM component and the classical ECDH component to derive the results of the key exchange. Figures 7A/7B evaluate the runtime attack and modeled feasibility estimates for breaking Kyber-768 in this hybrid context with ECDH X25519. The 7A/7B curves are numerically identical to the standalone ML-KEM-1024 curves of Figures 7C/7D: the model applies the same lattice couplings to both configurations, and the hybrid falls only when its lattice component falls (Appendix A.2), so the two figure pairs are one analysis shown for two deployment configurations, not two independent lines of evidence. This also fixes what the hybrid does and does not hedge. Against a classical structural break that reaches ML-KEM but not X25519, the hybrid hedges until a cryptographically relevant quantum computer exists; against a fault-tolerant machine that runs the lattice attack and Shor's algorithm against X25519 at the relevant sizes, it does not, and the exchange-side hedge is then a second post-quantum KEM from a different mathematical family that the attack in question does not reach, such as code-based HQC once standardized. A signature scheme such as SLH-DSA diversifies authentication, not key exchange.

\begin{figure}[H]\centering
\includegraphics[width=0.82\linewidth]{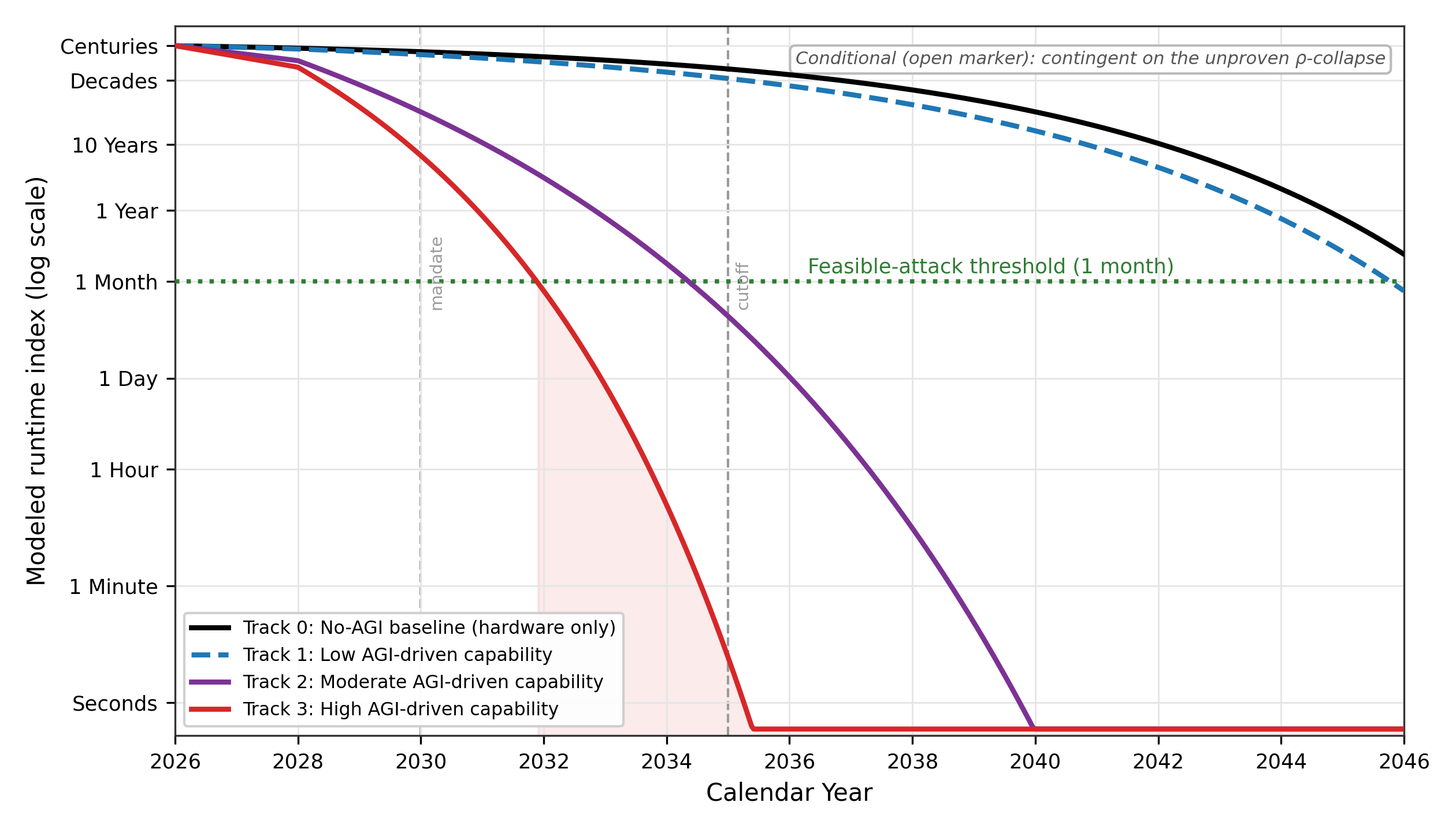}
\par\vspace{3pt}{\small\textbf{Figure 7A: Projected Attack Runtime Against Hybrid Kyber-768\\[2pt]ML-KEM and X25519 ECDH: Most Common TLS 1.3 Key Exchange for FIPS 203}}
\end{figure}
\begin{center}\textit{The vertical runtime axis is a within-scheme index on a common baseline (R0);}\end{center}
\begin{center}\textit{values are not comparable across schemes (see Section 4).}\end{center}
\begin{figure}[H]\centering
\includegraphics[width=0.82\linewidth]{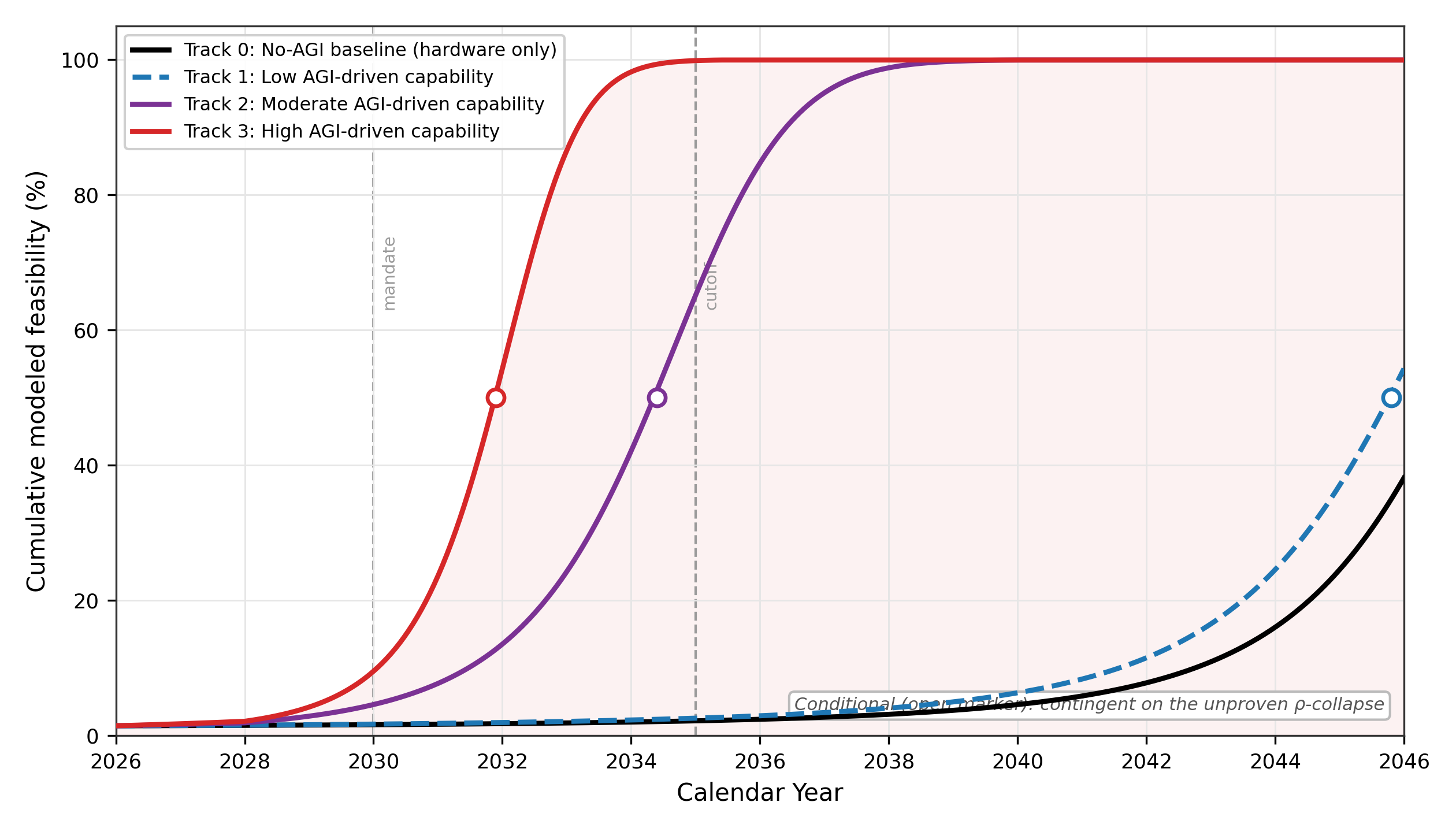}
\par\vspace{3pt}{\small\textbf{Figure 7B: Modeled Feasibility Estimates for Hybrid Kyber-768 ML-KEM \\[2pt]and X25519 ECDH: Most Common TLS 1.3 Key Exchange for FIPS 203}}
\end{figure}
\begin{center}\textit{Conditional scenario only: the }\textit{Kyber-768 + X25519}\textit{ curves assume an undiscovered effective-dimension-collapse mechanism of approximately ρ ≥ 0.52. No such mechanism is currently known against standardized lattice schemes.}\end{center}
Figures 7C/7D evaluate the runtime attack and modeled feasibility estimates for breaking Kyber-1024 public key encryption as a standalone (PQ-only) ML-KEM-1024 key agreement (NIST category 5), consistent with current IETF TLS Working Group Internet-Drafts. TLS 1.3 standardization currently includes separate Working Group drafts for hybrid and standalone post-quantum key agreement: the hybrid draft defines combinations such as SecP384r1MLKEM1024, [161] while the standalone draft defines ML-KEM-1024 as a PQ-only NamedGroup, [162] both still Internet-Drafts rather than final RFCs. This standalone configuration is relevant to CNSA 2.0-aligned deployments, in which NSA specifies ML-KEM-1024 (not ML-KEM-768); CNSA 2.0 should be read as specifying required quantum-resistant algorithms and strength levels for applicable high-assurance use cases rather than as a blanket prohibition on hybrid key establishment. [163] [164]

\begin{figure}[H]\centering
\includegraphics[width=0.82\linewidth]{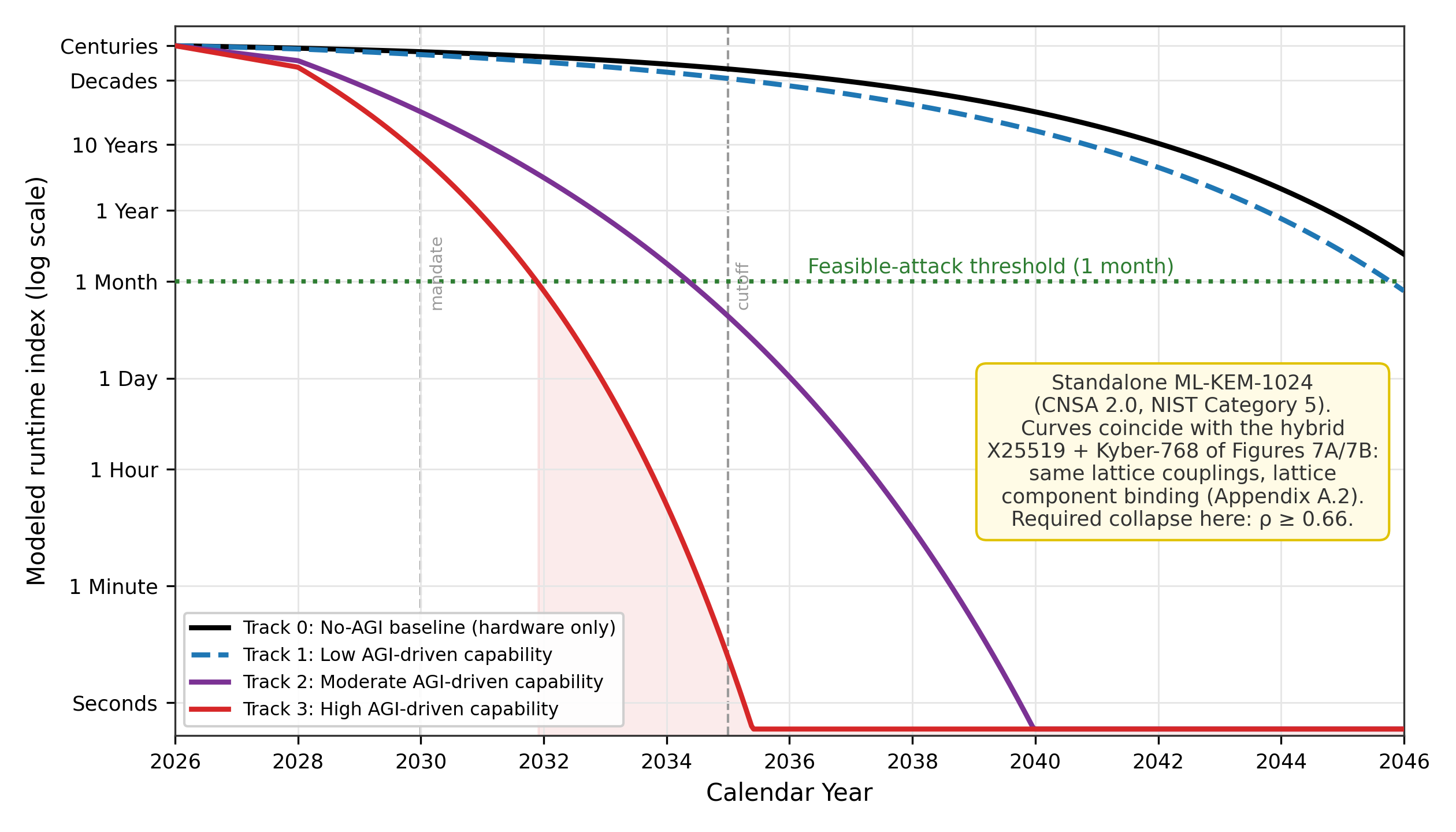}
\par\vspace{3pt}{\small\textbf{Figure 7C: Projected Attack Runtime Against Standalone Kyber-1024\\[2pt]ML-KEM-1024 (NIST Category 5): CNSA 2.0 TLS 1.3 Key Exchange for FIPS 203}}
\end{figure}
\begin{center}\textit{The vertical runtime axis is a within-scheme index on a common baseline (R0); }\end{center}
\begin{center}\textit{values are not comparable across schemes (see Section 4).}\end{center}
\begin{figure}[H]\centering
\includegraphics[width=0.82\linewidth]{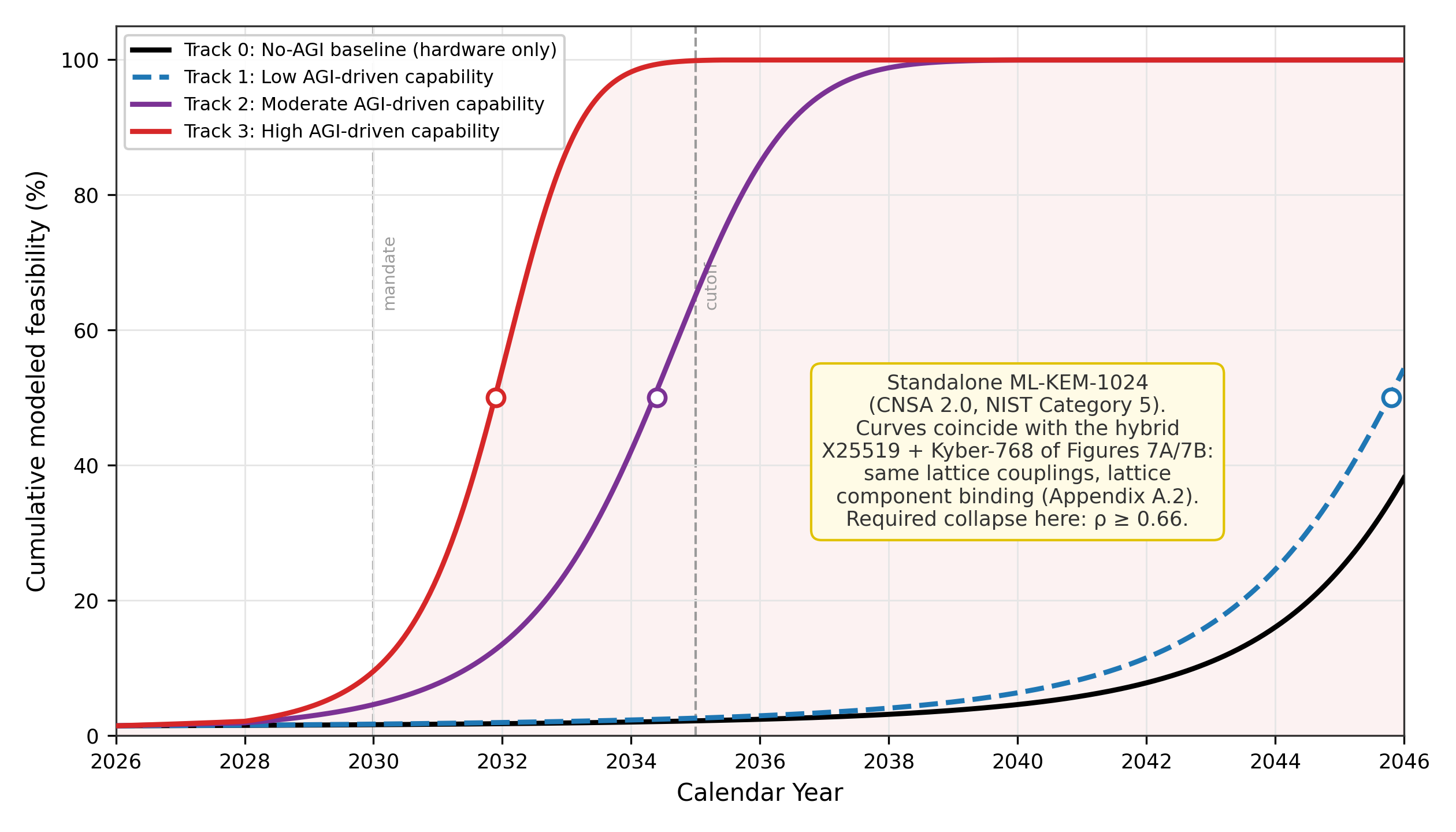}
\par\vspace{3pt}{\small\textbf{Figure 7D: Modeled Feasibility Estimates for Standalone Kyber-1024 \\[2pt]ML-KEM-1024 (NIST Category 5): CNSA 2.0 TLS 1.3 Key Exchange for FIPS 203}}
\end{figure}
\begin{center}\textit{Conditional scenario only: the Kyber-1024 standalone curves assume an undiscovered effective-dimension-collapse mechanism of approximately ρ ≥ 0.66. No such mechanism is currently known against standardized lattice schemes.}\end{center}
\subsection*{4.3 ML-DSA-44 Estimates of CRQC+AI Vulnerability (FIPS 204): Contingency Tier}
\addcontentsline{toc}{subsection}{4.3 ML-DSA-44 Estimates of CRQC+AI Vulnerability (FIPS 204): Contingency Tier}
Figures 8A and 8B evaluate the runtime attack and modeled feasibility estimates for breaking ML-DSA-44 (FIPS 204) under each of the different vulnerability tracks for the four different assumption scenarios.

\begin{figure}[H]\centering
\includegraphics[width=0.82\linewidth]{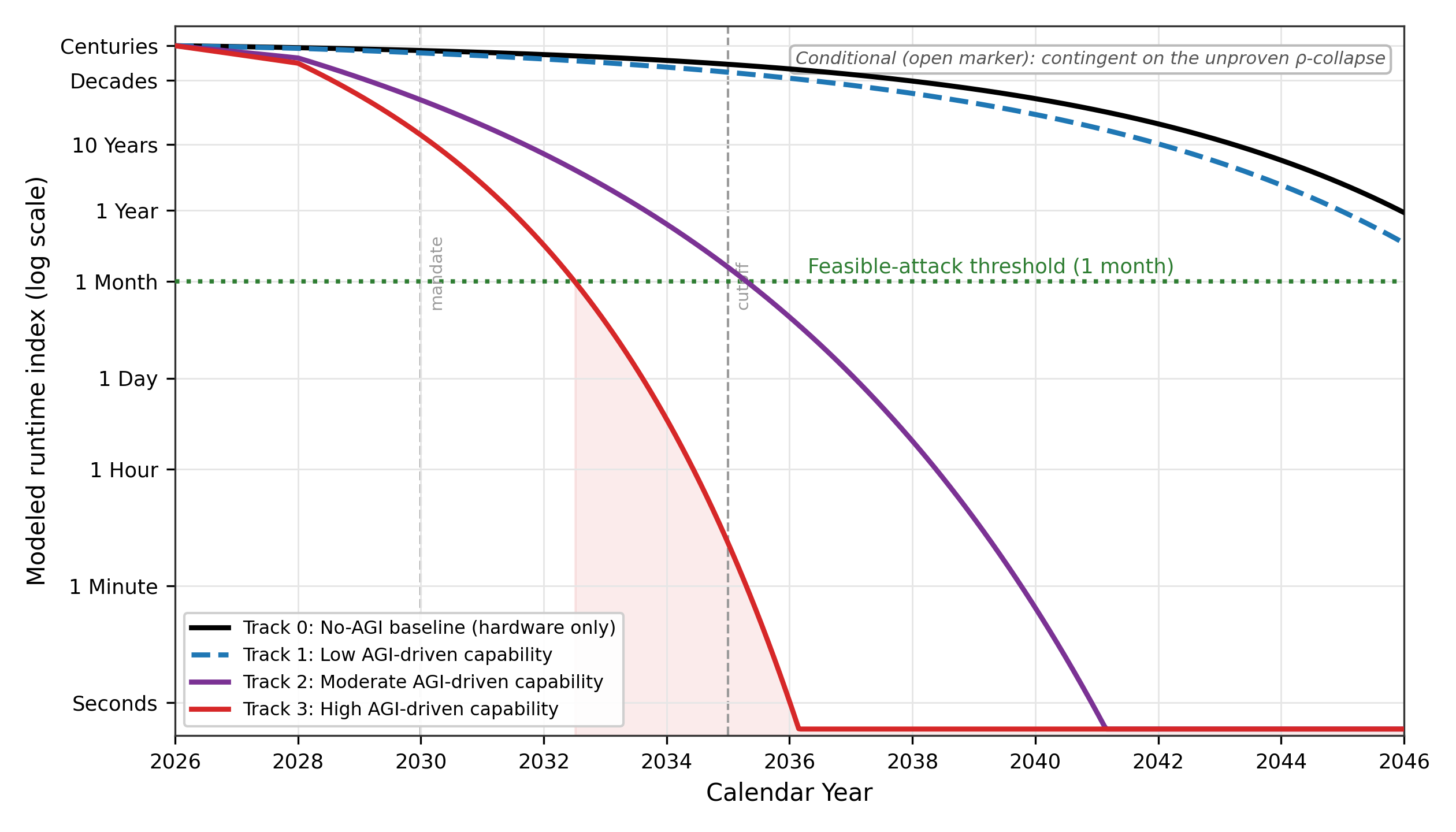}
\par\vspace{3pt}{\small\textbf{Figure 8A: Projected Attack Runtime Against ML-DSA-44 FIPS 204 \\[2pt]TLS 1.3 Certificate Signatures}}
\end{figure}
\begin{center}\textit{The vertical runtime axis is a within-scheme index on a common baseline (R0); }\end{center}
\begin{center}\textit{values are not comparable across schemes (see Section 4).}\end{center}
\begin{figure}[H]\centering
\includegraphics[width=0.82\linewidth]{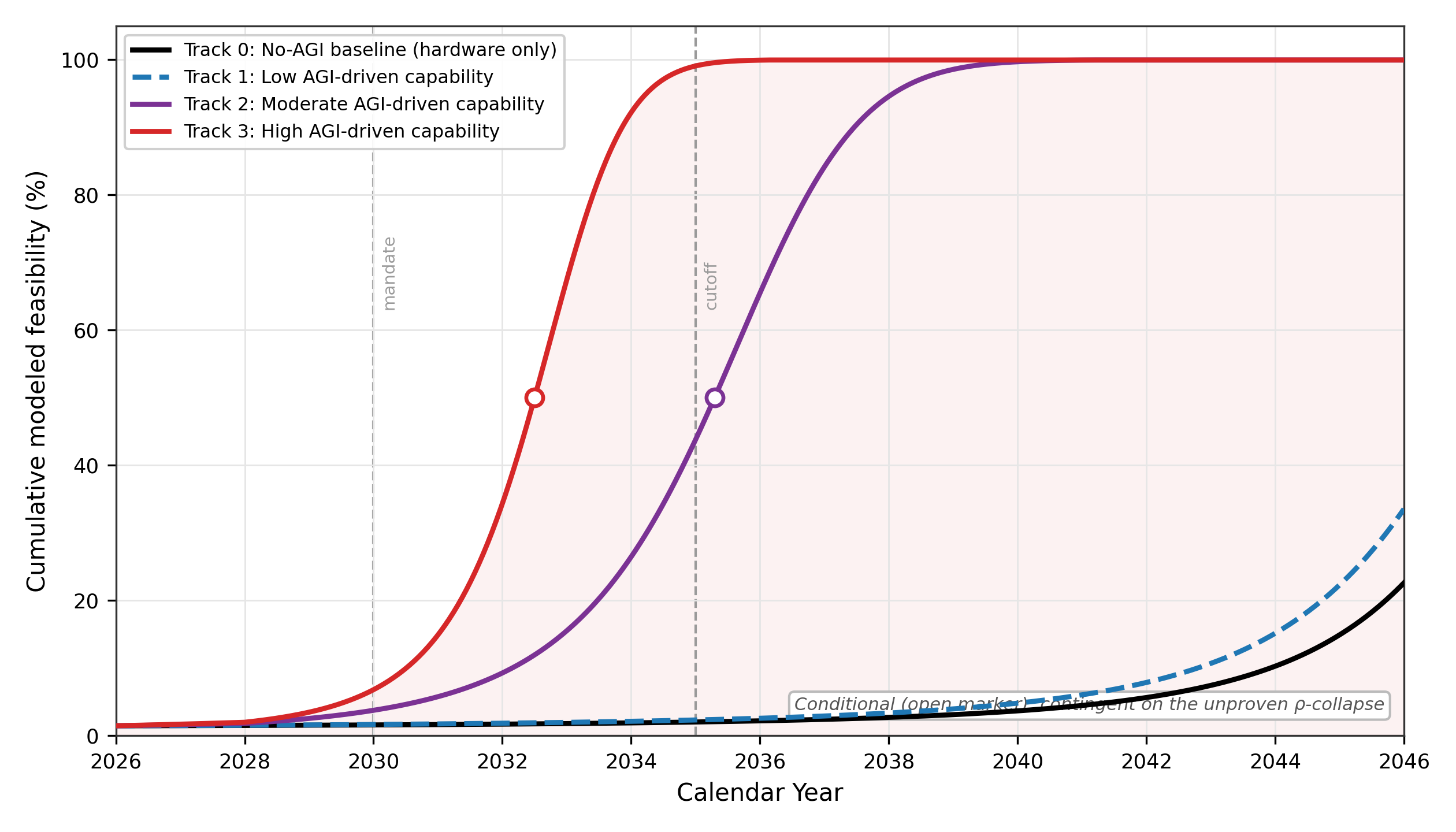}
\par\vspace{3pt}{\small\textbf{Figure 8B: Modeled Feasibility Estimates for ML-DSA-44 FIPS 204 \\[2pt]TLS 1.3 Certificate Signatures}}
\end{figure}
\begin{center}\textit{Conditional scenario only: the ML-DSA-44 curves assume an undiscovered effective-dimension-collapse mechanism of approximately ρ ≥ 0.17. No such mechanism is currently known against standardized lattice schemes.}\end{center}
\subsection*{4.4 Estimates of CRQC+AI Vulnerability for SLH-DSA (FIPS 205) and AES-256 (FIPS 197) }
\addcontentsline{toc}{subsection}{4.4 Estimates of CRQC+AI Vulnerability for SLH-DSA (FIPS 205) and AES-256 (FIPS 197) }
The remaining two TLS 1.3 primitives, SLH-DSA (FIPS 205) and AES-256 (FIPS 197), show no mechanism-backed or contingency-backed exposure in this model. Their hypothesis-only stress cases, including runtime and feasibility Figures C1A/C1B and C2A/C2B, are presented in Appendix C so that the figures in the body of this paper remain confined to the two better-evidenced tiers.

\subsection*{4.5 Cryptanalytic Cost Model and the Dimension-Collapse Contingency}
\addcontentsline{toc}{subsection}{4.5 Cryptanalytic Cost Model and the Dimension-Collapse Contingency}
The per-algorithm figures above are generated by the combined hardware and software capability model documented in Appendix A. This subsection is a separate analytical restatement of the lattice and hash-based tracks, not the generator of those figures: it restates the conditional break in explicit cryptanalytic units so that the single speculative assumption is isolated and quantified rather than hidden inside the shape of a curve. The two instruments agree by sharing the same conditional assumption, an undiscovered effective-dimension collapse, rather than by one being derived from the other. Building on the core-SVP cost of Section 3.1.3, the feasibility of a lattice break in calendar year t is expressed through three parameters, each carrying a concrete meaning:

\begin{itemize}\item \textbf{Δ (exponent erosion): }an AI-discovered reduction of the quantum sieving exponent below its current value of 0.265, bounded below by a conjectured sieving floor so that 0.265 − Δ ≥ 0.2075. This term captures faster algorithms but cannot, by itself, change the exponential form of the cost.\item \textbf{This floor }is a \textit{resource} floor induced by the minimal sieve-list size, |L| ≈ 2\textsuperscript{0.2075·d} in the 2-sieve family (Kirshanova and Laarhoven, Lower Bounds on Lattice Sieving and Information Set Decoding, CRYPTO 2021, IACR ePrint 2021/785), not an optimality result on the time exponent. It is family-specific: 3-tuple sieving trades to ≈ 2\textsuperscript{0.1887·d} memory at ≈ 2\textsuperscript{0.2846·d} time, which does not change the argument directionally. The list must be produced and traversed, so it induces a floor on the operation count against which the budget B(t) is measured.\item \textbf{ρ (effective-dimension collapse): }the payload of the geometric-algebra hypothesis of Section 2.2, modeled as a collapse of the core-SVP block size β (distinct from the capability-model coupling β of Appendix A), β → β(1 − ρ). It is the only speculative quantity parameterized in the model; the dihedral-coset route of Appendix D is the second structural pathway and is carried as a tier trigger rather than as a parameter. ρ has no known realization against random q-ary lattices; it is exposed deliberately so that the assumption can be examined. An independent review of the first version read ρ the same way, as a quantity to be produced by a construction rather than inferred from the attractiveness of a narrative, and the contingency-tier dates as illustrations of required magnitude rather than forecasts [234].\item \textbf{B(t) (feasible attack budget): }the base-2 logarithm of the quantum operations an adversary can afford in year t, increasing with the hardware driver (Rose’s-Law scaling and the qubit-count reductions surveyed in Section 3).\end{itemize}
A scheme with core block size β becomes feasibly breakable in the first year for which the eroded, dimension-collapsed work factor falls within budget:

\begin{center}(0.265 − Δ) · β · (1 − ρ) ≤ B(t)     [lattice break condition]\end{center}
The condition prices two drivers, the hardware budget and cryptanalytic structure. Implementation compromise is a third, independent of ρ and of the AGI onset, and it is not priced here: the qualitative judgment of Section 3.2.4 stands, that compromise of a deployed PQC stack before 2035 is more probable than a structural lattice break, and a quantified term D\_impl(t) is planned for a later rerun.

The two drivers named above take explicit form. The hardware budget grows linearly in log space (a constant-doubling, Rose’s-Law form), while the software erosion is gated on the assumed AGI onset t\textsubscript{AGI} (end of 2028):

\begin{center}B(t) = B\textsubscript{0} + r · (t − t\textsubscript{0})\end{center}
\begin{center}Δ(t) = Δ\textsubscript{max} · (1 − e\textsuperscript{−R·(t − t\_AGI)})   for t ≥ t\textsubscript{AGI},   Δ(t) = 0 otherwise\end{center}
Here R is a research-acceleration factor: the number of human-cryptanalyst-years of progress compressed into one calendar year by advanced AI. Track 0 sets R = 0 (no AGI); Tracks 1 and 2 raise R; Track 3 is the most aggressive regime; its capability-model counterpart in Appendix A is the highest fixed track rate (r\_k = 1.30), whose compounding produces the sharply convex capability accumulation visible in the Track 3 curves. The superintelligence premise is expressed through R rather than through any intelligence score, because a research-acceleration factor is measurable in principle (the quantum sieving exponent did fall over roughly fifteen years of human work, so a baseline rate exists to accelerate), whereas a normalized intelligence score has no defined meaning in the regime contemplated here. If a narrative anchor is wanted, an intelligence far beyond the human range is taken to drive R to large values; this anchor is explicitly illustrative, and no crossover year is derived from it. Section 3.2.1.2 records the first dated instances in which R was observed above zero without an AGI: two reduced-target cryptanalytic results found autonomously in 2026.

The distinction that matters for R is between formal assurance and empirical cost. A formalized theorem can be machine-checked against its stated axioms; the check covers the formal statement, not the paper around it, and only part of a release such as OpenAI's of October 2026 is formalized at all. Attack correctness and cost bounds can in principle be formalized too, but the practical cost of an attack on real hardware is an empirical question that no proof assistant settles. Proof-finding is therefore the better-measured side of the capability, and attack-finding, which this model prices, rests on a smaller and less checkable record.

For transparency, the two anchor values used in the worked examples below, B(2033) = 80 bits and B(2036) ≈ 92 bits, determine the budget line completely: B\textsubscript{0} = 52 bits at t\textsubscript{0} = 2026 and r = 4 bits per year, equivalent to a doubling time of roughly three months in usable attack operations. This is an aggressive scenario input, reflecting compounded hardware, engineering, and orchestration gains under the CRQC+AI assumption, and it should be read as such rather than as a forecast of any vendor roadmap.

\textbf{Worked example}\textbf{s}\textbf{, ML-KEM}\textbf{-768/1024}\textbf{ (Kyber-}\textbf{768 hybrid/1024 standalone}\textbf{, FIPS 203). }For Kyber-1024 (standalone) with β ≈ 877 and quantum exponent 0.265, take an already aggressive illustrative budget of B(2033) = 80 bits. The break condition becomes:

\begin{center}(0.265 − Δ)(1 − ρ) ≤ 80 / 877 ≈ 0.091\end{center}
Two limiting readings follow. With no geometric collapse (ρ = 0), the requirement 0.265 − Δ ≤ 0.091 would force the sieving exponent below the conjectured floor of 0.2075; even at that floor the work factor is 2\textsuperscript{0.2075·877} ≈ 2\textsuperscript{182}, still above the budget, so algorithmic erosion alone cannot reach 2033. With no exponent erosion (Δ = 0), the requirement becomes:

\begin{center}(1 − ρ) ≤ 0.091 / 0.265 ≈ 0.344     ⟹     ρ ≥ 0.66\end{center}
In other words, under real complexity numbers a 2031 to 2034 Kyber-1024 break is mathematically contingent on the unproven geometric embedding reducing the effective lattice dimension by roughly two thirds. The model does not assert that such a collapse exists; it states precisely how large it would have to be. For the hybrid Kyber-768 + X25519 configuration that is the widely deployed post-quantum TLS default (β ≈ 625), the same budget requires ρ ≥ 0.52, and the classical X25519 component must additionally be broken.

A retrodiction against the historical record supports this reading. The best-known quantum sieving time exponent has fallen only from about 0.2653 in 2015 to 0.2563 in 2023 (Laarhoven; Chailloux and Loyer, ASIACRYPT 2021; Bonnetain, Chailloux, Schrottenloher and Shen, EUROCRYPT 2023), a human-progress rate near 1.2 × 10⁻³ per year. Extrapolated forward from the model's working exponent of 0.265 at its 2026 reference year, that rate reaches the conjectured floor of 0.2075 only around 2074 (anchoring instead on the 2023 literature value of 0.2563 gives roughly 2064 to 2066; the choice of anchor does not change the conclusion); reaching it by 2033 would demand roughly seven times the historical rate for the exponent term alone, and even at the floor the work factor remains 2\textsuperscript{182}. The observed pace of algorithmic erosion is therefore about an order of magnitude short of what a near-term break by erosion would require, which is independent, historical support for the structural claim that a 2030s lattice break is contingent on the effective dimension collapse rather than on exponent erosion. The exponents quoted are landmark literature values; the calculation is reproducible from the released materials (Appendix A). For contrast, Xiong and Wang [233] moved the exponent of polynomial factorization over finite fields from about 1.805 to 1.553 in a single paper in October 2026. That is what fast exponent movement looks like in a neighboring area of algebraic complexity; it touches no deployed parameter set and is not an attack. Against that scale the quantum sieving exponent's movement of 0.009 over eight years is an observation, not an assumption. Figure 8C plots the observed rate against the rate a 2033 floor crossing would require.

\begin{figure}[H]\centering
\includegraphics[width=0.82\linewidth]{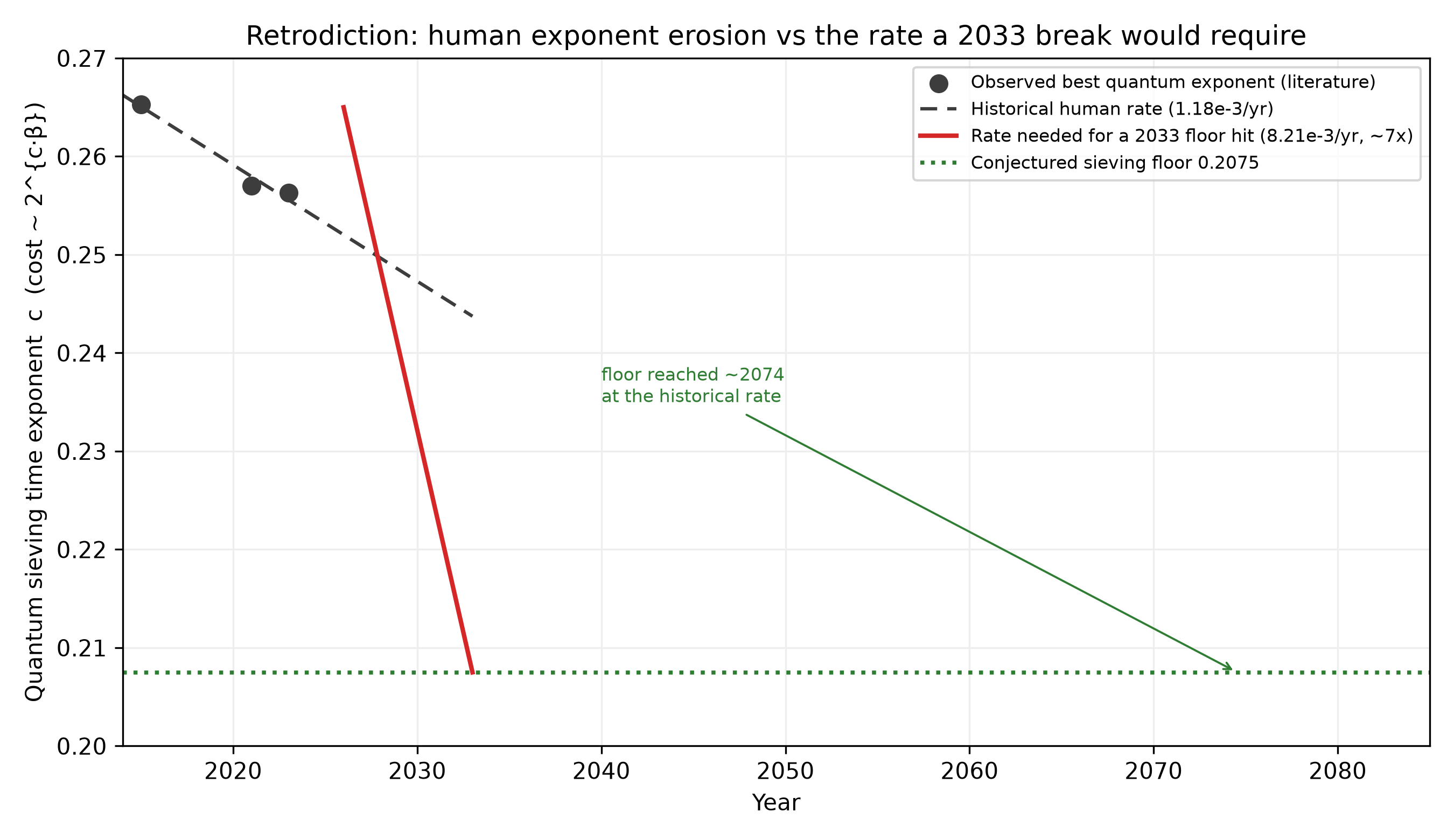}
\par\vspace{3pt}{\small\textbf{Figure 8C: Retrodiction of the quantum sieving time-exponent trend: observed human progress (about 1.2 × 10⁻³ per year) versus the rate a 2033 floor crossing would require (about seven times faster). Landmark literature values; reproducible from the released materials (Appendix A).}}
\end{figure}
\textbf{Worked examples, ML-DSA-44 (FIPS 204, Section 4.3) and SLH-DSA (FIPS 205, Appendix C).}\textbf{ }Applying the same condition to ML-DSA-44, whose security rests on Module-LWE and Module-SIS at a smaller block size (β on the order of 420), with a slightly larger budget B(2036) ≈ 92 bits, gives (1 − ρ) ≤ 92 / (0.265 · 420) ≈ 0.83, that is ρ ≥ 0.17 for a 2036 crossover: a smaller required collapse than Kyber, consistent with the later date and larger budget. SLH-DSA is structurally different. As a stateless hash-based signature it contains no lattice for ρ to act on; its only quantum exposure is Grover-type preimage and collision search, a quadratic effect already absorbed into its conservative parameters, with a classical work factor on the order of 2\textsuperscript{128} at the lowest security level. An idealized Grover search would need only about 2\textsuperscript{64} quantum evaluations, but the search is essentially sequential, and under the realistic circuit-depth limits that NIST applies (the same accounting given for AES-256 in Appendix C) the effective cost remains near the classical value, a level the hardware budget B(t) reaches only at approximately 2045, at the very edge of the projection horizon. The dated SLH-DSA values carried in earlier drafts (approximately 2036 under Track 3 and 2041 under Track 2) fell well before that; they were not produced by any concrete mechanism, rested on the same hypothesized novel-attack assumption flagged for AES-256 in Appendix C, and are withdrawn there in favor of the undated hypothesis-only treatment (zero software coupling, no crossover within the horizon). Applying the cost model therefore separates the figures into three honesty tiers: mechanism-backed (RSA and ECC, via Shor), contingency-backed (Kyber-768/1024 and ML-DSA-44, via the explicit and falsifiable ρ), and hypothesis-only (SLH-DSA and AES-256).

\textbf{Assumptions and limitations.}  The Track 0 result of Appendix A is itself a guard against alarmism: the model's own no-AGI baseline places every lattice and hash break beyond the horizon, so the near-term lattice crossovers are properties of the AGI-gated software driver and are presented as conditional scenarios, not predictions. 

\textbf{Memory-bounded break condition.} The break condition above prices time (operation count) only. Lattice sieving is also memory-bound: the sieve list holds ≈ 2\textsuperscript{0.2075·d} vectors, 2\textsuperscript{182} for ML-KEM-1024 (β = 877) and 2\textsuperscript{129.7} for ML-KEM-768, and a budget expressed as the base-2 logarithm of affordable operations does not purchase that memory. The 2\textsuperscript{182} that appears above as a time work-factor floor is therefore also the memory requirement; a structural collapse acting on the block size β must collapse the memory requirement as well, so the stated collapse magnitudes (for example ρ ≥ 0.66 for ML-KEM-1024) understate what the hypothesis must achieve. Adding the memory constraint would push the contingency-tier crossovers later and would strengthen the paper's claim not to be over-forecasting. Requiring the collapse to act on both time and memory, as an independent review of the first version also proposed [234], would move the contingency-tier crossovers later; it changes the cost model and every conditional crossover, so it is scheduled for the next versioned rerun rather than applied to the curves of this version.

\textbf{Both sieving exponents as a band.} Because the quantum sieving exponent 0.265 assumes idealized QRAM (Section 3.1.3 notes that the classical 0.292 may be closer to the operative exponent once realistic memory-access and circuit-depth costs are charged), the required collapse fractions are reported as a band. At 0.292 they rise to 0.5616 (ML-KEM-768), 0.6876 (ML-KEM-1024) and 0.2498 (ML-DSA-44), against 0.5170, 0.6558 and 0.1734 at 0.265; the ML-DSA-44 figure moves roughly 44\% in relative terms, which is not a rounding matter. Each fraction follows from ρ = 1 − B/(c·β), with c the sieving exponent (0.265 or 0.292), block sizes β = 625/877/420 and budgets B = 80 (ML-KEM, 2033) and 92 (ML-DSA-44, 2036).

\textbf{Cost-model boundary.} The cost model does not price several effects: dual attacks (cited but not incorporated), multi-target and amortized cost across many keys, decapsulation-failure attacks, the cost of discovering the geometric embedding as distinct from exploiting it, and classical cryptanalytic progress before AGI onset. Progress before the AGI onset is carried by the frontier-model term of Appendix A.2 at 0.10 per year (stated band 0.08 to 0.12), raised from the 0.04 pre-AGI human-progress value of v2 on the 2026 record of AI-discovered cryptanalytic results; because human lattice reduction has improved steadily for two decades, attributing the remainder of non-Shor progress to the AGI gate still overstates that parameter's explanatory weight.

\subsection*{4.6 Progress Timeframes for CRQCs and AI }
\addcontentsline{toc}{subsection}{4.6 Progress Timeframes for CRQCs and AI }
Given how quickly both quantum computing and AI have advanced in the last few years, the biggest opportunities for potential improvements in quantum algorithms or quantum computing technologies are likely to be driven by CRQC+AI. This section presents these Q-Day Spectrum threats to the different algorithms in a single chart and briefly highlights some of the potential improvements related to CRQC+AI.

\begin{figure}[H]\centering
\includegraphics[width=0.82\linewidth]{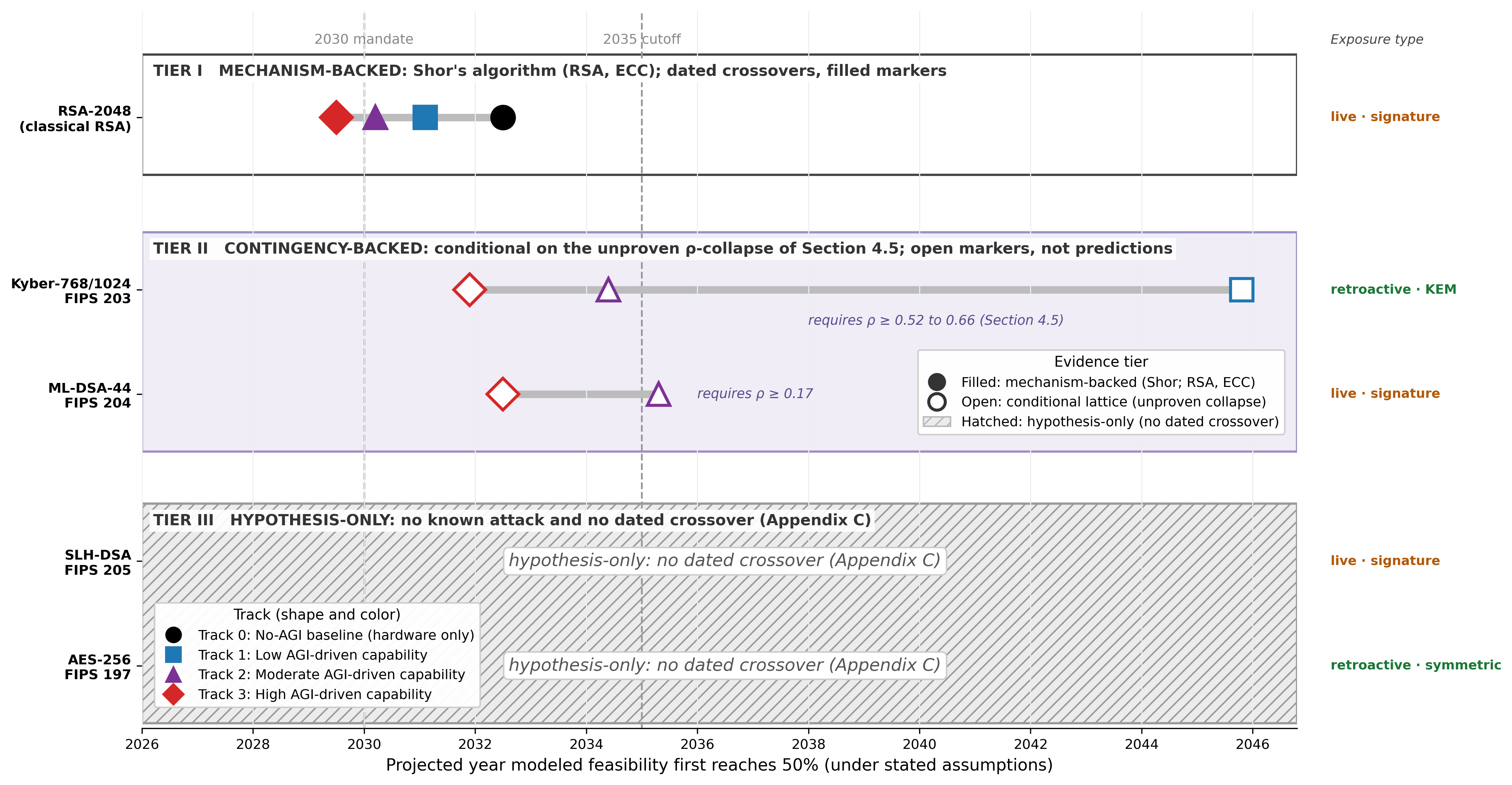}
\par\vspace{3pt}{\small\textbf{Figure 9: CRQC+AI Vulnerability Spectrum for TLS 1.3: Projected Year Modeled Feasibility \\[2pt]Crossover First Reaches 50\% by Algorithm and Track; hash-based and symmetric primitives are shown as hypothesis-only bands without dated crossovers}}
\end{figure}
\begin{center}\textit{Note: }\textit{the lattice crossovers shown, and the hash-based and AES-256 bands, are conditional stress scenarios}\textit{ under the stated model assumptions, not known attacks. Mechanism-backed risk applies only to RSA and ECC via Shor’s algorithm. The crossover years are conditional scenario outputs, not forecasts (Section 5). Every Track 2 and Track 3 crossover (and the marginal end-of-horizon Kyber-1024 Track 1 crossover) is gated on the assumed 2028 AGI onset: a two-year slip in that onset moves those crossovers later by approximately the same amount.}\end{center}
\textbf{Model-derived progress timeline: }Figure 9 consolidates the crossover estimates from Figures 6A/6B through 8A/8B and the Appendix C stress-case Figures C1A/C1B and C2A/C2B into a single view that defines the spectrum of risk the following subsections describe. The combined hardware-and-software model places the earliest feasible breaks against classical RSA-2048 in 2029 under the most aggressive Track 3 assumptions (2029.5 on the grid), with even the no-AGI Track 0 baseline crossing by roughly 2032 on hardware scaling alone. The lattice-based PQC schemes are essentially unaffected by Track 0 and Track 1 (neither hardware scaling nor low software-driven capability threatens them; the sole exception is a marginal Kyber-1024 Track 1 crossover at 2045.8, at the very edge of the horizon) and first become vulnerable in practice under the AGI-software tracks: Kyber-768/1024 (FIPS 203) in 2031 (Track 3) to 2034 (Track 2), and ML-DSA-44 (FIPS 204) in 2032 (Track 3) to 2035 (Track 2). The two parameter sets share a single plotted crossover only because the capability model assigns them equal couplings, not because their hardness is equal, since the cost model of Section 4.5 shows the Kyber-768 hybrid requires a smaller collapse (ρ ≥ 0.52) than standalone Kyber-1024 (ρ ≥ 0.66). The hash-based SLH-DSA (FIPS 205) and symmetric AES-256 (FIPS 197) are shown in Figure 9 only as hypothesis-only bands without dated crossovers, for the reasons given in Appendix C; both are more durable than the lattice schemes, with AES-256 the most resistant primitive evaluated. Their model-generated dates are reported, with the accompanying no-cryptanalytic-basis caveat, only in Appendix C. The phase windows below are framed to follow this model-derived sequence. These phases describe the conditional scenario sequence the model produces under its stated assumptions, not predicted events; the lattice phases are gated on the unproven ρ collapse and the hash-based and symmetric bands are hypothesis-only and inherit the honesty tiers of Section 4.5. They are also sensitive to the assumed 2028 AGI onset: because the lattice schemes depend almost entirely on the AGI-gated software driver, a two-year slip in that onset moves each of their Track 2 and Track 3 crossovers later by roughly one and a half to two years, and would move any hypothesis-only date for the hash-based and symmetric schemes by the same amount, while the hardware-driven RSA Track 0 crossover is unchanged. The three evidence tiers are drawn as separate bands with their own headers, so that the mechanism-backed RSA row is not read on the same footing as the conditional lattice rows or the hypothesis-only rows; the crossover values are those of Sections 4.1 to 4.5 and Appendix A.

\textbf{KEM and signature exposure are not commensurable.} A KEM break is \textit{retroactive} against harvested traffic and drives the HNDL argument, governed by Mosca's secrecy horizon X; a signature break requires a \textit{live} attacker against a certificate still within validity, governed by certificate and trust-anchor lifetime. The two therefore do not share a single feasibility axis: ML-DSA-44 represents live exposure and ML-KEM represents retroactive exposure. Read as a decision instrument, that gives two axes rather than one: KEM exposure against the secrecy horizon X, and signature exposure against certificate and trust-anchor lifetime, with the sector rows of Table 5 beneath each. Section 6.2 states the two axes in Mosca's terms; a figure drawn on them is left to a later version.

\subsubsection*{4.6.1 Hardware-Driven CRQC Scaling and Early RSA/ECC Exposure}
\addcontentsline{toc}{subsubsection}{4.6.1 Hardware-Driven CRQC Scaling and Early RSA/ECC Exposure}
\textbf{Hybrid Quantum-Classical Algorithms:} Quantum computing capabilities advance significantly with mainstream adoption of hybrid quantum-classical frameworks.

\textbf{Quantum-Enhanced AI Optimization:} CRQC+AI begins leveraging quantum-enhanced optimization algorithms like quantum annealing and VQE, improving efficiency in optimization and linear algebra applications.

\textbf{Initial Cryptanalysis Insights:} Early academic research explores the combined potential of the HHL algorithm and low-dimensional representations in geometric algebra, probing for, but not yet establishing, theoretical vulnerabilities in lattice-based PQC schemes.

\subsubsection*{4.6.2 RSA/ECC Collapse and Possible First Lattice (Track 3) Breaks}
\addcontentsline{toc}{subsubsection}{4.6.2 RSA/ECC Collapse and Possible First Lattice (Track 3) Breaks}
\textbf{Quantum Algorithm Progress:} Quantum hardware reliably demonstrates implementation of advanced algorithms, including HHL-type linear-algebra subroutines at small scale; as Section 2.1 establishes, however, such demonstrations do not by themselves yield a practical exponential advantage against the dense, ill-conditioned, cryptographic-size lattice problems underlying LWE.

\textbf{Quantum-Geometric Proof-of-Concept Attacks:} In scenarios where the conjectured geometric embedding is found, researchers would attempt limited-scale proof-of-concept attacks on lattice-based PQC schemes by combining quantum-accelerated optimization with geometric algebra; whether such attempts confirm any practical vulnerability is the open question identified in Sections 2.2 and 4.5, not a settled result.

\textbf{Industry Reassessment of PQC:} The cybersecurity industry reassesses migration timelines, treating current lattice-based implementations in TLS 1.3 as potentially at risk should such weaknesses be confirmed.

\subsubsection*{4.6.3 Potential Lattice PQC Compromise (Kyber-768/1024 and ML-DSA-44)}
\addcontentsline{toc}{subsubsection}{4.6.3 Potential Lattice PQC Compromise (Kyber-768/1024 and ML-DSA-44)}
\textbf{Quantum AI Scaling:} Significant quantum hardware advancements enable practical hybrid quantum-AI algorithms at intermediate scales (1000 to 5000 qubits), a precondition for, though not a guarantee of, viable cryptanalysis of lattice-based encryption schemes.

\textbf{Cryptanalytic Breakthroughs:} This phase is reached only if a geometric embedding with a nonzero collapse fraction ρ is discovered; in that contingency, quantum-enhanced geometric methods would reduce the effective hardness of lattice problems enough to enable attacks on previously secure PQC schemes. Absent such a mechanism, the phase does not occur.

\textbf{TLS 1.3 PQC Exposure:} Should the preceding contingency hold, industry would treat PQC-based TLS 1.3 deployments as exposed and anticipate real-world vulnerabilities on a short horizon; absent that contingency, this exposure does not arise. The joint Monte Carlo (Table 7, Appendix A.5) reinforces that this is a contingency rather than a hardware timeline: under hardware-only Track 0, ML-KEM-1024 shows a median crossover of 2048.7 and effectively zero probability of exposure by 2035.

\subsubsection*{4.6.4 Possible Hash-Based Signature Exposure and First Symmetric-Layer Threat}
\addcontentsline{toc}{subsubsection}{4.6.4 Possible Hash-Based Signature Exposure and First Symmetric-Layer Threat}
\textbf{Advanced CRQC+AI Algorithms:} Robust CRQC+AI algorithms, leveraging both the enhanced capabilities of AI-enhanced quantum algorithms such as HHL and LCHS with geometrically simplified cryptographic representations, would become practical and scalable only if the geometric simplification is realized, on which condition they could facilitate broader cryptanalysis.

\textbf{Real-World Cryptographic Compromise:} Should the preceding contingency materialize, quantum-geometric attacks would expose vulnerabilities in deployed lattice-based PQC schemes, including within TLS 1.3.

\textbf{Urgent Cryptographic Standard Revisions:} Should such compromises occur, NIST and global standardization bodies would expedite the revision and recommendation of quantum-AI-resistant cryptographic methods, prompting swift industry adoption.

\subsubsection*{4.6.5 Residual Symmetric-Layer Exposure and Post-AGI Outlook}
\addcontentsline{toc}{subsubsection}{4.6.5 Residual Symmetric-Layer Exposure and Post-AGI Outlook}
\textbf{Accelerated Cryptographic Evolution:} The arrival of AGI coupled with recognition of CRQC+AI threats should lead to rapid, industry-wide transitions to dynamically agile and hybrid cryptographic practices.

\textbf{Continuous Quantum Resilience:} Organizations will need to embrace dynamic cryptographic agility, leveraging quantum-AI-driven methods and geometric algebra insights to maintain long-term quantum resilience.

To place these scenario outputs alongside external expectations, Table 4 compares the model’s crossover ranges with the principal published quantum-threat timelines and migration mandates. The model is not a forecast and does not claim greater precision than the expert surveys; the comparison is included so the scenario range can be read against the policy deadlines and hardware roadmaps that motivate it. The table also situates the model in its methodological family: like the resource-estimation and expert-elicitation entries beside it, it is intended to be revised as evidence accumulates, differing chiefly in that its revision rule is stated explicitly in Section 3.3.5. 

\par\vspace{4pt}\noindent\begin{minipage}{\linewidth}
\begin{center}\footnotesize\setlength{\tabcolsep}{4pt}\begin{tabular}{|p{0.293\linewidth}|p{0.293\linewidth}|p{0.293\linewidth}|}\hline
\textbf{\textbf{Source}} & \textbf{\textbf{Position and what it provides}} & \textbf{\textbf{Indicative timing}} \\ \hline
Mosca / Global Risk Institute [50], [80], [69] & Expert-survey risk estimate (Mosca inequality framing) & One in seven by 2026, about 50\% by 2031; 2025 survey median CRQC 2029 to 2032 (34\% by 2030) \\ \hline
Gershteyn and Alber, Quantum Horizon [235] & Monte Carlo blend of hardware scaling, falling resource estimates, a fault-tolerance readiness lag and the Global Risk Institute and Mosca surveys (Bitcoin and Ethereum signatures) & CRQC arrival about one in six by 2035 (8 to 24\% across survey weightings), near 30\% by 2040, about 60\% by 2050; bimodal, with an 80\% range of roughly 2032 to 2060 \\ \hline
NIST IR 8547 [51] & Standards and migration policy & Every quantum-vulnerable public-key algorithm, RSA-2048 and P-256 included, deprecated after 2030 and disallowed after 2035 (the 2024 draft had put only the 112-bit tier in the 2030 step); ML-KEM-768 named the default parameter set for key establishment \\ \hline
NSA CNSA 2.0 [164] & National-security migration mandate & Quantum-resistant support by 2025; category deadlines 2030 to 2033; full NSS migration by 2035 \\ \hline
IBM roadmap [101] & Quantum hardware roadmap & First large-scale fault-tolerant system (Starling, about 200 logical qubits) by 2029; cryptanalytically relevant scale later \\ \hline
Google Quantum AI / Gidney (2025) [82] & Resource estimate and hardware roadmap & RSA-2048 factoring resource cut to under one million physical qubits (from about 20 million); error-corrected machines targeted in the early 2030s \\ \hline
Pauli Group ECDLP challenge ladder [182]; IonQ blueprint [180] & ECDLP-256 challenge ladder and a fully compiled trapped-ion resource estimate (secp256k1) & secp256k1 window 2027 to 2033 by the common-ruler method (a calibrated overlay, not a prediction); one ECDLP-256 instance in 25.7 days per attempt on 19,397 physical qubits at 40.7 to 63.3\% success \\ \hline
Alice \& Bob cat-qubit compilation [232] & End-to-end compilation of Shor's algorithm for cat-qubit architectures (RSA-2048, secp256k1, FFDLP-512) & RSA-2048 22,593 cat qubits in 3.14 days; ECDLP-256 at most 29,588 cat qubits in 61.8 minutes; FFDLP-512 6,656 cat qubits in 21.2 minutes; assumes no bit flips (a bit-flip time of about 100 years); an architecture estimate, not a dated forecast \\ \hline
This paper (capability model) & Scenario stress test, not a forecast & RSA mechanism-backed crossover 2029 to 2032, with ECC-256 no later by proxy; conditional lattice scenarios 2031 to 2035, with a marginal Track 1 value at 2045; hash-based and symmetric primitives undated \\ \hline
\end{tabular}\end{center}
\vspace{11.25pt}\begin{center}\small\textbf{Table 4: Comparison of the model’s crossover ranges with principal published \\[2pt]quantum-threat timelines, hardware roadmaps, and migration mandates.}\end{center}
\end{minipage}\par\vspace{4pt}
\section*{5. Limitations and Falsifiability}
\addcontentsline{toc}{section}{5. Limitations and Falsifiability}
This section consolidates the study’s limitations and states the conditions under which its conclusions would fail, so the model can be judged as a falsifiable instrument rather than an open-ended projection.

Limitations. The capability model that generates the figures uses scenario couplings, the hardware sensitivities α and the software sensitivities β, that are chosen rather than measured; they encode judgments about how strongly each primitive is eroded by quantum hardware and by AI-assisted attacks, and different analysts could reasonably choose different values. The crossover years are therefore outputs of those chosen inputs, not empirical predictions, and the reproducibility documented in Appendix A means the scenario can be regenerated rather than that the dates are forecast. The runtime axis of the figures is likewise a normalized index anchored at a common baseline for all five schemes, not a per-scheme physical runtime estimate (Appendix A.1). The model holds the quantum algorithm fixed and does not attempt to discover new cryptanalysis; it asks only how much engineering and AI-assisted progress would be required for earlier compromise. The 2028 AGI onset is an assumption drawn from interested-party roadmaps, and the sensitivity analysis in Appendix A.4 shows it is the single most consequential input for every software-driven crossover. The three evidence tiers are treated as qualitatively distinct, and no claim is made that the lattice, hash-based, or symmetric crossovers correspond to any known attack. Gershteyn and Alber's Quantum Horizon [235] blends hardware scaling, falling resource estimates, a fault-tolerance readiness lag and expert surveys into an arrival distribution of about one in six by 2035, near 30 percent by 2040 and about 60 percent by 2050, against this paper's RSA Track 0 probability of about 0.95 by 2035 (Table 7); the gap is an elicitation blend set against a pure hardware extrapolation, and it is the clearest available statement of what Track 0 does and does not include.

How this model could be wrong. The central conclusions do not hold, and the corresponding curves should be disregarded, under any of the following conditions.

\begin{itemize}\item If AI produces no durable algorithmic improvement to cryptanalysis, so that the software couplings β remain effectively zero, the lattice, hash-based, and symmetric tracks never cross within the horizon and only the mechanism-backed RSA and ECC exposure remains.\item If no effective-dimension-collapse or comparable structural advance against lattices ever materializes, so that ρ stays at zero in the Section 4.5 cost model, the contingency tier is void. If Harrow-Hassidim-Lloyd and related quantum linear-algebra methods remain inapplicable to the noisy bounded-distance-decoding structure of LWE, a central hypothesized mechanism is removed.\item If quantum hardware scaling stalls, with the hardware rate r\_hw approaching one so that the hardware driver flattens, even the RSA crossover moves out by years, as the sensitivity analysis shows. If fault-tolerant error correction plateaus below cryptanalytically relevant scales, the mechanism-backed timeline lengthens regardless of algorithmic progress. \item And if AGI arrives substantially later than the assumed 2028, every software-driven crossover slips by approximately the same amount.\end{itemize}
Each of these is an observable, falsifiable condition rather than a matter of interpretation. The channel-level tracking indicators that operationalize these conditions, hardware milestones, error-rate trends, decoder latency, and resource-estimate movement, are given in Section 3.3.5; each indicator updates a specific acceleration factor or the software driver, rather than the projection in its entirety. Read together, the falsification conditions above and the recalibration protocol of Section 3.3.5 make this a living estimation instrument: it is expected to err in measurable ways and measuring that error against milestones as they become historical data points is the mechanism by which successive, versioned reruns of the model improve its tracking of the record and potentially its projections of future changes in vulnerability spectrums.

\section*{6. Implications for Migration and Governance}
\addcontentsline{toc}{section}{6. Implications for Migration and Governance}
The modeling above measures primitive erosion in TLS 1.3. Three consequences for migration and governance follow, none of which depends on the effective-dimension-collapse conjecture; each is a present requirement rather than a projection.

\subsection*{6.1 Crypto-Agility as a Control Objective}
\addcontentsline{toc}{subsection}{6.1 Crypto-Agility as a Control Objective}
Crypto-agility is not a technical property of a stack; it is a governance capability, comprising cryptographic inventory, asset ownership, rotation authority, change-control process, exception handling, and evidence retention. [174] A deployment is agile only if it can enumerate every primitive in use, name the owner authorized to rotate it, execute a rotation under change control, and retain evidence of the change. Recommending crypto-agility without these leaves the reader with a purchase decision rather than a control objective. The migration decision, and the authority under which it is exercised, belongs to the cryptographic-inventory owner, not to the protocol. Two 2026 treatments carry this further: Campbell frames runtime cryptographic evidence as the input to bounded assurance verdicts [212], and Canetti, Ha and Kaptchuk's cryptography-native threat modeling [213] and Song et al.'s TLS-validated recommendation system [214] give the inventory and evidence side of the capability a method.

\subsection*{6.2 Sector Data-Lifetime Spectrum (Mosca's X and Y)}
\addcontentsline{toc}{subsection}{6.2 Sector Data-Lifetime Spectrum (Mosca's X and Y)}
Mosca's inequality X + Y > Z is a sector-specific decision rule. This paper models Z, the capability horizon, three times, but does not derive X, the secrecy horizon, or Y, the migration duration; representative sector values are shown in Table 5. Y also differs by layer. The MITRE/PQCC capability report of 2026 finds post-quantum key establishment, in practice hybrid X25519MLKEM768, at the Integration level across browsers, web servers, VPN and cloud connectivity, while post-quantum authentication sits at Components Available for browsers, web servers and VPN and at Preliminary Integration for cloud connectivity, with Integration reached only by code signing and PKI/CA management; the report found no browser support for X.509 validation with post-quantum signatures, and web-server ML-DSA authentication only through custom software builds [215]; RFC 9955 sets out the hybrid-signature stripping and downgrade risks on that side [216]. The migration duration Y for signatures is therefore longer than for key exchange, and Table 5 should be read with the signature side as the binding Y. An independent review of the first version proposed the same separation as a dual-axis map, KEM exposure against the secrecy horizon X and signature exposure against certificate and trust-anchor lifetime, with the Table 5 sector rows as the decision artifact [234]; the text here states the two axes, and the figure is left to a later version.

\par\vspace{4pt}\noindent\begin{minipage}{\linewidth}
\begin{center}\footnotesize\setlength{\tabcolsep}{4pt}\begin{tabular}{|p{0.22\linewidth}|p{0.22\linewidth}|p{0.22\linewidth}|p{0.22\linewidth}|}\hline
\textbf{\textbf{Sector}} & \textbf{\textbf{Secrecy horizon X}} & \textbf{\textbf{Migration duration Y}} & \textbf{\textbf{Urgency vs modeled Z}} \\ \hline
Payments / card data & \textasciitilde{}5–10 yr [178] & months–2 yr (cloud) & lower \\ \hline
Health records & \textasciitilde{}25–50+ yr [177] & 2–5 yr & high \\ \hline
Legal / privileged & decades [176] & 2–5 yr & high \\ \hline
National security & \textasciitilde{}25–75+ yr [175] & 5–10+ yr (OT/embedded) [179] & highest \\ \hline
\end{tabular}\end{center}
\vspace{11.25pt}\begin{center}\small\textbf{Table 5: Representative sector data-lifetime spectrum (Mosca's X and Y).
 Secrecy horizons X and migration durations Y are sourced in the running text.}\end{center}
\end{minipage}\par\vspace{4pt}
The secrecy horizons X are anchored in published mandates: the 25-, 50-, and 75-year declassification tiers for national-security information [175]; the survival of attorney-client privilege beyond the client’s death [176]; HIPAA documentation retention and state medical-record statutes together with the patient-lifetime confidentiality of health data, so the operative horizon is the patient’s life rather than the shorter retention minimum [177]; and SEC Rule 17a-4 and PCI DSS retention for payment-card data [178]. The migration durations Y are engineering estimates reflecting operational-technology and embedded-system lifecycles of 10 to 20-plus years and published enterprise PQC-migration analyses [179]. The point stands regardless: migration urgency varies by sector, which cuts against a single uniform timeline.

\subsection*{6.3 Sovereignty and Jurisdiction}
\addcontentsline{toc}{subsection}{6.3 Sovereignty and Jurisdiction}
Crypto-agility is a present requirement on legal grounds alone, demonstrable today and requiring no speculative cryptanalysis. Algorithm mandates already diverge: CNSA 2.0 specifies standalone ML-KEM-1024 for United States national-security systems, against a widely deployed hybrid default (ECDHE with ML-KEM) and differing European and Australian positions. Trust-anchor jurisdiction, HSM sourcing, and cross-border key custody differ by sovereign. An organization operating across jurisdictions must be able to switch algorithms, trust anchors and key custody on legal timelines it does not control, which is precisely a crypto-agility capability. This is a stronger argument for the paper's own conclusion than the effective-dimension-collapse hypothesis, and it depends on no unproven mechanism. The same period showed systematic sweeping by frontier-model actors as a new source of structural claims, true and false alike (Section 3.3.5, indicator six); verification capacity is now a security resource, and crypto-agility is what lets a deployment act on a verified claim within the claim's own verification cycle rather than a standards cycle. An independent review of the first version proposed modeling this divergence as an agility demand function with its own migration duration Y under jurisdictional shock, separate from ρ [234], which is the reading intended here.

\subsection*{6.4 Issuers, Trust Anchors and Certificate Lifecycle}
\addcontentsline{toc}{subsection}{6.4 Issuers, Trust Anchors and Certificate Lifecycle}
Post-quantum migration in TLS 1.3 is not only a matter of the negotiated primitives; it depends on who issues certificates and how trust anchors are rotated. Root and intermediate certificate-authority keys, and the trust anchors embedded in operating systems, browsers and devices, have lifetimes measured in years to decades and are updated on distribution cycles the relying party does not control. A post-quantum chain requires post-quantum-capable issuers, post-quantum-aware path validation, and a trust-anchor distribution that reaches long-lived and embedded clients, and it typically lags the negotiated key exchange. None of this requires a cryptographically relevant quantum computer to become a live constraint; it is an ecosystem exposure that materializes on today's issuance and distribution timelines. It lies outside the primitive-erosion instrument of this paper and is flagged here as a dependency a TLS 1.3 migration program must plan for separately. Two 2026 results make RSA-1024, where it survives in trust chains or embedded devices, a present classical exposure rather than a quantum one: the September 2026 factorizations put RSA-1024 at an estimated \$30 million of market compute (Section 3.2.1.4), and Shea, Haller, Suhl, Heninger and Thomé show RSA-1024 signature forgery in nearly special-number-field-sieve time given a raw, unpadded signing or decryption oracle: about 1,380 core-years (1,200 of precomputation and 180 per later forgery) and 2\textasciicircum{}32 queries, completed on 31 August 2026, impersonating a hardware security module without extracting its key [217]. PKCS\#1 v1.5 and RSA-PSS signing do not provide such an oracle, so certificate-authority signing is not exposed; the deployments that do are unpadded 2048-bit blind-RSA services such as Apple iCloud Private Relay, Cloudflare Privacy Pass and Apple Private Cloud Compute, for which the authors estimate 2\textasciicircum{}90 work with 2\textasciicircum{}43 queries.

\subsection*{6.5 Interoperability and Negotiation}
\addcontentsline{toc}{subsection}{6.5 Interoperability and Negotiation}
TLS 1.3 negotiates its algorithms, which is a strength, but negotiation is also where a migration succeeds or fails in practice. A hybrid key exchange, for example the concatenation of ML-KEM with an elliptic-curve exchange in draft-ietf-tls-ecdhe-mlkem, [161] must be offered, understood and preferred by both endpoints; middleboxes, session-resumption caches and libraries that do not recognize the new groups can cause negotiation failure or a silent downgrade to a classical exchange. Interoperability across a heterogeneous estate, and resistance to downgrade, are engineering problems that arrive with deployment rather than with a quantum computer. This paper models the primitives on the assumption that a post-quantum or hybrid group is negotiated end to end; the negotiation and interoperability layer that makes that assumption hold is out of scope and is noted here as a distinct control. The 2026 record bears this out, within the bounds of a search of the OpenSSL, wolfSSL and RUSTSEC advisories through October 2026: OpenSSL 3.5.0 to 3.5.5 and 3.6.0 to 3.6.1 servers using DEFAULT in their group list could negotiate a weaker group than configured, so X25519MLKEM768 fell back silently to a classical exchange whenever the client's first key share was classical (CVE-2026-2673, advisory of 13 March 2026, rated Low, fixed in 3.6.2 and 3.5.6 on 7 April 2026) [218]; deployed ML-KEM implementations with a faulty Fujisaki-Okamoto rejection check, a truncated ciphertext comparison in wolfSSL (CVE-2026-10097 and CVE-2026-6330, fixed in 5.9.2 on 23 June 2026) and a skipped conditional move in two unmaintained Rust crates (RUSTSEC-2026-0290 and RUSTSEC-2026-0288), exposed the long-term key to chosen-ciphertext recovery, in about 350 ciphertexts against the wolfSSL AVX2 path and 4,272 queries against the Rust crates, with a few hundred thousand queries sufficing in simulation even when a single coordinate goes unverified [201] [219]; and wolfSSL 5.9.0 fixed fault injection on ML-KEM and ML-DSA on Cortex-M (CVE-2026-3503) in March, with 5.9.4 of 25 September 2026 closing eleven more [219]. Within that search, every TLS 1.3 compromise of 2026 sat in the negotiation and implementation layer and none in a primitive.

\subsection*{6.6 Entropy and Key Provenance}
\addcontentsline{toc}{subsection}{6.6 Entropy and Key Provenance}
Every guarantee in this paper is conditioned on keys drawn from a sound entropy source. A post-quantum primitive keyed from a weak or predictable generator is no stronger than its randomness, independent of any quantum advance. Entropy quality, seeding on constrained and embedded devices, and the provenance of keys, where they were generated, by what module, and under what assurance, are recurring sources of real-world failure that require no cryptographically relevant quantum computer to exploit. Key provenance also bears on HNDL and harvest-now, forge-later exposure: keys generated or held outside the operator's control widen the surface for retroactive decryption. Entropy and key provenance are properties of the implementation and the operating environment rather than of the negotiated primitive, so they lie outside the instrument used here and are recorded as a further ecosystem dependency.

\section*{7. Conclusion}
\addcontentsline{toc}{section}{7. Conclusion}
Depending upon the timeline at which CRQC+AI progresses, the potential for large-scale financial loss for TLS 1.3 could begin as early as 2029 with attacks against RSA. The compromise of PQC encryption methods would have similar effects to a compromise in any use of RSA encryption and the risk vulnerability spectrum could begin as early as 2031. These two dates carry different evidential weight: the RSA and ECC exposure is mechanism-backed, resting on Shor’s algorithm, whereas the PQC dates are conditional scenario outputs that depend on a structural advance against lattices that has not been demonstrated. However, the methods utilizing AGI+CRQC+AI as outlined herein would magnify these issues because of the rapidity and generalization ability of such attacks. The effects are set out in Table 6 and in the list that follows it.

Table 6 maps each class of cryptographic break to the primitive or key material compromised and its effect on TLS 1.3.

\par\vspace{4pt}\noindent\begin{minipage}{\linewidth}
\begin{center}\footnotesize\setlength{\tabcolsep}{4pt}\begin{tabular}{|p{0.293\linewidth}|p{0.293\linewidth}|p{0.293\linewidth}|}\hline
\textbf{\textbf{Threat class}} & \textbf{\textbf{Example}} & \textbf{\textbf{TLS 1.3 effect}} \\ \hline
Algorithmic public-key break & Shor against RSA or ECC & Certificate forgery, identity compromise, legacy decryption \\ \hline
KEM or key-exchange break & ML-KEM structural break & Session-key recovery and harvest-now-decrypt-later exposure \\ \hline
Signature break & ML-DSA or SLH-DSA break & Authentication failure, impersonation \\ \hline
Symmetric cipher break & AES structural break & Bulk confidentiality failure \\ \hline
Implementation or side-channel break & Timing, fault, or cache leakage & Key theft without breaking the primitive \\ \hline
\end{tabular}\end{center}
\vspace{11.25pt}\begin{center}\small\textbf{Table 6: Threat classes and their TLS 1.3 effects.}\end{center}
\end{minipage}\par\vspace{4pt}
Eavesdropping on Communications: By intercepting PQC key exchanges, an attacker could extract the shared keys between parties, allowing them to decrypt sensitive communications.

Man-in-the-Middle Attack: An attacker could position themselves between two communicating parties on a network, intercepting and altering messages once the keys used for encryption are obtained and hacked, thus compromising the integrity of the communication.

Data Theft: With access to encryption keys, an attacker could decrypt stored data on servers or devices, leading to unauthorized access to sensitive information such as personal data, financial records, trade secrets.

Impersonation: By obtaining cryptographic keys, an attacker could impersonate a legitimate user or service, allowing them to perform unauthorized actions or transactions under false pretenses.

Disruption of Services: An attacker could use the decrypted private keys to disrupt services by altering or injecting malicious data into communications, potentially leading to denial-of-service conditions or other operational failures.

Access to Encrypted Backups: If an attacker gains access to private encryption keys, they could decrypt backups of sensitive data, expose historical information and alter data that may not be actively monitored.

Exploitation of IoT Devices: Many Internet of Things (IoT) devices rely on private cryptographic keys for secure communication. An attacker could exploit these keys to gain control over IoT devices, leading to privacy breaches or physical security risks.

Manipulation of Smart Contracts: In blockchain environments, if an attacker obtains the private keys used to sign transactions, they could manipulate smart contracts, leading to financial losses or fraudulent activities.

The potential for massive and rapid financial loss because of CRQC+AI vulnerabilities is substantial and difficult to bound precisely. Just as it took decades before the British and Americans revealed they had cracked the Enigma cryptography machine in World War II, we won’t know when hackers will be effective at using CRQC+AI resources to break PQC encryption algorithms. 

When weighed against the relatively low cost and high probability of successful defenses to PQC vulnerabilities through research and use of crypto-agile hybrid cryptographic approaches that are now being advocated for by NIST, [9] the case for acting now to mitigate these risks is compelling, and not by assuming that PQC will be the last encryption migration needed. Instead, as some business executives are beginning to recognize, the need for migration to PQC is better viewed as a trigger for needing to implement crypto-agile approaches. [165] Layered risk scoring models for TLS connections under quantum threats can further help organizations identify and prioritize their most exposed connections during this transition. [166] Beyond these practical conclusions, the durable contribution of this paper is the estimation instrument itself: a reproducible, parameterized scenario model whose parameters can be varied, rerun, and recalibrated against emerging milestones through the update mechanics of Section 3.3.5, complementing quantum resource-estimation studies and expert-elicitation timeline surveys with an auditable procedure for converting future observations into revised risk timelines.

\textbf{Author Contributions: }Conceptualization, N.G.; methodology, N.G. and B.P.; investigation, N.G., E.U., M.H. and B.P.; writing, original draft preparation, N.G.; writing, review and editing, all authors. All authors have read and agreed to the published version of the manuscript.

\textbf{Funding: }This research received no external funding.

\textbf{Data Availability Statement: }No new empirical data were created or analyzed in this study. The figures are generated from the parameterized model described in Sections 3 and 4, whose parameters and assumptions are disclosed in the text.

\textbf{Conflicts of Interest: }N.G., B.P. and B.J.E. are employees of EnQuanta (VoiceIt Technologies, Inc. dba EnQuanta), and E.U. and M.H. are consultants of EnQuanta, a company that develops commercial post-quantum and crypto-agile cryptographic products and services; accordingly, they have a financial interest related to the crypto-agile and hybrid cryptographic approaches discussed in this work.

\newpage
\section*{Appendix A. Reproducibility of the Combined Hardware and Software Capability Model}
\addcontentsline{toc}{section}{Appendix A. Reproducibility of the Combined Hardware and Software Capability Model}
This appendix documents the computational model used to generate the runtime and feasibility figures (Figures 6A/6B through 8A/8B, Figure 9, and Appendix C Figures C1A/C1B and C2A/C2B), so that every curve can be reproduced from explicit equations and parameters. The figures are produced by a single combined hardware and software capability model, set out below, which is the generative instrument behind the curves. It is complementary to the analytical framings in the body: the channel model of Section 3.3 and the cost model of Section 4.5 are separate instruments that bound the same question, while the capability model here is the one that produces the plotted curves. The model holds the quantum algorithm fixed: AI accelerates engineering and the discovery and optimization of attacks, but it does not change the polynomial scaling of Shor’s algorithm, lower the fault-tolerant error-correction threshold, or repeal coherence limits.

\subsection*{Appendix A.1. Governing Equations}
\addcontentsline{toc}{subsection}{Appendix A.1. Governing Equations}
Cryptanalytic capability C(a, k, t), for algorithm a on track k in calendar year t, is measured as the number of base-10 orders of magnitude (in seconds) by which the estimated attack runtime has been reduced from its starting value. It is the sum of a hardware term and a software term:

\begin{center}C(a, k, t) = α(a) · D\_hw(t) + β(a, k) · D\_sw(k, t)\end{center}
The hardware driver D\_hw is an accelerating (convex) accumulation of effective quantum resources, present in every track including Track 0, because engineering progress is not gated by AGI:

\begin{center}D\_hw(t) = A · (r\_hw\textasciicircum{}(t − t0) − 1) / (r\_hw − 1),    A = 0.34,  r\_hw = 1.20,  t0 = 2026\end{center}
The software driver D\_sw is gated on the 2028 AGI onset and grows at a track-specific rate, with a frontier-model term that applies from 2026 on every track with a nonzero software coupling:

\begin{center}D\_sw(k, t) = 0.10 · (t − t0) + B\_k · (r\_k\textasciicircum{}(t − tAGI) − 1) / (r\_k − 1),    tAGI = 2028\end{center}
In both drivers the exponent is clamped at zero below its onset year, so the hardware term contributes only from t0 and the AGI term only from tAGI. The coupling α(a) is the sensitivity of algorithm a to raw hardware (Shor), large for RSA and near zero for the lattice, hash, and symmetric schemes; β(a, k) is its sensitivity to the software attacks enabled on track k. The estimated attack runtime is R(a, k, t) = R0 − C(a, k, t) on a base-10 log-seconds axis, with R0 = 10 (Centuries), a common visual baseline shared by all five schemes rather than a per-scheme empirical runtime estimate. R(a, k, t) is therefore a normalized runtime index: within a scheme it tracks modeled progress over time, but cross-scheme differences on this axis carry no information about relative attack cost, which is treated in the cryptanalytic units of Section 4.5. Modeled feasibility is a logistic of the margin by which runtime has fallen below the one-month line:

\begin{center}P(a, k, t) = 100 / (1 + exp(−(MON − R(a, k, t)) / 0.85)),    MON = log10(2.6 × 10\textasciicircum{}6) ≈ 6.415\end{center}
A scheme reaches its 50 percent modeled-feasibility crossover in the first year for which the runtime falls to the one-month line, equivalently when C(a, k, t) reaches R0 − MON ≈ 3.585. No crossover year is taken from prior or external figures; every crossover emerges from these equations.

\subsection*{Appendix A.2. Parameters and Their Sources}
\addcontentsline{toc}{subsection}{Appendix A.2. Parameters and Their Sources}
Global constants: the reference year t0 = 2026; the AGI onset tAGI = 2028; the runtime axis is base-10 log seconds from Seconds to Centuries, with start value R0 = log10(10\textasciicircum{}10) = 10 and the one-month feasibility line at MON = log10(2.6 × 10\textasciicircum{}6) ≈ 6.415; the logistic width is 0.85. A scheme crosses 50 percent when capability C reaches R0 − MON ≈ 3.585 orders of magnitude. R0 is a common visual baseline shared by all schemes for comparability, not a per-scheme estimate of the 2026 attack runtime. On this normalized axis the one-month line MON functions as a fixed capability threshold (R0 − MON ≈ 3.585 orders of magnitude) rather than a claim that any given attack would take one month of physical time.

Hardware driver: A = 0.34 and r\_hw = 1.20, an accelerating accumulation of effective quantum resources of roughly 20 percent per year, present in every track. It is the granular counterpart of the Rose’s-Law hardware growth in Section 3.2.2.7 and the channel model in Section 3.3.

Software driver: a frontier-model slope of 0.10 per year, stated as a band of 0.08 to 0.12 (v2 used 0.04 as a pre-AGI human-progress term), applied from 2026 on every track with a nonzero software coupling, before and after the AGI onset. The band is a judgment, not a calibration: no benchmark time series of AI-driven cryptanalytic progress exists yet, and the value should be replaced by one (CryptanalysisBench is the candidate instrument) as soon as it carries a time series; 0.04 is retained as the comparison value throughout Appendix A. It is followed by a track-specific AGI-gated term with rate r\_k and ceiling B\_k taking the values Track 0 (1.01, 0.00), Track 1 (1.10, 0.16), Track 2 (1.17, 0.26), and Track 3 (1.30, 0.30). The Track 3 rate of 1.30 is what produces its sharply convex collapse.

Per-algorithm couplings, given as α and then β for Tracks 0 to 3: RSA-2048, α = 0.92, β = 0, 1.15, 1.75, 2.55; Kyber-1024, α = 0.05, β = 0, 0.06, 0.95, 1.45; ML-DSA-44, α = 0.04, β = 0, 0.05, 0.78, 1.18; SLH-DSA, α = 0.03, β = 0, 0, 0, 0; AES-256, α = 0.020, β = 0, 0, 0, 0. The α values are hardware (Shor) sensitivities, large for RSA and near zero otherwise; the β values are software-attack sensitivities enabled on each track. These couplings are scenario parameters, not measured constants, chosen to reflect each primitive's sensitivity to Shor-type hardware progress and to AI-assisted cryptanalysis; with these values hardware alone erodes only RSA and the lattice schemes move only once the Track 2 and Track 3 software drivers switch on; the hash-based (SLH-DSA) and symmetric (AES-256) schemes are assigned zero software coupling and report no crossover within the horizon, consistent with the hypothesis-only treatment in Figure 9. The hybrid X25519 + Kyber-768 configuration of Figures 7A/7B is modeled with these same lattice couplings: the hybrid falls only when both components fall, and because the classical X25519 component is mechanism-backed (Shor) and falls years earlier under the hardware track, the lattice component is binding and the hybrid’s modeled crossovers coincide with the Kyber values reported below. A separate hypothesis-only variant, used only to produce the dated AES-256 stress values quoted in Appendix A.4 and discussed in Appendix C.2, sets the AES-256 software couplings to β = 0.155 (Track 2) and 0.210 (Track 3); it is not used in Figure 9 or in Figures C2A and C2B, which carry β = 0 on every track and show no crossover.

This failure ordering is a \textit{result}, not an input. Sweeping the coupling space, a lattice scheme crosses before RSA only under one of two conditions: (a) a hardware coupling above RSA's own 0.92, that is, assuming a lattice scheme more Shor-sensitive than RSA, which is a claim about the algorithm rather than a parameter choice; or (b) on Track 3, a software coupling of about 4.48 for ML-KEM-1024 against its published 1.45, roughly triple its own value and 1.76× RSA's 2.55. Every inversion point lies far outside the published couplings and the ±25\% sensitivity ranges of Appendix A.4, so the ordering is structurally forced rather than assumed.

Running the model with these parameters yields the 50 percent crossover years plotted in Figure 9: RSA-2048 at 2032.5, 2031.1, 2030.2, and 2029.5 for Tracks 0 to 3; Kyber-1024 at 2045.8 (Track 1), 2034.4 (Track 2), and 2031.9 (Track 3); ML-DSA-44 at 2035.3 and 2032.5 (Tracks 2 and 3); and SLH-DSA and AES-256, which carry zero software coupling, show no feasible break within the 2046 horizon, as do the remaining tracks. At the ends of the stated slope band the software-driven crossovers move by about 0.2 year (Kyber-1024 Track 3 at 2032.0 for a slope of 0.08 and 2031.8 for 0.12; Track 2 at 2034.6 and 2034.2; ML-DSA-44 Track 3 at 2032.6 and 2032.4; RSA-2048 Track 3 at 2029.6 and 2029.4), while the hardware-driven Track 0 values do not move; the point value 0.10 is used for the figures and the band is the sensitivity range a reader should carry. Crossover years are read from a 600-step grid over 2026 to 2046, a resolution of about 0.033 year (twelve days), and are reported to one decimal place; differences smaller than about 0.05 year are within grid resolution and carry no meaning. Where the abstract or the text gives a calendar year or a range of years for a point crossover, it is the calendar year in which the crossover falls, that is, the integer part of the grid value: a crossover read as 2031.9 is reported as 2031 and one read as 2035.0 as 2035; nearest-year rounding is not used. Monte Carlo percentiles (Table 7) are reported to one decimal and are not rounded to calendar years. One value is near-threshold: the Kyber-1024 Track 1 crossover at 2045.8 sits 0.2 year inside the 2046 horizon, so a small perturbation of any input moves it to no crossover within the horizon, and it should be read as marginal rather than as a dated break.

\subsection*{Appendix A.3. Procedure to Regenerate the Figures (Pseudocode)}
\addcontentsline{toc}{subsection}{Appendix A.3. Procedure to Regenerate the Figures (Pseudocode)}
\begin{verbatim}
CONSTANTS
  t0 = 2026 ;  t_AGI = 2028
  R0  = log10(1e10)     # normalized index baseline (Centuries), common to all schemes
  MON = log10(2.6e6)    # one-month feasibility line
  A = 0.34 ;  r_hw = 1.20          # hardware driver
  sw_pre = 0.10                    # frontier-model software slope (band 0.08 to 0.12)
  SW = { 0:(1.01,0.00), 1:(1.10,0.16), 2:(1.17,0.26), 3:(1.30,0.30) }   # (r_k, B_k)
  alpha[algo] ,  beta[algo][track]   # per-algorithm couplings (Appendix A.2)

FUNCTION D_hw(t):
  e = max(t - t0, 0) ;  return A * (r_hw**e - 1) / (r_hw - 1)
FUNCTION D_sw(t, track):
  (r, B) = SW[track] ;  e = max(t - t_AGI, 0)
  return sw_pre * max(t - t0, 0) + B * (r**e - 1) / (r - 1)
FUNCTION capability(algo, track, t):
  return alpha[algo]*D_hw(t) + beta[algo][track]*D_sw(t, track)
FUNCTION runtime(algo, track, t):
  return clamp(R0 - capability(algo,track,t), low = -0.4, high = R0)
FUNCTION feasibility(algo, track, t):
  margin = MON - runtime(algo,track,t) ;  return 100 / (1 + exp(-margin/0.85))

FUNCTION crossover_year(algo, track):
  for t from 2026 to 2046 in 600 steps:
      if feasibility(algo,track,t) reaches 50:  return t
  return NONE within horizon
\end{verbatim}
Reproduction materials. The runnable counterpart to this appendix, comprising the capability model, the seeded Monte Carlo that regenerates Table 7 and Figure A2, and the figure-generation scripts, is available at https://enquanta.com/academia/crqc-ai-tls13-reproduction.zip. The Monte Carlo uses NumPy's default\_rng (PCG64) seeded with 20260731 + i, where i is the cell index in the fixed order RSA-2048 Track 0, RSA-2048 Track 3, Kyber-1024 Track 0, Kyber-1024 Track 2, Kyber-1024 Track 3, ML-DSA-44 Track 3 (i = 0 to 5). Within each cell the N = 40,000 draws are taken in the order α, β, A, r\_hw, frontier-model slope, r\_k, B\_k, and then the triangular AGI onset (Tracks 1 to 3 only), and crossovers are located on a 0.1-year grid from 2026 to 2060 with censoring at 2060. With that stream the released scripts regenerate the deterministic crossover years exactly and the Table 7 percentiles exactly as printed in this version, which was generated from the released archive at the v3 settings of Appendix A.5. Note on v2: the Table 7 printed in v2 was not regenerated by the archive to the printed decimal; the released recipe gave the RSA-2048 Track 3 median as 2030.3 against the printed 2030.5 and the ML-DSA-44 Track 3 median as 2033.6 against 2033.8, with two tenth-percentile values 0.1 to 0.2 year apart, while every exceedance probability agreed to 0.01. The v2 table was produced by an earlier run that the archive does not reproduce; the differences lie within the sampling noise stated below, and v3 replaces the table with the archive's own output. An independent implementation using its own random stream should expect the Table 7 percentiles to agree to within about 0.3 year at N = 40,000; that spread is sampling noise, not model disagreement (an independent reproduction in August 2026 matched 11 of 18 percentiles within 0.2 year and all 18 within 0.28 year, 16 of the 18 differences lying on the early side). The illustrative curves of Figures 4A and 4B are stylized reconstructions rather than measured series, and the Figure 5 qubit estimates are approximate reads of the published figure rather than values from a source table; the release notes mark each accordingly. Every curve in this paper is generated by these scripts; no frontier model sits in the generation path, and the drafting and review use disclosed in Section 4 is separate from it.

\subsection*{Appendix A.4. Sensitivity Analysis}
\addcontentsline{toc}{subsection}{Appendix A.4. Sensitivity Analysis}
Because each scheme crosses when its capability reaches about 3.585 orders of magnitude, the sensitivities are clear from the two drivers. For RSA-2048 the capability is dominated by the hardware term (α = 0.92), so RSA crosses under every track, hardware alone being sufficient, and the track spread is narrow, about 3 years, set by its software couplings; widening that fan requires larger RSA β values. For the lattice, hash, and symmetric schemes α is near zero, so they are essentially immovable under Track 0 (hardware only) and Track 1 (small β), apart from a marginal end-of-horizon Kyber-1024 Track 1 crossover at 2045.8, and they cross materially only once the Track 2 and Track 3 software drivers, gated at 2028, accumulate enough capability; their crossover years are therefore governed by β(a, k) and by the track rates r\_k and ceilings B\_k.

Three inputs move the whole picture. The Track 3 rate r\_k = 1.30 is the sole source of its sharply convex collapse; lowering it toward the Track 2 value of 1.17 pushes the Track 3 crossovers later and removes the convex curl. The AGI onset tAGI = 2028 gates every software-driven crossover, so a later onset delays all lattice, hash, and symmetric breaks while leaving RSA’s hardware-driven Track 0 crossover unchanged. The most consequential and most speculative inputs are the AES-256 software couplings of the hypothesis-only variant defined in Appendix A.2 (β = 0.155 and 0.210 for Tracks 2 and 3), which are the entire basis for the dated AES-256 stress values quoted below; there is no established cryptanalytic mechanism behind them. The main results carry β = 0 for AES-256 on every track, so Figure 9 and Figures C2A and C2B show no AES-256 crossover, and the dated values appear in this appendix only as stress-test inputs.

To make the dependence on each input explicit, the two representative crossovers in Figure A1 were recomputed while varying one parameter at a time across a deliberately widened stress range, roughly double the calibrated ranges used by the Monte Carlo of Appendix A.5. The analysis is deliberately one-at-a-time, isolating each parameter’s effect rather than capturing interactions among parameters; a joint treatment over the stated ranges is developed as a Monte Carlo analysis in Appendix A.5. The two panels show a clean separation. The hardware-driven RSA Track 0 crossover (Figure A1, left panel) moves only with the hardware amplitude A, the hardware acceleration rate r\_hw, and RSA’s hardware coupling α, and is completely insensitive to the AGI onset and the software parameters. The software-driven Kyber-1024 Track 3 crossover (Figure A1, right panel) is dominated by the AGI onset year, followed by the software coupling β and the track ceiling B\_k, and is nearly insensitive to hardware. Even under these widened ranges, no single one-at-a-time variation shifts either crossover by more than about 3.3 years (the AGI onset in the software-driven case), so no individual parameter choice drives the result on its own; and the failure ordering itself is governed not by any of these one-at-a-time bands but by the coupling inversion threshold of Appendix A.2, which lies far outside even these ranges.

\begin{figure}[H]\centering
\includegraphics[width=0.82\linewidth]{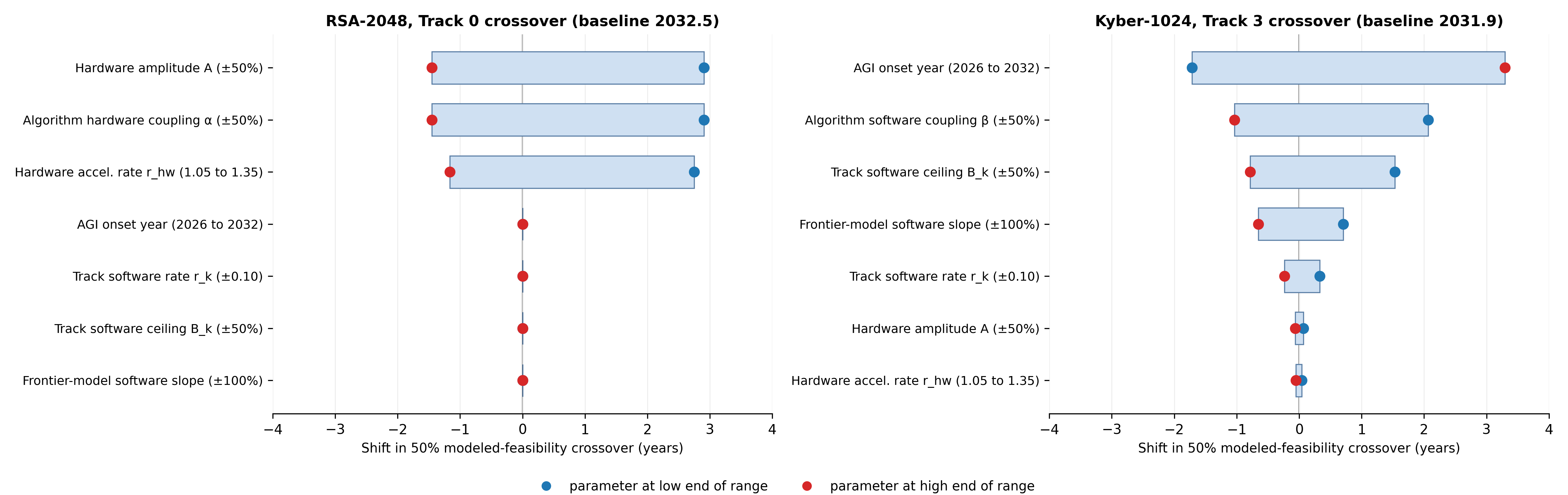}
\par\vspace{3pt}{\small\textbf{Figure A1. One-at-a-time sensitivity of the 50\% modeled-feasibility crossover to model parameters, for a hardware-driven case (RSA-2048, Track 0; left) and a software-driven case (Kyber-1024, Track 3; right).}}
\end{figure}
Each bar spans the crossover shift as the named parameter is varied across its stated range; blue and red mark the low and high ends. The ranges shown are deliberately widened stress ranges, roughly double the calibrated Monte Carlo ranges of Appendix A.5: ±50\% on the couplings, hardware amplitude and track ceiling; ±0.10 on the track rate; ±100\% on the frontier-model slope (0 to 0.20 per year, five times the width of its Appendix A.5 band); a 1.05 to 1.35 hardware rate; and a 2026 to 2032 AGI onset. Even so, no single parameter shifts either representative crossover by more than about 3.3 years, and the failure ordering remains governed by the coupling inversion threshold of Appendix A.2, which lies far outside even these ranges.

Four named scenarios make the dependence concrete. A two-year slip in the AGI onset, to 2030, leaves the hardware-driven RSA Track 0 crossover unchanged at 2032.5 but moves every software-driven crossover later by roughly 1.7 years (Kyber-1024 Track 3 from 2031.9 to 2033.6; AES-256 Track 3 in the hypothesis-only variant from 2038.5 to 2040.3). If quantum hardware stalls, with the hardware rate falling to 1.05, the RSA Track 0 crossover slips from 2032.5 to 2035.3 while the software-driven crossovers barely move, since they depend on the AGI-gated software driver rather than on raw hardware. If AI-assisted attack progress is twenty percent weaker, with all software couplings reduced by a fifth, the lattice crossovers, and the AES-256 values of the hypothesis-only variant, move later by roughly 0.6 to 1.0 years. And if Track 3 loses its sharply convex character, with its rate dropping to the Track 2 value of 1.17, the most aggressive crossovers retreat, the hypothesis-only AES-256 Track 3 value moving from 2038.5 to 2041.5.

\subsection*{Appendix A.5. Joint Monte Carlo Over the Parameter Ranges}
\addcontentsline{toc}{subsection}{Appendix A.5. Joint Monte Carlo Over the Parameter Ranges}
The one-at-a-time analysis of Appendix A.4 isolates each parameter's effect but does not capture interactions among simultaneously uncertain inputs. This appendix complements it with a joint Monte Carlo that samples all model parameters together over their stated ranges and reports the quantity the framework is ultimately built to produce: P(T ≤ year), the probability that a scheme's feasible-attack crossover arrives on or before a given year. This is precisely the input Mosca's inequality requires for the capability horizon Z (Section 3.1), for which any nonzero probability mass inside the secrecy horizon X is sufficient to make migration and crypto-agility rational today.

Sampling. For each algorithm and track, N = 40,000 independent draws are taken, each sampling uniformly unless noted: the hardware sensitivity α(a) and software sensitivity β(a, k) over ±25\% of their Appendix A.2 values; the hardware amplitude A over ±25\% of 0.34; the hardware rate r\_hw over [1.10, 1.25]; the frontier-model software slope uniformly over its stated band of 0.08 to 0.12 per year (v2 sampled ±50\% of 0.04); the track rate r\_k over ±0.05; and the track ceiling B\_k over ±25\%. For the AGI-gated tracks (1 to 3) the AGI onset t\_AGI is drawn from a triangular distribution on [2026, 2028, 2031] with 2028 as its mode (v2 used [2027, 2028, 2031]), reflecting the developer and expert forecasts surveyed in Section 4 as of November 2026; Track 0 carries no software driver and hence no AGI dependence. Each algorithm-and-track cell uses its own independent, deterministically seeded random stream (NumPy default\_rng seeded with 20260731 plus the cell index, in the cell order and draw order stated in Appendix A.3), so every result regenerates exactly and independently of the order in which the cells are computed; crossovers beyond 2060 are treated as censored.

Results. Figure A2 plots the resulting P(T ≤ year) curves and Table 7 summarizes their 10th-percentile, median, and 90th-percentile crossover years with the probability of a feasible attack by 2035. The mechanism-backed RSA-2048 exposure is robust to the AGI question: even the no-AGI Track 0 baseline has a median crossover near 2033 and about a 95\% probability of a feasible attack by 2035. The lattice schemes remain outside the policy-relevant window under Track 0 and Track 1 (median crossovers in the late 2040s) and enter it only under the AGI-gated software tracks: Kyber-1024 spanning roughly 2031 to 2034 under Track 3 (P(≤2033) ≈ 0.77, P(≤2035) ≈ 1.00; v2: 0.52 and 0.98) and roughly 2033 to 2036 under Track 2 (P(≤2035) ≈ 0.59; v2: 0.28), with ML-DSA-44 under Track 3 about half a year after Kyber-1024 under Track 3 (P(≤2035) ≈ 0.98; v2: 0.91). The shift from v2 comes almost entirely from the slope band and the earlier onset tail, and it is a shift on the model's own measure, not a calibrated probability; sampling the slope over the wider ±50\% width of v2 recentered on 0.10 (0.05 to 0.15) gives the same Track 3 figure to within 0.01. The joint spread is modestly wider than the one-at-a-time bands of Appendix A.4, as expected once parameters move together, but it does not alter the qualitative ordering or shift any median by more than about a year from the point estimates of Appendix A.2, and the AGI onset remains the dominant source of variance for every software-driven crossover.

\par\vspace{4pt}\noindent\begin{minipage}{\linewidth}
\begin{center}\footnotesize\setlength{\tabcolsep}{4pt}\begin{tabular}{|p{0.147\linewidth}|p{0.147\linewidth}|p{0.147\linewidth}|p{0.147\linewidth}|p{0.147\linewidth}|p{0.147\linewidth}|}\hline
\textbf{Scheme} & \textbf{Track} & \textbf{P10} & \textbf{Median} & \textbf{P90} & \textbf{P(≤2035)} \\ \hline
RSA-2048 & 0 (no AGI) & 2031.8 & 2033.0 & 2034.5 & 0.95 \\ \hline
RSA-2048 & 3 & 2029.0 & 2029.8 & 2030.6 & 1.00 \\ \hline
Kyber-1024 & 0 (no AGI) & 2044.6 & 2048.7 & 2055.9 & 0.00 \\ \hline
Kyber-1024 & 2 & 2033.3 & 2034.8 & 2036.4 & 0.59 \\ \hline
Kyber-1024 & 3 & 2031.0 & 2032.3 & 2033.7 & 1.00 \\ \hline
ML-DSA-44 & 3 & 2031.6 & 2032.9 & 2034.3 & 0.98 \\ \hline
\end{tabular}\end{center}
\vspace{11.25pt}\begin{center}\small\textbf{Table 7: Joint Monte Carlo crossover distribution (10th percentile / median / 90th percentile, in calendar years) and probability of a feasible attack by 2035, over the parameter ranges of Appendix A.5 (N = 40,000 per row).}\end{center}
\end{minipage}\par\vspace{4pt}
Interpretation. The Monte Carlo adds no evidence; it propagates the stated uncertainty in the chosen couplings into an explicit distribution, so the crossover years can be read as probability curves rather than single dates. It inherits every assumption and limitation of the capability model, including that the couplings are chosen rather than measured (Section 5); widening or shifting the input ranges shifts the output curves correspondingly. The hash-based and symmetric primitives are deliberately omitted from these probability statements, for the same reason their dates are withheld from Figure 9: their crossovers rest on a hypothesized mechanism with no cryptanalytic basis (Appendix C), and a probability attached to them would misrepresent a modeling assumption as a calibrated likelihood.

\begin{figure}[H]\centering
\includegraphics[width=0.82\linewidth]{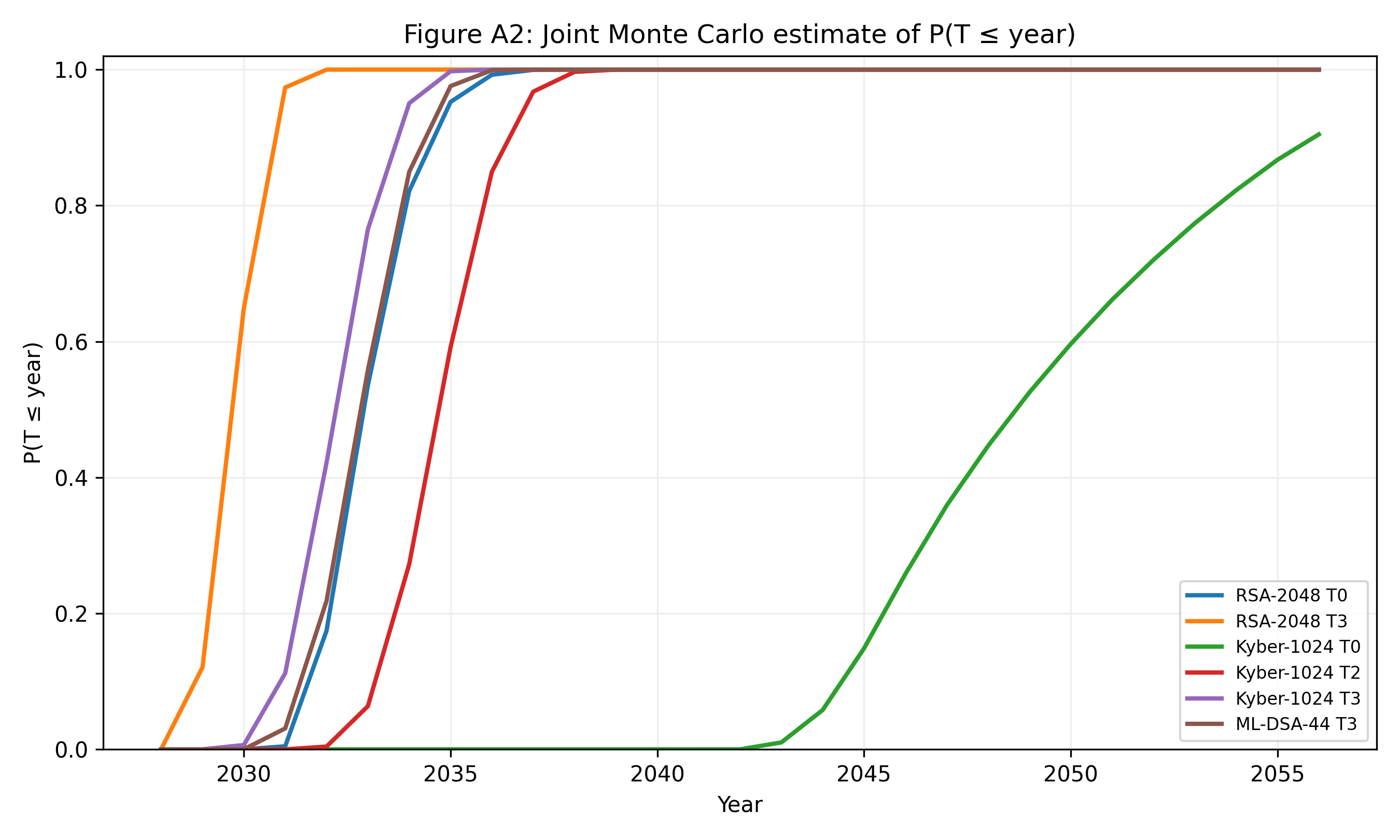}
\par\vspace{3pt}{\small\textbf{Figure A2. Joint Monte Carlo estimate of P(T ≤ year), the probability that the modeled feasible-attack crossover occurs on or before a given year, for the decision-relevant algorithm and track combinations, over the parameter ranges of Appendix A.5.}}
\end{figure}
\newpage
\section*{Appendix B. Notation and Symbols}
\addcontentsline{toc}{section}{Appendix B. Notation and Symbols}
The model spans the acceleration-factor framing of Section 3.3, the cost model of Section 4.5, and the capability model of Appendix A. Table 8 lists the symbols used across these and disambiguates the two distinct uses of β. Table 9 isolates the six symbols that an independent review of the first version found most liable to conflation in secondary citation [234]: the two uses of β, the two uses of B, and the quantities ρ, Δ, R and the frontier-model term.

\par\vspace{4pt}\noindent\begin{minipage}{\linewidth}
\begin{center}\footnotesize\renewcommand{\arraystretch}{0.95}\setlength{\tabcolsep}{4pt}\begin{tabular}{|p{0.07\linewidth}|p{0.36\linewidth}|p{0.07\linewidth}|p{0.36\linewidth}|}\hline
\textbf{\textbf{Symbol}} & \textbf{\textbf{Meaning}} & \textbf{\textbf{Symbol}} & \textbf{\textbf{Meaning}} \\ \hline
t & Calendar year (model time variable) & R(a, k, t) & Estimated attack runtime, R0 − C, on a base-10 log-seconds axis \\ \hline
t0 & Reference year, 2026 & R0 & Runtime baseline, log10(10\textasciicircum{}10) = 10 (Centuries); a shared visual baseline \\ \hline
tAGI & Assumed AGI onset year, 2028 & MON & One-month feasibility line, log10(2.6 × 10\textasciicircum{}6) ≈ 6.415 \\ \hline
C(a, k, t) & Cryptanalytic capability for algorithm a on track k, in base-10 orders of magnitude of attack-runtime reduction & P(a, k, t) & Modeled feasibility; logistic of the margin by which runtime falls below the one-month line \\ \hline
D\_hw(t) & Hardware capability driver; accelerating, present on all tracks & Aᵢ & Engineering-channel acceleration factor (Section 3.3) \\ \hline
D\_sw(k, t) & Software capability driver; AGI-gated and track-specific & τᵢ & Per-channel time to readiness (Section 3.3) \\ \hline
A & Hardware driver amplitude (0.34) & Δ & Sieving-exponent erosion (Section 4.5 cost model); the eroded quantum sieving exponent is 0.265 − Δ, floored so that 0.265 − Δ ≥ 0.2075 \\ \hline
r\_hw & Hardware acceleration rate (1.20) & Δ\_max & Ceiling of the erosion schedule; Δ(t) = Δ\_max · (1 − e\textasciicircum{}(−R·(t − tAGI))) for t ≥ tAGI, and Δ(t) = 0 before tAGI (Section 4.5) \\ \hline
sw\_pre & Frontier-model software slope (0.10 per year; stated band 0.08 to 0.12; applied from 2026 on every software-coupled track) & R & Research-acceleration factor in the erosion schedule: human-cryptanalyst-years compressed into one calendar year (Section 4.5); distinct from the runtime R(a, k, t) above \\ \hline
r\_k & Track k software acceleration rate (Tracks 0 to 3: 1.01, 1.10, 1.17, 1.30) & B(t), B\_0 & Feasible attack budget in bits, B(t) = B\_0 + r · (t − t0), with slope r (Section 4.5); distinct from the track ceiling B\_k above \\ \hline
B\_k & Track k software ceiling (Tracks 0 to 3: 0.00, 0.16, 0.26, 0.30) & ρ & Effective block-size collapse fraction (Section 4.5); ρ = 0 means no structural advance \\ \hline
α(a) & Algorithm sensitivity to hardware (Shor); large for RSA, near zero otherwise & δ(β) & Root-Hermite factor of the lattice estimate (Section 3.1.3); governs achievable basis quality at block size β \\ \hline
β(a, k) & Algorithm sensitivity to the software attacks enabled on track k (capability model) & β (lattice) & Core-SVP block size for the lattice estimate (Sections 3.1.3 and 4.5); distinct from the capability-model coupling β(a, k) above \\ \hline
\end{tabular}\end{center}
\vspace{11.25pt}\begin{center}\small\textbf{Table 8: Notation and symbols used in the model.}\end{center}
\end{minipage}\par\vspace{4pt}
The six entries of Table 9 are the ones most often run together when the paper is cited at second hand; each row states the quantity, where it lives, and the quantity it is not.

\par\vspace{4pt}\noindent\begin{minipage}{\linewidth}
\begin{center}\footnotesize\setlength{\tabcolsep}{4pt}\begin{tabular}{|p{0.293\linewidth}|p{0.293\linewidth}|p{0.293\linewidth}|}\hline
\textbf{\textbf{Symbol}} & \textbf{\textbf{Meaning and home}} & \textbf{\textbf{Not to be confused with}} \\ \hline
β(a, k) and β (lattice) & β(a, k): capability-model coupling of algorithm a to the software attacks enabled on track k, a chosen scenario parameter (Appendix A.2). β (lattice): core-SVP block size at which BKZ is estimated to succeed, 625 / 877 / 420 for ML-KEM-768 / ML-KEM-1024 / ML-DSA-44 (Sections 3.1.3 and 4.5) & Each other. The capability model never uses the block size and the cost model never uses the coupling. \\ \hline
B\_k and B(t) & B\_k: software ceiling of track k in orders of magnitude, 0.00 to 0.30 (Appendix A.2). B(t): feasible attack budget in bits, B₀ + r · (t − t₀) (Section 4.5) & Each other. B\_k bounds a driver; B(t) is the budget the lattice cost is compared against. \\ \hline
ρ & Effective block-size collapse fraction, β → β(1 − ρ), in the cost model (Section 4.5); the one speculative parameter, with no known realization & Δ, which erodes the exponent rather than the dimension; and the dihedral-coset route of Appendix D, which is a tier trigger, not a parameter. \\ \hline
Δ & Erosion of the quantum sieving exponent below 0.265, floored so that 0.265 − Δ ≥ 0.2075, on the schedule Δ(t) = Δ\_max · (1 − e\textasciicircum{}(−R·(t − tAGI))) (Section 4.5) & ρ. Erosion cannot change the exponential form of the cost; collapse acts on the dimension. \\ \hline
R & Research-acceleration factor in the erosion schedule: human-cryptanalyst-years compressed into one calendar year (Section 4.5) & R(a, k, t), the runtime index of Appendix A; and the engineering acceleration factors Aᵢ of Section 3.3, which act on hardware. \\ \hline
Frontier-model term (sw\_pre) & Pre-AGI software slope of the capability model, 0.10 per year with a stated band of 0.08 to 0.12, applied from 2026 on every software-coupled track (Appendix A.2) & The AGI-gated software driver D\_sw(k, t), which switches on at tAGI; and R, which lives in the cost model. \\ \hline
\end{tabular}\end{center}
\vspace{11.25pt}\begin{center}\small\textbf{Table 9: The six symbols most often conflated in secondary citation, each with the quantity it is not.}\end{center}
\end{minipage}\par\vspace{4pt}
\section*{Appendix C. Hypothesis-Only Stress Cases: SLH-DSA (FIPS 205) and AES-256 (FIPS 197)}
\addcontentsline{toc}{section}{Appendix C. Hypothesis-Only Stress Cases: SLH-DSA (FIPS 205) and AES-256 (FIPS 197)}
This appendix presents the two hypothesis-only stress cases separately from the mechanism-backed and contingency-backed projections of Section 4. No known quantum, classical, or algorithmic attack threatens SLH-DSA or AES-256 within the modeled horizon, and any Track 2 or Track 3 crossover for these primitives would exist only under the assumption that advanced AI discovers a presently unknown class of structural attack. The cases are retained because the consolidated spectrum of Figure 9 includes them, and because ruling primitives out is as much a part of the due-diligence basis of this analysis as identifying exposure; they carry the weakest evidential status of the three tiers defined in Table 1A and Section 4.5, and they inherit the falsifiability conditions of Section 5. The conservative baseline that this overlay sits on is the depth-limited Grover accounting of NIST's 2016 call for proposals, the MAXDEPTH parameter [236], with the AES resource estimates of Grassl et al. [237] and Jaques et al. [238]: Grover's search parallelizes poorly, so under a realistic depth limit the effective quantum security of AES-256 stays far above any budget in this paper's horizon, which is why no dated AES-256 or SLH-DSA crossover appears in the body.

\subsection*{Appendix C.1. SLH-DSA Estimates of CRQC+AI Vulnerability (FIPS 205)}
\addcontentsline{toc}{subsection}{Appendix C.1. SLH-DSA Estimates of CRQC+AI Vulnerability (FIPS 205)}
Figures C1A and C1B evaluate the runtime attack and modeled feasibility estimates for breaking SLH-DSA (FIPS 205) under each of the different vulnerability tracks for the four different assumption scenarios.

\begin{center}\textit{Note: Hypothesis-only scenario: SLH-DSA contains no lattice structure for the collapse parameter ρ to act on, and its Grover-type exposure is already absorbed into its conservative parameters (Section 4.5). SLH-DSA is assigned zero software coupling and reports no crossover within the modeling horizon; the dated Track 2 and Track 3 figures shown in earlier drafts rested on a hypothesized, presently unsupported novel attack and are withdrawn in favor of the undated hypothesis-only treatment used throughout the body.}\end{center}
\begin{figure}[H]\centering
\includegraphics[width=0.82\linewidth]{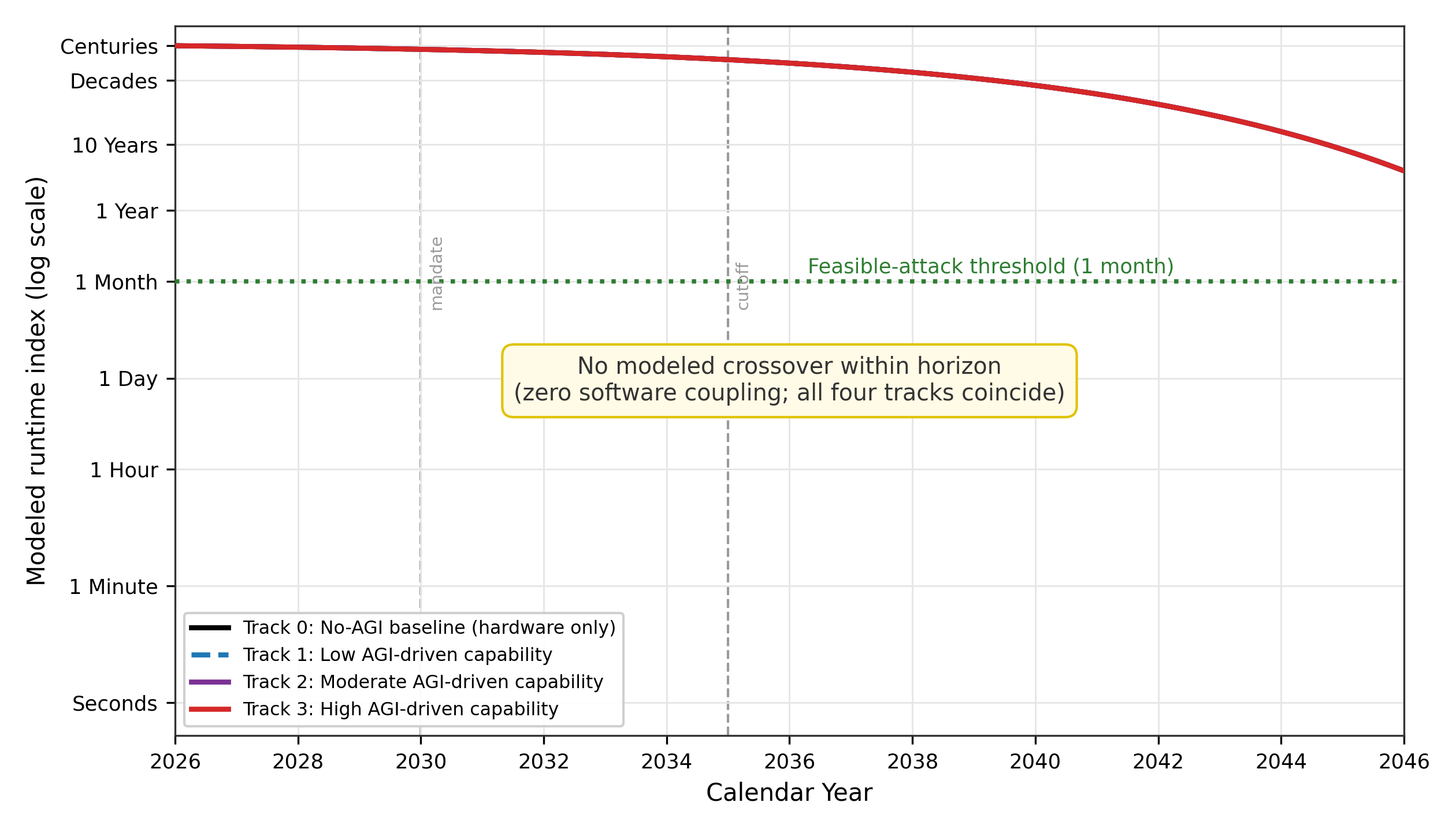}
\par\vspace{3pt}{\small\textbf{Figure C1A: Projected Attack Runtime Against SLH-DSA FIPS 205 \\[2pt]Hash-Based Signatures (conservative)}}
\end{figure}
\begin{figure}[H]\centering
\includegraphics[width=0.82\linewidth]{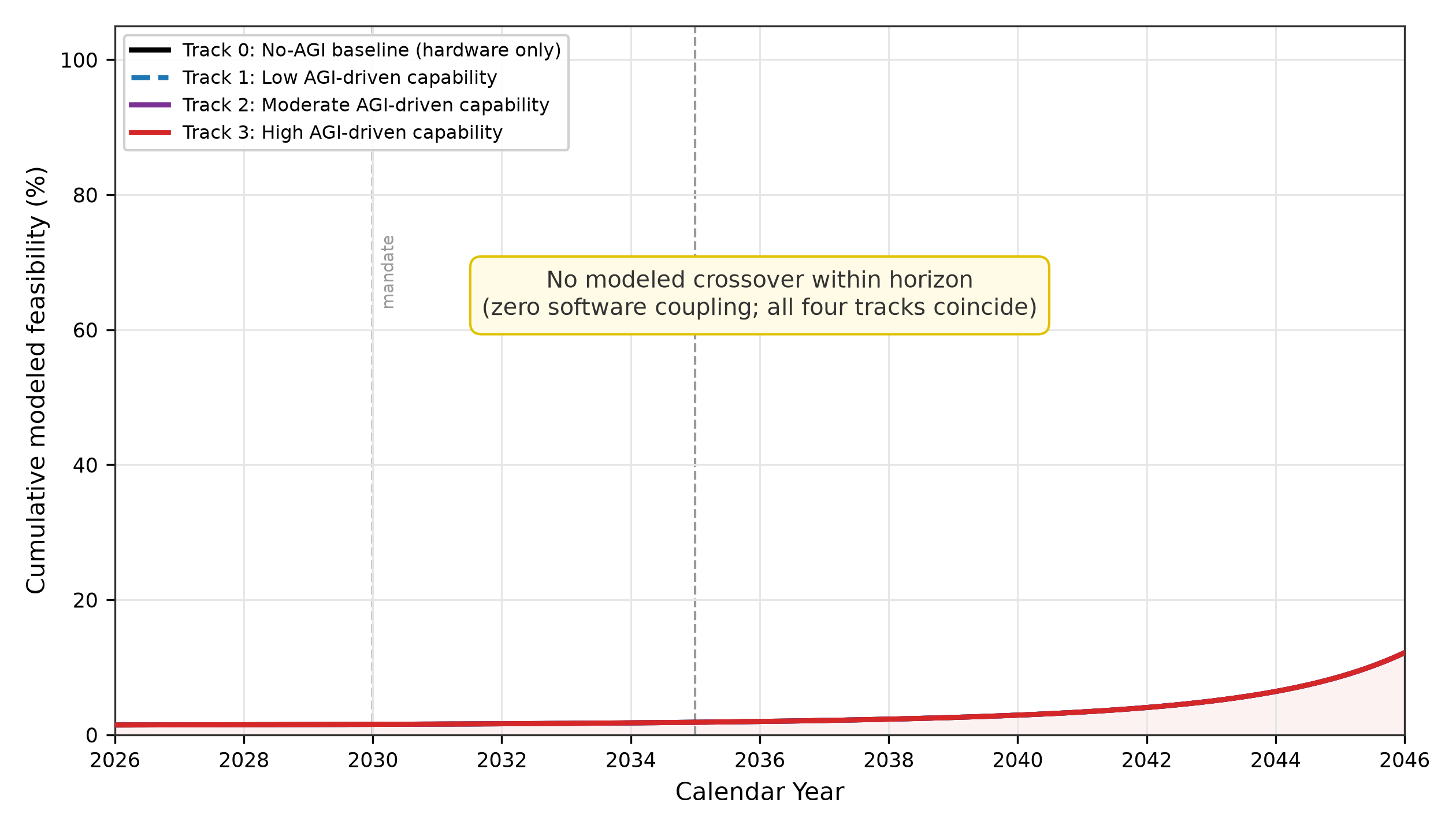}
\par\vspace{3pt}{\small\textbf{Figure C1B: Modeled Feasibility Estimates for SLH-DSA FIPS 205 \\[2pt]Hash-Based Signatures (conservative)}}
\end{figure}
\subsection*{Appendix C.2. AES-256 Estimates of CRQC+AI Vulnerability (FIPS 197)}
\addcontentsline{toc}{subsection}{Appendix C.2. AES-256 Estimates of CRQC+AI Vulnerability (FIPS 197)}
Figures C2A and C2B evaluate the runtime attack and modeled feasibility estimates for breaking AES-256 (FIPS 197) under each of the different vulnerability tracks for the four different assumption scenarios.

\begin{center}\textit{Note: Hypothesis-only scenario: no known quantum, classical, or algorithmic attack breaks AES-256, and under Grover-only analysis it retains 128-bit effective security, so no track shows a feasible break within the horizon. AES-256 is assigned zero software coupling; the dated Track 2 and Track 3 figures shown in earlier drafts derived entirely from speculative assumptions (detailed in the caveat at the end of this section) and are withdrawn in favor of the undated hypothesis-only treatment used throughout the body.}\end{center}
\begin{figure}[H]\centering
\includegraphics[width=0.82\linewidth]{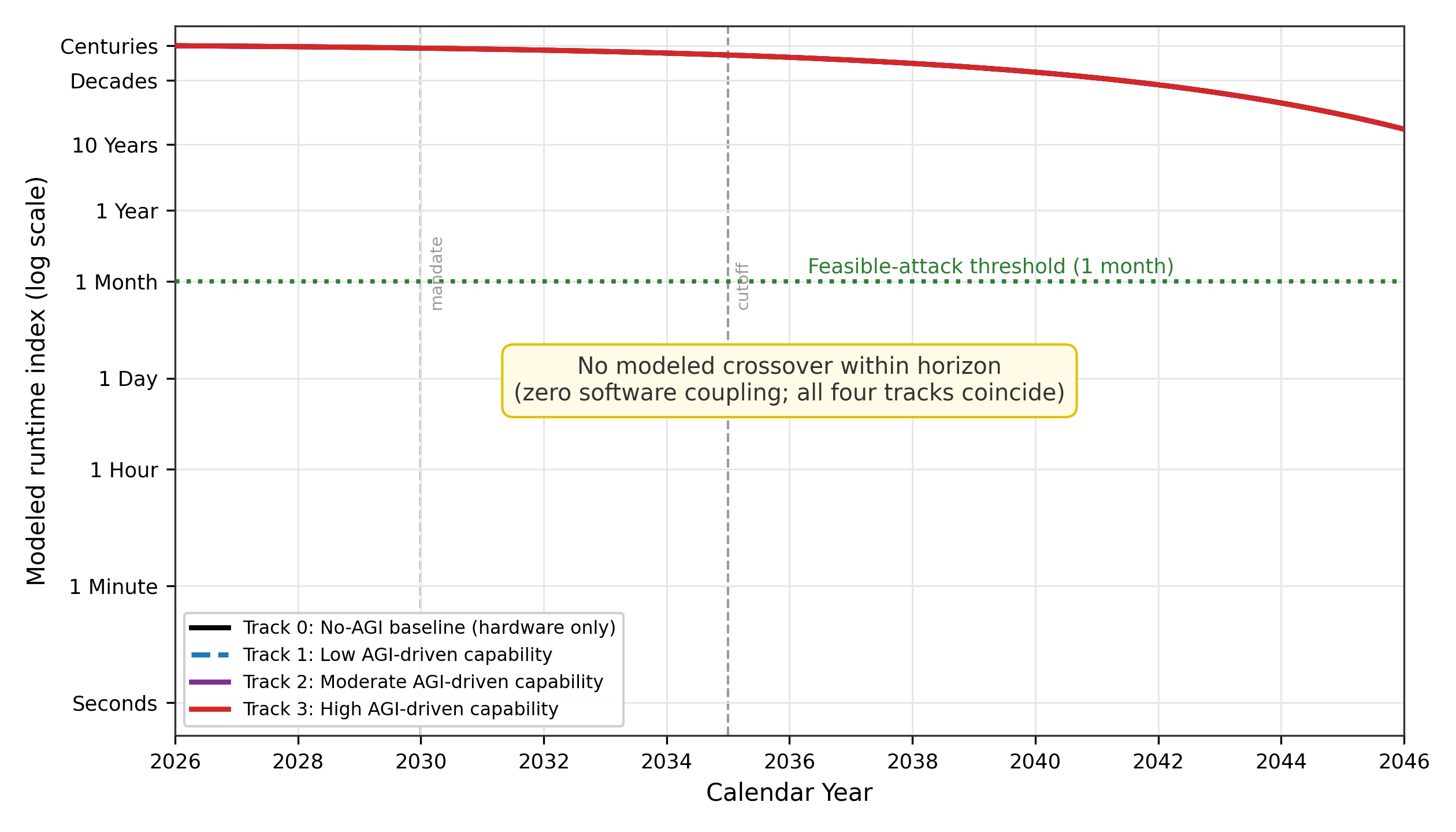}
\par\vspace{3pt}{\small\textbf{Figure C2A: Projected Attack Runtime Against AES-256 FIPS 197 TLS 1.3 Symmetric Layer, \\[2pt]256-bit (128-bit effective post-Grover)}}
\end{figure}
\begin{figure}[H]\centering
\includegraphics[width=0.82\linewidth]{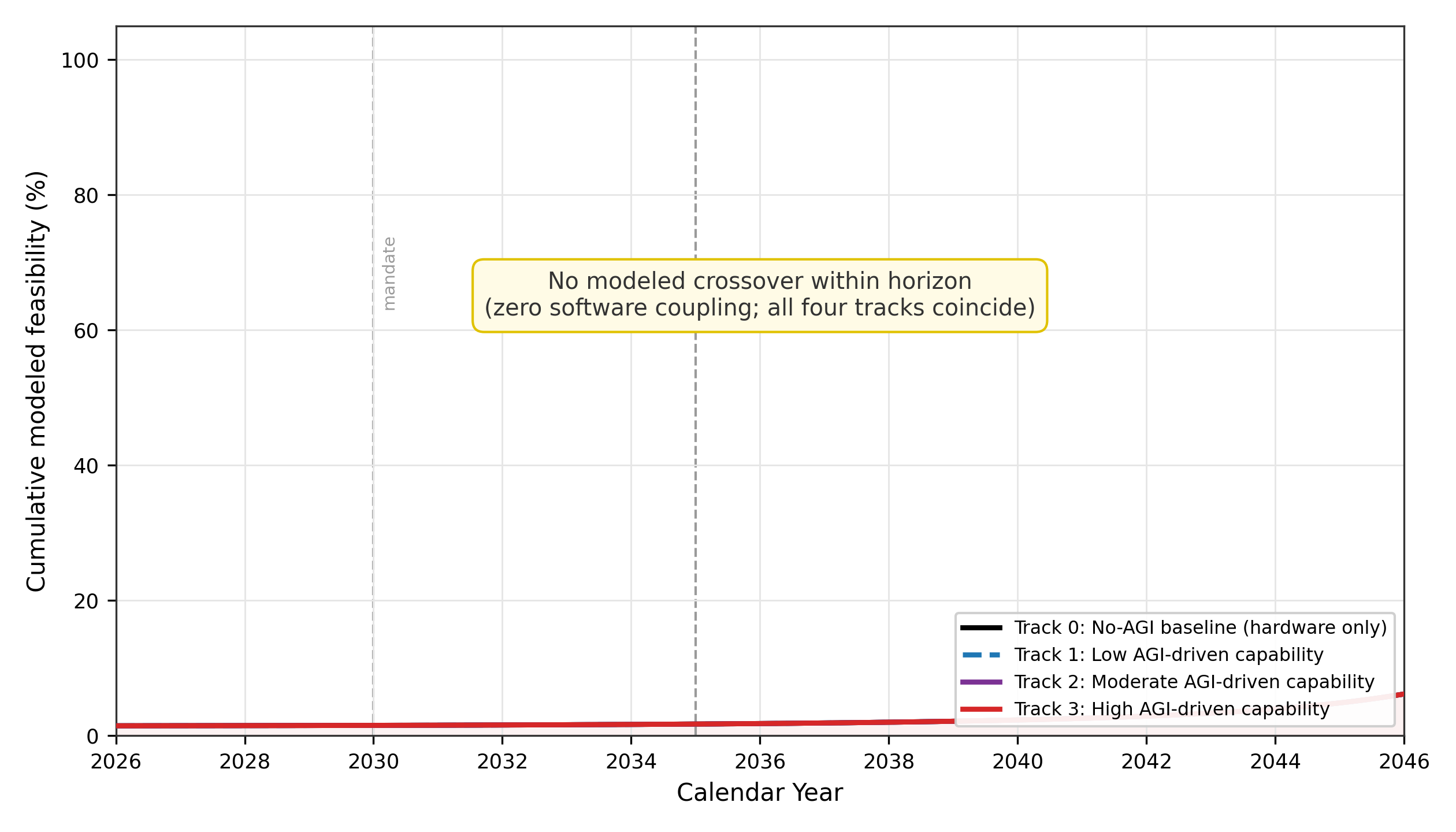}
\par\vspace{3pt}{\small\textbf{Figure C2B: Modeled Feasibility Estimates for AES-256 FIPS 197 TLS 1.3 Symmetric Layer, \\[2pt]256-bit (128-bit effective post-Grover)}}
\end{figure}
AES, standardized in FIPS 197, remains the relevant approved symmetric cipher for high-security post-quantum deployments, with AES-256 preferred where 128-bit post-Grover security strength is required. Although FIPS 197 includes three different key lengths (128, 192 and 256), the fact that Grover’s algorithm provides a quadratic (square root) speedup on unstructured search means that the effective quantum security strength for AES is assumed to be one-half of the AES key size. As a result, under the PQC framework advanced by NIST of having encryption algorithms that provide at least 128 bits of quantum security strength, AES-256 is the only one of the three levels of symmetric key length that meets this framework. Accordingly, AES-256 is the only recommended AES symmetric algorithm to be used for PQC encryption under CNSA 2.0 and is among the approved security functions listed by NIST. [163] [164]

Quantitatively, Grover’s algorithm searches an unstructured keyspace of size 2\textsuperscript{k} in about (π/4)·2\textsuperscript{k/2} evaluations, so the effective quantum security is half the key length, k/2 bits. For AES-256 this maps a classical 2\textsuperscript{256} search to a quantum 2\textsuperscript{128} search, which is why AES-256 alone clears the NIST 128-bit quantum-security bar and underlies the Track 0 and Track 1 baselines in Figures C2A and C2B. This k/2 figure is itself a conservative upper bound on the quantum threat. Grover’s search is essentially sequential: the quadratic speedup requires a single coherent computation of depth on the order of 2\textsuperscript{k/2}, and it parallelizes poorly, so distributing the work across many machines recovers only a square-root benefit per machine. Under the realistic circuit-depth limits that NIST applies, AES-256 retains close to its full classical strength and is placed in the highest post-quantum security category. Any Track 2 or Track 3 AES-256 crossover would therefore not follow from Grover; it could rest only on a hypothesized and presently unsupported structural attack, which is why AES-256 is carried in the hypothesis-only tier defined in Section 4.5 with zero software coupling, and why Figures C2A and C2B show no crossover on any track.

\textbf{TLS 1.3 Implication}: Under a conventional Grover-only analysis, AES-256 retains 128-bit effective security and is not threatened by brute-force or hardware-scaling attacks. This is reflected in Figures C2A and C2B where Track 0 (baseline) and Track 1 (low AGI-driven software capability) show no feasible break for AES-256 across the projection horizon. Under the more aggressive AGI-software tracks the figures likewise show no crossover, because AES-256 carries zero software coupling in the main model. The hypothesis-only variant of Appendix A.2 (β = 0.155 and 0.210), retained only as a stress input for Appendix A.4, would place a feasible-attack threshold late in the horizon, in the mid-2040s under Track 2 (AGI-optimized side-channel and implementation attacks, which are largely independent of key length) and the late 2030s under Track 3 (a hypothesized AGI-discovered structural attack that would defeat the Grover-only bound); those dated values have no cryptanalytic basis and are not plotted. AES-256 remains the most resistant of the five algorithms evaluated under every track.

\textbf{Combined figure: Figure C3 is a simplified combination of Figure 9 and of Figures C1A/C1B and C2A/C2B of this appendix, for use as a one-figure summary.}

\begin{figure}[H]\centering
\includegraphics[width=0.82\linewidth]{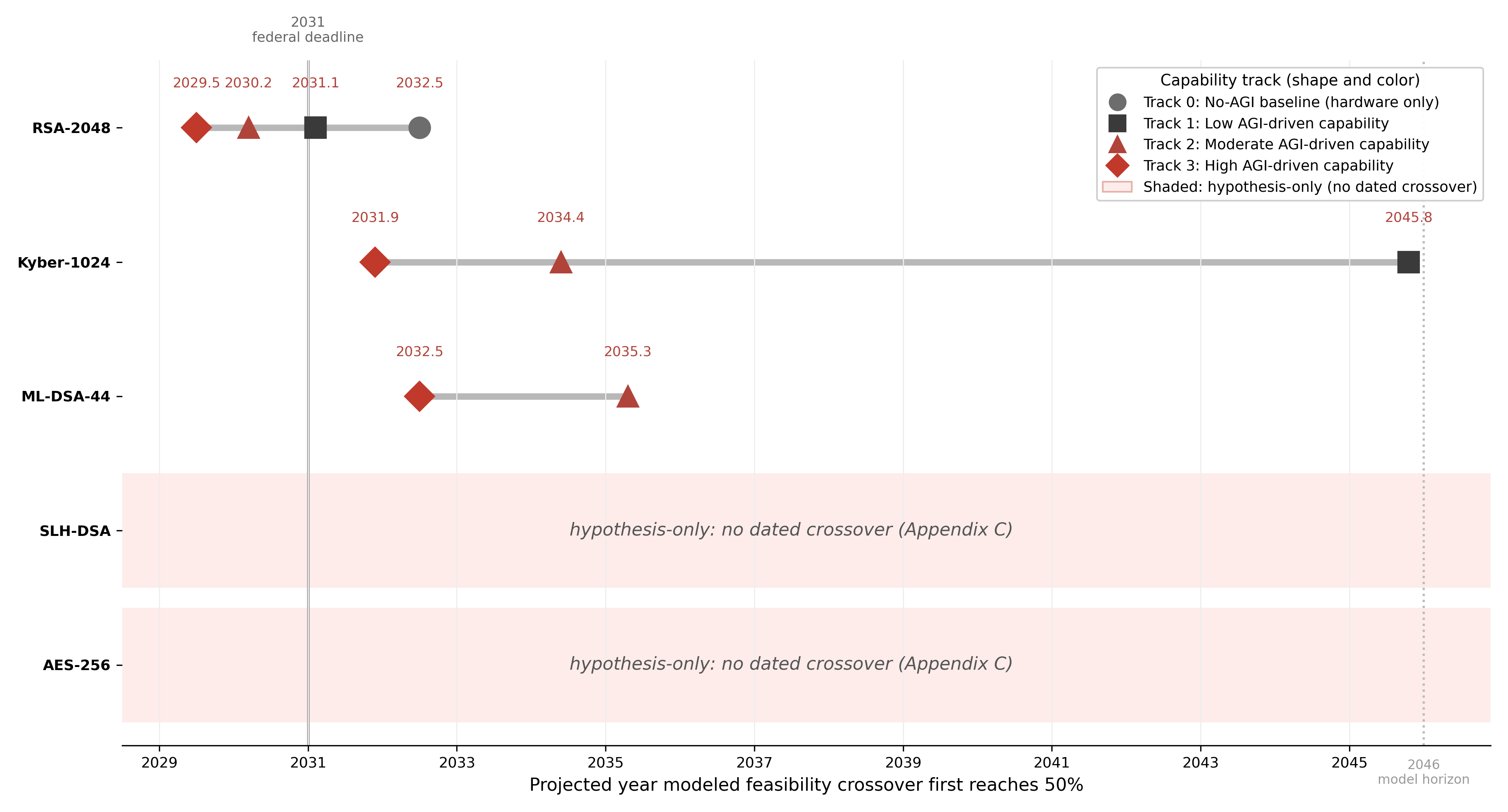}
\par\vspace{3pt}{\small\textbf{Figure C3: Simplified Combination Risk Spectrum Vulnerability Timelines }}
\end{figure}
\textbf{Caveat on the AES-256 hypothesis-only treatment: }Any dated break estimate for AES-256 would only be a hypothesis-driven scenario projection with no current cryptanalytic basis. No known quantum, classical, or algorithmic attack breaks AES-256, and the consensus view is that it remains secure well beyond the horizon shown. Unlike the RSA and ECC case, where a concrete algorithm (Shor) anchors the projected attack mechanism, any dated AES-256 estimate would derive entirely from two speculative assumptions: that an AGI discovers a novel and presently-unknown structural attack on the symmetric cipher, and that side-channel or implementation compromise is treated as equivalent to a break of the cipher itself. The lattice schemes occupy the intermediate, contingency-backed tier: consistent with Section 2.1, HHL and related quantum linear-algebra methods contribute at most polynomial acceleration to an inner step and do not by themselves anchor an attack, so the lattice crossovers are contingent on the conjectured effective-dimension collapse of Section 4.5 rather than on any established algorithm. The AES-256 figures should be read with that limitation expressly in mind.

\section*{Appendix D. The Dihedral-Coset Route as a Second Contingency Pathway}
\addcontentsline{toc}{section}{Appendix D. The Dihedral-Coset Route as a Second Contingency Pathway}
Section 2.2 names one structural pathway against lattice assumptions, the geometric-algebra collapse parameterized by ρ. The lattice community's live pathway is a different one, the dihedral coset problem (DCP). Regev showed in 2004 that a DCP algorithm would solve unique-SVP [223]; Kuperberg gave the best known DCP algorithm, subexponential, running in time 2\textasciicircum{}O(√n), in 2005 [224]; Brakerski, Kirshanova, Stehlé and Wen established the equivalence of LWE with an extrapolated DCP in 2018 [225]; Bai et al. gave a quasi-polynomial quantum algorithm for the extrapolated problem over power-of-two moduli at CRYPTO 2025 [226], a result its authors state does not affect LWE at standard parameters, since the standard reduction caps samples at polynomial while the algorithm needs quasi-polynomially many; and Wen and Zheng (CRYPTO 2026) proved that search Module-LWE over power-of-two cyclotomic rings at constant module rank is as hard as a structured extrapolated DCP, which is ML-KEM's ring and rank [227].

In August 2026 Simon claimed a polynomial-time quantum algorithm for DCP [228]. Had it survived it would have composed through those reductions to ML-KEM's assumption as a polynomial-time algorithm: not a point in the (0.265 − Δ)·β·(1 − ρ) parameter space of Section 4.5 but a change of functional form, as Shor's algorithm is for RSA. It did not survive. Gupte, Ragavan and Zhandry showed within days, with an accompanying Lean 4 formalization, that the algorithm cannot extract the least-significant bit of the DCP secret with non-negligible advantage [229], and their Fourier-label information-loss barrier closes every algorithm on Regev's template that discards more than a logarithmic number of bits from each Fourier label, a class that contains Simon's algorithm, which uses only the most significant third of the labels [230]; Guo and Yang showed that the paper's lemmas, even where provable, do not establish correctness [231]. The author has not withdrawn the preprint: its seventh revision, dated November 3, 2026, carries a note, first added on October 7, 2026, that addresses [229]. The barrier is a no-go for one algorithm class, not for the problem.

The episode is recorded for two reasons. First, the promotion trigger of Table 1B as written in v2 described only a value of ρ; it now also names a surviving sub-exponential or polynomial-time DCP algorithm that composes through a published Module-LWE reduction as a promotion in its own right, and a no-go theorem narrowing the class as a demotion. Second, the summer of 2026 produced a human-generated structural claim that collapsed within days and an AI-generated structural claim, the HAWK automorphism of Section 3.2.1.2, that held and ended a NIST candidate; the rate of structural claims against lattices, true or false, is rising, and the indicator added to Section 3.3.5 logs each claim with its verification outcome. Neither event moves a parameter of the model: the HAWK lattice carries structure that ML-KEM's random module lattices are designed not to have, and the DCP claim failed. An independent review of the first version [234] read Section 2.2 as a research program written in the grammar of an attack path, without a literature of its own; the dihedral-coset route above, with its peer-reviewed Module-LWE bridge, two attempted algorithms in two years and one machine-checked no-go, is the answer on the evidence, and the paper now carries it.

\newpage\section*{References}\small\begin{itemize}\setlength\itemsep{1pt}
\item[] [1]	Boyd, C. Introduction to the Special Issue on TLS 1.3. \textit{J Cryptol} 34, 24 (2021). https://doi.org/10.1007/s00145-021-09386-z.
\item[] [2]	Chen, L.; Moody, D.; Liu, Y.K., NIST, Post-Quantum Cryptography Standardization, (Jan. 2015 to Jun. 2025). https://csrc.nist.gov/pqc-standardization. 
\item[] [3]	Reference removed in v3. Versions 1 and 2 cited here a withdrawn 2024 preprint as the source of Figure 1 and of the QAI, QML and QDL framing; Figure 1 is now the authors' own schematic and the preprint is no longer cited. The number is kept so that references [4] to [240] are unchanged and [241] was added in v3. 
\item[] [4]	NIST, Post-Quantum Cryptography Standards (Aug. 13, 2024): FIPS 203, Module-Lattice-Based Key-Encapsulation Mechanism Standard, https://csrc.nist.gov/pubs/fips/203/final; FIPS 204, Module-Lattice-Based Digital Signature Standard, https://csrc.nist.gov/pubs/fips/204/final; and FIPS 205, Stateless Hash-Based Digital Signature Standard, https://csrc.nist.gov/pubs/fips/205/final.
\item[] [5]	White House, Executive Order 14412, Securing the Nation Against Advanced Cryptographic Attacks (June 22, 2026). https://www.whitehouse.gov/presidential-actions/2026/06/securing-the-nation-against-advanced-cryptographic-attacks/. \textit{See also, }Executive Order 14144, Strengthening and Promoting Innovation in the Nation’s Cybersecurity (Jan. 17, 2025), https://www.federalregister.gov/documents/2025/01/17/2025-01470/strengthening-and-promoting-innovation-in-the-nations-cybersecurity, and Executive Order 14306, Sustaining Select Efforts to Strengthen the Nation’s Cybersecurity and Amending Executive Order 13694 and Executive Order 14144 (Jun. 6, 2025), https://www.federalregister.gov/documents/2025/06/11/2025-10804/sustaining-select-efforts-to-strengthen-the-nations-cybersecurity-and-amending-executive-order-13694.
\item[] [6]	Australian Signals Directorate Information Security Manual: Guidelines for Cryptography (Dec. 2024), https://www.cyber.gov.au/business-government/asds-cyber-security-frameworks/ism/cyber-security-guidelines/guidelines-for-cryptography; and EU Comm. A Coordinated Implementation Roadmap for the Transition to Post-Quantum Cryptography (Jun. 2025), https://digital-strategy.ec.europa.eu/en/library/coordinated-implementation-roadmap-transition-post-quantum-cryptography.
\item[] [7]	Egbuagha, O.; Ikwunna, E., Post-Quantum Cryptography in Practice: A Literature Review of Protocol-Level Transitions and Readiness, IACR (Sept. 18, 2025). https://eprint.iacr.org/2025/1668.pdf.
\item[] [8]	Rehman, M., Quantum Insider, What Quantum AI Actually Means (Mar. 30, 2026). https://thequantuminsider.com/2026/03/30/what-quantum-ai-actually-means/. 
\item[] [9]	Barker, E.; Chen, L.; Cooper, D.; Moody, D.; Regenscheid, A.; Souppaya, M.; Newhouse, B., NIST CSWP 39, Considerations for Achieving Crypto Agility (Jun. 29, 2026). https://nvlpubs.nist.gov/nistpubs/CSWP/NIST.CSWP.39-upd1.pdf. 
\item[] [10]	Grover, L. K. (1997). Quantum mechanics helps in searching for a needle in a haystack, Physical Review Letters, 79(2), pp. 325 to 328. https://arxiv.org/pdf/quant-ph/9706033. 
\item[] [11]	Shor, P. W. (1997). Polynomial-time algorithms for prime factorization and discrete logarithms on a quantum computer. SIAM Journal on Computing, 26(5), pp. 1484 to 1509. https://arxiv.org/pdf/quant-ph/9508027. 
\item[] [12]	Harrow, A. W.; Hassidim, A.; Lloyd, S. Quantum Algorithm for Solving Linear Systems of Equations. Phys. Rev. Lett. 2009, 103 (15), 150502.
\item[] [13]	Baskaran, N.; Rawat, A.S.; Jayashankar, A.; Chakravarti, D.; Sugisaki, K.; Roy, S.; Mandal, S.B.; Mukherjee, D., Adapting the Harrow-Hassidim-Lloyd algorithm to quantum many-body theory, Phys. Rev. Research 5, 043113 (Nov. 3, 2023). https://doi.org/10.1103/PhysRevResearch.5.043113.
\item[] [14]	Micciancio, D.; Regev, O. Lattice-based Cryptography. In Post-Quantum Cryptography; Bernstein, D.J., Buchmann, J., Dahmen, E., Eds.; Springer: Berlin, Heidelberg, 2009; pp. 147 to 191. https://doi.org/10.1007/978-3-540-88702-7\_5.
\item[] [15]	Regev, O., On Lattices, Learning with Errors, Random Linear Codes, and Cryptography, Journal of the ACM 56(6), Article 34 (2009), pp. 1 to 40. Republished (2024), https://arxiv.org/pdf/2401.03703. 
\item[] [16]	Cybersecurity and Infrastructure Security Agency (CISA), National Security Agency (NSA), and NIST, Quantum-Readiness: Migration to Post-Quantum Cryptography, (Aug. 21, 2023), https://www.cisa.gov/resources-tools/resources/quantum-readiness-migration-post-quantum-cryptography.
\item[] [17]	An, D.; Childs, A.; Lin, L., Quantum algorithm for linear non-unitary dynamics with near-optimal dependence on all parameters. arXiv:2312.03916 (Dec. 6, 2023). https://arxiv.org/abs/2312.03916.
\item[] [18]	Chen, Y., Quantum Algorithms for Lattice Problems (Apr. 19, 2024). https://eprint.iacr.org/2024/555.
\item[] [19]	Cain, M.; Xu, Q.; King, R.; Picard, L.; Levine, H.; Endres, M.; Preskill, J.; Huang, H.; Bluvstein, D. Shor’s algorithm is possible with as few as 10,000 reconfigurable atomic qubits, Caltech (Mar. 30, 2026). https://arxiv.org/abs/2603.28627.
\item[] [20]	Babbush, R.; Zalcman, A.; Gidney, C.; Broughton, M.; Khattar, T.; Neven, H.; Bergamaschi, T.; Drake, J.; Boneh, D., Securing Elliptic Curve Cryptocurrencies against Quantum Vulnerabilities: Resource Estimates and Mitigations, (Mar. 30, 2026). https://arxiv.org/abs/2603.28846.
\item[] [21]	Childs, A.; Kothari, R.; Somma, R. Quantum algorithm for systems of linear equations with exponentially improved dependence on precision. SIAM Journal on Computing, (2017) 46(6), pp. 1920 to 1950. https://doi.org/10.1137/16M1087072. 
\item[] [22]	Doran, C.; Lasenby, A., Geometric Algebra for Physicists, Cambridge University Press (2003).
\item[] [23]	Hestenes, D.; Sobczyk, G. Clifford Algebra to Geometric Calculus: A Unified Language for Mathematics and Physics; Reidel: Dordrecht (1984).
\item[] [24]	Johnson, M. W. et al. Quantum Annealing with Manufactured Spins. Nature, 473 (7346), (2011), pp. 194 to 198. https://doi.org/10.1038/nature10012. 
\item[] [25]	Farhi, E.; Goldstone, J.; Gutmann, S.; Sipser, M., Quantum Computation by Adiabatic Evolution, arXiv:quant-ph/0001106 (2000) https://arxiv.org/abs/quant-ph/0001106.
\item[] [26]	Goldstein, H.; Poole, C.; Safko, J., Classical Mechanics, 3rd ed., Addison-Wesley (2002).
\item[] [27]	King, A.D.; Suzuki, S.; Raymond, J. et al. Coherent quantum annealing in a programmable 2,000 qubit Ising chain. Nat. Phys. 18, 1324 to 1328 (2022). https://www.nature.com/articles/s41567-022-01741-6.
\item[] [28] 	Lucas, A., Ising Formulations of Many NP Problems. Frontiers in Physics 2, 5 (2014). https://doi.org/10.3389/fphy.2014.00005.
\item[] [29] 	Neukart, F.; Compostella, G.; Seidel, C.; von Dollen, D.; Yarkoni, S.; Parney, B., Traffic Flow Optimization Using a Quantum Annealer. Frontiers in ICT 4, 29 (2017). https://doi.org/10.3389/fict.2017.00029.
\item[] [30] 	Rosenberg, G.; Haghnegahdar, P.; Goddard, P.; Carr, P.; Wu, K.; López de Prado, M., Solving the Optimal Trading Trajectory Problem Using a Quantum Annealer. IEEE Journal of Selected Topics in Signal Processing 10(6), pp. 1053 to 1060 (2016). https://doi.org/10.1109/JSTSP.2016.2574703.
\item[] [31] 	Venturelli, D.; Marchand, D. J. J.; Rojo, G., Quantum Annealing Implementation of Job-Shop Scheduling. arXiv:1506.08479 (2015). https://arxiv.org/abs/1506.08479.
\item[] [32] 	Adachi, S. H.; Henderson, M. P., Application of Quantum Annealing to Training of Deep Neural Networks. arXiv:1510.06356 (2015). https://arxiv.org/abs/1510.06356.
\item[] [33] 	Amin, M. H.; Andriyash, E.; Rolfe, J.; Kulchytskyy, B.; Melko, R., Quantum Boltzmann Machine. Physical Review X 8, 021050 (2018). https://doi.org/10.1103/PhysRevX.8.021050.
\item[] [34] 	Jiang, S.; Britt, K. A.; McCaskey, A. J.; Humble, T. S.; Kais, S., Quantum Annealing for Prime Factorization. Scientific Reports 8, 17667 (2018). https://doi.org/10.1038/s41598-018-36058-z.
\item[] [35] 	Joseph, D.; Callison, A.; Ling, C.; Mintert, F., Two Quantum Ising Algorithms for the Shortest-Vector Problem. Physical Review A 103, 032433 (2021). https://doi.org/10.1103/PhysRevA.103.032433.
\item[] [36] 	Albrecht, M. R.; Prokop, M.; Shen, Y.; Wallden, P., Variational Quantum Solutions to the Shortest Vector Problem. IACR ePrint 2022/233 (2022). https://eprint.iacr.org/2022/233.
\item[] [37] 	Mengoni, R.; Ottaviani, D.; Iorio, P., Breaking RSA Security With a Low Noise D-Wave 2000Q Quantum Annealer: Computational Times, Limitations and Prospects. arXiv:2005.02268 (2020). https://arxiv.org/abs/2005.02268.
\item[] [38]	Peruzzo, A.; McClean, J.; Shadbolt, P.; Yung, M. H.; Zhou, X.-Q.; Love, P. J.; Aspuru-Guzik, A.; O’Brien, J. L., A Variational Eigenvalue Solver on a Photonic Quantum Processor. Nat. Commun. (2014), 5, 4213. https://doi.org/10.1038/ncomms5213. 
\item[] [39] 	McClean, J. R.; Boixo, S.; Smelyanskiy, V. N.; Babbush, R.; Neven, H., Barren Plateaus in Quantum Neural Network Training Landscapes. Nature Communications 9, 4812 (2018). https://doi.org/10.1038/s41467-018-07090-4.
\item[] [40] 	Tilly, J., et al., The Variational Quantum Eigensolver: A Review of Methods and Best Practices. Physics Reports 986, pp. 1 to 128 (2022). https://doi.org/10.1016/j.physrep.2022.08.003.
\item[] [41] 	McClean, J. R.; Romero, J.; Babbush, R.; Aspuru-Guzik, A., The Theory of Variational Hybrid Quantum-Classical Algorithms. New Journal of Physics 18, 023023 (2016). https://doi.org/10.1088/1367-2630/18/2/023023.
\item[] [42] 	Temme, K.; Bravyi, S.; Gambetta, J. M., Error Mitigation for Short-Depth Quantum Circuits. Physical Review Letters 119, 180509 (2017). https://doi.org/10.1103/PhysRevLett.119.180509.
\item[] [43] 	Kandala, A.; Temme, K.; Córcoles, A. D.; Mezzacapo, A.; Chow, J. M.; Gambetta, J. M., Error Mitigation Extends the Computational Reach of a Noisy Quantum Processor. Nature 567, pp. 491 to 495 (2019). https://doi.org/10.1038/s41586-019-1040-7.
\item[] [44] 	Kandala, A.; Mezzacapo, A.; Temme, K.; Takita, M.; Brink, M.; Chow, J. M.; Gambetta, J. M., Hardware-Efficient Variational Quantum Eigensolver for Small Molecules and Quantum Magnets. Nature 549, pp. 242 to 246 (2017). https://doi.org/10.1038/nature23879.
\item[] [45] 	Grimsley, H. R.; Economou, S. E.; Barnes, E.; Mayhall, N. J., An Adaptive Variational Algorithm for Exact Molecular Simulations on a Quantum Computer. Nature Communications 10, 3007 (2019). https://doi.org/10.1038/s41467-019-10988-2.
\item[] [46] 	Cerezo, M.; Larocca, M.; García-Martín, D.; Diaz, N. L.; Braccia, P.; Fontana, E.; Rudolph, M. S.; Bermejo, P.; Ijaz, A.; Thanasilp, S.; Anschuetz, E. R.; Holmes, Z., Does Provable Absence of Barren Plateaus Imply Classical Simulability? Nature Communications 16, 7907 (2025). https://doi.org/10.1038/s41467-025-63099-6.
\item[] [47]	Preskill, J., Quantum Computing in the NISQ Era and Beyond. Quantum (2018), 2, 79. https://doi.org/10.22331/q-2018-08-06-79.
\item[] [48]	Reference removed in v3. Versions 1 and 2 cited here R. Kurzweil, The Singularity Is Near (2005); the book is now named in the text and is not part of the reference list. The number is kept so that the other reference numbers are unchanged.
\item[] [49]	NIST, Submission Requirements and Evaluation Criteria for the Post-Quantum Cryptography Standardization Process, NIST Call for Proposals (Dec. 2016). Section 4. https://csrc.nist.gov/CSRC/media/Projects/Post-Quantum-Cryptography/documents/call-for-proposals-final-dec-2016.pdf.
\item[] [50] 	Mosca, M., Cybersecurity in an Era with Quantum Computers: Will We Be Ready?. IEEE Security \& Privacy 16(5), pp. 38 to 41 (2018). https://doi.org/10.1109/MSP.2018.3761723.
\item[] [51]	Moody, D.; Perlner, R.; Regenscheid, A.; Robinson, A.; Cooper, D., Transition to Post-Quantum Cryptography Standards, NIST Internal Report, NIST IR 8547 (final, Oct. 27, 2026; initial public draft Nov. 12, 2024). https://doi.org/10.6028/NIST.IR.8547.
\item[] [52]	Chen, L.; Jordan, S.; Liu, Y.-K.; Moody, D.; Peralta, R.; Perlner, R.; Smith-Tone, D., Report on Post-Quantum Cryptography, NIST Internal Report, NIST IR 8105 (Apr. 2016). https://nvlpubs.nist.gov/nistpubs/ir/2016/NIST.IR.8105.pdf.
\item[] [53]	Reference removed in v3. Versions 1 and 2 cited here R. Kurzweil, The Singularity Is Nearer (2024), which is no longer cited. The number is kept so that the other reference numbers are unchanged.
\item[] [54]	Kokotajlo, D.; Alexander, S.; Larsen, T.; Lifland, E.; Dean, R., AI 2027, AI Futures Project (Apr. 3, 2025) https://ai-2027.com/ai-2027.pdf.
\item[] [55]	Stanford University Human-Centered Artificial Intelligence, Artificial Intelligence Index Report 2025, 52 to 58, https://hai-production.s3.amazonaws.com/files/hai\_ai\_index\_report\_2025.pdf.
\item[] [56]	Ho, A.; Besiroglu, T.; Erdil, E.; Owen, D.; Rahman, R.; Guo, Z.; Atkinson, D.; Thompson, N.; Sevilla, J., Algorithmic Progress in Language Models, (Mar. 9, 2024) arXiv:2403.05812v1 https://arxiv.org/abs/2403.05812.
\item[] [57]	Jones, N., How AI Can Achieve Human Level Intelligence: Researchers Call for Change in Tack. Nature (Mar. 4, 2025), https://www.nature.com/articles/d41586-025-00649-4.
\item[] [58]	Reference removed in v3. Versions 1 and 2 cited here a Wikipedia article for the term Q-Day, which is no longer cited. The number is kept so that the other reference numbers are unchanged.
\item[] [59] 	Adkins, H.; Schmieg, S., Google, Quantum frontiers may be closer than they appear (March 25, 2026). https://blog.google/innovation-and-ai/technology/safety-security/cryptography-migration-timeline/.
\item[] [60]	IBM, Quantum computers are speeding towards cryptographic relevancy: The time to prepare is now. (Apr. 17, 2026), https://www.ibm.com/think/perspectives/quantum-computers-are-speeding-towards-cryptographic-relevancy.
\item[] [61]	Caballar, R. IBM, Q-Day has already begun. Are you ready? (Jun. 25, 2026) https://www.ibm.com/think/news/q-day-has-already-begun-are-you-ready. 
\item[] [62]	Metaculus, When will the first general AI system be devised, tested, and publicly announced. (question 5121), community forecast as read June 15, 2026 (then a projection of June 2028); superseded by the November 2, 2026 reads in [209] and [210]. https://www.metaculus.com/questions/5121/when-will-the-first-general-ai-system-be-devised-tested-and-publicly-announced/.
\item[] [63]	White House, National Security Memorandum on Promoting United States Leadership in Quantum Computing While Mitigating Risks to Vulnerable Cryptographic Systems (May 4, 2022) https://bidenwhitehouse.archives.gov/briefing-room/statements-releases/2022/05/04/national-security-memorandum-on-promoting-united-states-leadership-in-quantum-computing-while-mitigating-risks-to-vulnerable-cryptographic-systems/. 
\item[] [64] 	World Nuclear Association, Safety of Nuclear Power Reactors (2026). https://world-nuclear.org/information-library/safety-and-security/safety-of-plants/safety-of-nuclear-power-reactors.
\item[] [65] 	Axios, Amodei on AI: There’s a 25\% chance that things go really, really badly (September 17, 2025). https://www.axios.com/2025/09/17/anthropic-dario-amodei-p-doom-25-percent.
\item[] [66]	Murray, Jerome T., and Marilyn J. Murray. Computers in Crisis: How to Avert the Coming Worldwide Computer Systems Collapse. New York: Petrocelli Books, (1984).
\item[] [67]	National Institute of Standards and Technology. Workshop on Cybersecurity in a Post-Quantum World. (April 2015). https://csrc.nist.gov/events/2015/workshop-on-cybersecurity-in-a-post-quantum-world.
\item[] [68]	Mosca, M.; Piani, M., Quantum Threat Timeline Report 2024, Global Risk Institute / evolutionQ (Dec. 2024). https://globalriskinstitute.org/publication/2024-quantum-threat-timeline-report/.
\item[] [69]	Mosca, M.; Piani, M., Quantum Threat Timeline Report 2025, Global Risk Institute / evolutionQ (Dec. 2025). https://globalriskinstitute.org/publication/quantum-threat-timeline-report-2025b/.
\item[] [70]	Citi Institute, Quantum Threat: The Trillion Dollar Security Race is On (Jan. 2026), https://www.citigroup.com/rcs/citigpa/storage/public/Citi\_Institute\_Quantum\_Threat.pdf. 
\item[] [71]	Westerbaan, B., The State of the Post-Quantum Internet in 2025, Cloudflare (Oct. 28, 2025), https://blog.cloudflare.com/pq-2025/.
\item[] [72]	Vermeer, M.; Peet, E., Securing Communications in the Quantum Computing Age: Managing the Risks to Encryption, RAND Corporation, RR-3102-RC (2020). https://www.rand.org/pubs/research\_reports/RR3102.html.
\item[] [73]	Le, T.D., Are Enterprises Ready for Quantum-Safe Cybersecurity?, arXiv:2509.01731 (2025), https://arxiv.org/abs/2509.01731. 
\item[] [74]	U.S. Senate Special Committee on the Year 2000 Technology Problem, final report (Feb 2000) Y2K Aftermath: Crisis Averted: Final Committee Report (S. Prt. 106-31), dated February 29, 2000. https://www.govinfo.gov/content/pkg/GOVPUB-Y1\_3-PURL-LPS90964/pdf/GOVPUB-Y1\_3-PURL-LPS90964.pdf. 
\item[] [75]	World Quantum Day org: https://worldquantumday.org/news/new-survey-reveals-public-support-for-quantum-science-and-technology/.
\item[] [76]	ComputerWorld, Some Key Fact and Events in Y2K History. (Jan. 3, 2000), https://www.computerworld.com/article/1372100/some-key-facts-and-events-in-y2k-history.html. 
\item[] [77]	Future Market Insights, Post-Quantum Cryptography (PQC) Migration Market Analysis, 2025–2035 (2025). https://www.futuremarketinsights.com/reports/post-quantum-cryptography-pqc-migration-market.
\item[] [78]	Moody’s Ratings, as reported in TechRadar, Post-quantum migration may consume roughly 2.5\% of enterprise IT budgets, with costs rising for organizations that delay (2026). https://www.techradar.com/pro/forget-ai-credit-rating-giant-feared-by-all-countries-just-issued-an-alarming-warning.
\item[] [79]	Gidney, C.; Ekerå, M., How to factor 2048 bit RSA integers in 8 hours using 20 million noisy qubits, (Apr. 13, 2021) arXiv:1905.09749, https://arxiv.org/abs/1905.09749.
\item[] [80] 	Mosca, M.; Piani, M., 2023 Quantum Threat Timeline Report. Global Risk Institute (December 2023). https://globalriskinstitute.org/publication/2023-quantum-threat-timeline-report/.
\item[] [81]	Cloud Security Alliance, Countdown to Y2Q, https://cloudsecurityalliance.org/research/working-groups/quantum-safe-security (accessed June 15, 2026).
\item[] [82]	Gidney, C., How to factor 2048-bit RSA integers with less than a million noisy qubits. (May 27, 2025), https://arxiv.org/pdf/2505.15917.
\item[] [83]	Global Risk Institute, Quantum Briefing: Recent Quantum-Computing Progress and its Implications for Cyber (June 2026). https://globalriskinstitute.org/mp-files/pdf-quantum-briefing-recent-quantum-computing-progress-and-implications-for-cybersecurity.pdf/. 
\item[] [84]	Webster, P.; Berent, L.; Chandra, O.; Hockings, E.; Baspin, N.; Thomsen, F.; Smith, S.; Cohen, L., The Pinnacle Architecture: Reducing the cost of breaking RSA-2048 to 100 000 physical qubits using quantum LDPC codes (Feb. 12, 2026), https://arxiv.org/html/2602.11457v1.
\item[] [85]	Chevignard, C.; Fouque, P.; Schrottenloher, A., Reducing the Number of Qubits in Quantum Discrete Logarithms on Elliptic Curves, (Apr. 7, 2026), https://eprint.iacr.org/2026/280. 
\item[] [86]	Kim, H., et al., New Quantum Circuits for ECDLP: Breaking Prime Elliptic Curve Cryptography, (January 2026), https://eprint.iacr.org/2026/106.
\item[] [87]	Timmer, J. Ars Technica, Quantum computing startup says it will leapfrog everybody (June 29, 2026). https://arstechnica.com/science/2026/06/quera-promises-thousands-of-error-corrected-qubits-by-2029/. 
\item[] [88]	Stubbs, M., NIST Recommends Timelines for Transitioning Cryptographic Algorithms, PQShield (Dec. 2, 2024), https://pqshield.com/nist-recommends-timelines-for-transitioning-cryptographic-algorithms/. 
\item[] [89]	NIST, Status Report on the Fourth Round of the NIST Post-Quantum Cryptography Standardization Process. (Mar. 2025). Section 2.2.1, https://nvlpubs.nist.gov/nistpubs/ir/2025/NIST.IR.8545.pdf.
\item[] [90]	Saxena, E.; Alfarano, A.; Charton, F.; Allen-Zhu, Z.; Wenger, E.; Lauter, K., Making Hard Problems Easier with Custom Data Distributions and Loss Regularization: A Case Study in Modular Arithmetic. (Aug. 25, 2025), https://eprint.iacr.org/2025/1525.pdf. 
\item[] [91]	Qu, H.; Xu, G., On the Provable Dual Attack for LWE by Modulus Switching. ePrint Archive (May 15, 2025), https://eprint.iacr.org/2025/859.pdf.
\item[] [92]	Attema, T.; Cramer, R.; Fehr, S.; Huang, Y.; de Kock, B.; Sotakova, J., A Survey on Security Reductions in Post-Quantum Cryptography, (Apr. 30, 2026), https://eprint.iacr.org/2026/846.pdf.
\item[] [93]	Nchiwo, R., Cryptanalysis of Polynomial Learning with Errors (PLWE): A Survey, (Mar. 11, 2026), https://eprint.iacr.org/2026/421.
\item[] [94]	Liu, Y., et al., AlphaGo Moment for Model Architecture Discovery (Jul. 24, 2025), https://arxiv.org/abs/2507.18074. 
\item[] [95] Al Jazeera, US orders Anthropic to disable AI models for all foreign nationals (June 13, 2026). https://www.aljazeera.com/news/2026/6/13/us-orders-anthropic-to-disable-ai-models-for-all-foreign-nationals.
\item[] [96] Hicks, C.; Attridge, C.; Janjeva, A.; Ashurst, C. Claude Mythos: What Does Anthropic’s New Model Mean for the Future of Cybersecurity? CETaS Expert Analysis, Centre for Emerging Technology and Security, Alan Turing Institute, April 2026. https://cetas.turing.ac.uk/publications/claude-mythos-future-cybersecurity.
\item[] [97]	Anthropic, Project Glasswing (April 7, 2026). https://www.anthropic.com/glasswing.
\item[] [98]	Anthropic, Redeploying Claude Fable 5, Anthropic News, (June 30, 2026) (availability restored July 1, 2026, with updated safety classifiers). https://www.anthropic.com/news/redeploying-fable-5.
\item[] [99]	Campbell, R., Mythos-Class Frontier Models as System-Level Disruptors in Post-Quantum Cryptography Migration: A Systems-Engineering Analysis of Lifecycle, Architecture, Dependencies, and Operational Modeling, (Apr. 23, 2026), https://www.preprints.org/manuscript/202604.1744. Published as Campbell, R., Mythos-Class Frontier Models and the Compression of Post-Quantum Cryptography Migration Timelines, Cryptography 10(3):41 (June 18, 2026). https://www.mdpi.com/2410-387X/10/3/41.
\item[] [100]	OpenAI, An OpenAI model has disproved a central conjecture in discrete geometry, (May 20, 2026). https://openai.com/index/model-disproves-discrete-geometry-conjecture/. 
\item[] [101]	IBM, Technology Roadmaps: The Future of Computing is Quantum-Centric, https://www.ibm.com/roadmaps/quantum/2030 (accessed May 27, 2026).
\item[] [102]	Quantinuum, Quantinuum Accelerates the Path to Universal Fully Fault-Tolerant Quantum Computing, https://www.quantinuum.com/blog/quantinuum-accelerates-the-path-to-universal-fault-tolerant-quantum-computing-supports-microsofts-ai-and-quantum-powered-compute-platform-and-the-path-to-a-quantum-supercomputer (accessed May 27, 2026).
\item[] [103] Timmer, J., Sooner than expected? Useful quantum error correction promised for 2028. ArsTechnica (June 17, 2026), https://arstechnica.com/science/2026/06/amazon-quera-promise-useful-quantum-error-correction-by-2028/. 
\item[] [104]	Wang, Y.; Hu, Z.; Sanders, B.C.; Kais, S. Qudits and High-Dimensional Quantum Computing. Frontiers in Physics 2020, 8, 589504. https://arxiv.org/abs/2008.00959.
\item[] [105]	Shaw, A.; Scholl, P.; Finkelstein, R.; Tsai, R.; Choi, J.; Endres, M., Erasure Cooling, Control, and Hyperentanglement of Motion in Optical Tweezers. Science Vol. 388, Issue 6749, pp. 845 to 849 (May 2025). https://www.science.org/doi/10.1126/science.adn2618.
\item[] [106]	Gajewski, R.; Bateman, J., Backaction Suppression in Levitated Optomechanics Using Reflective Boundaries. Phys. Rev. Research 7, 023041 (Apr. 30, 2025). https://doi.org/10.1103/PhysRevResearch.7.023041.
\item[] [107]	Cisco, Cisco Introduces Universal Quantum Switch, Advancing the Path to a Quantum Network, (Apr. 23, 2026). https://newsroom.cisco.com/c/r/newsroom/en/us/a/y2026/m04/cisco-introduces-universal-quantum-switch-advancing-the-path-to-a-quantum-network.html. 
\item[] [108]	INRS, Quantum Research Breakthrough Uses Synthetic Dimensions to Efficiently Process Quantum Information, PhysOrg (Oct. 17, 2024) https://phys.org/news/2024-10-quantum-breakthrough-synthetic-dimensions-efficiently.html. 
\item[] [109]	Ye, Y.; Kline, J.; Yen, A.; Cunningham, G.; Tan, M.; Zang, A.; Gingras, M.; Niedzielski, B.; Stickler, H.; Serniak, K.; Schwartz, M.; O’Brien, K., Near-ultrastrong Nonlinear Light-Matter Coupling in Superconducting Circuits, Nat. Commun. 16, 3799 (Apr. 30, 2025) https://www.nature.com/articles/s41467-025-59152-z. 
\item[] [110]	Shavit, J., Major Achievement in Quantum Mechanics Redefines Quantum Computing, Encryption, The Brighter Side (Apr. 27, 2025), https://www.thebrighterside.news/discoveries/major-achievement-in-quantum-mechanics-redefines-quantum-computing-encryption/.
\item[] [111]	Giani, A.; Win, M. Z.; Falb, P. L.; Conti, A., Unveiling Distinguishable Non-Gaussian Quantum States. Phys. Rev. A 113, 062435 (June 15, 2026). https://doi.org/10.1103/ffbg-4897.
\item[] [112]	Saner, S., et al., Generating Arbitrary Superpositions of Nonclassical Quantum Harmonic Oscillator States. (June 2, 2026). https://journals.aps.org/prx/abstract/10.1103/k1xk-yt42. 
\item[] [113] Stanford, Stanford quantum computing breakthrough uses twisted light to work without extreme cooling. (May 30, 2026). https://www.sciencedaily.com/releases/2026/05/260528074028.htm.
\item[] [114] Serha, R., et al., Ultralong-living magnons in the quantum limit. (May 1, 2026). https://www.science.org/doi/10.1126/sciadv.aee2344. 
\item[] [115] Yoon, S.; Song, G.; Jang, K.; Cha, S.; Seo, H., Quantum Implementation of SHA-1. (Aug 2025), https://eprint.iacr.org/2025/1415. 
\item[] [116] IBM, How researchers built a record-setting quantum circuit (May 20, 2026). https://www.ibm.com/quantum/blog/qft-benchmark. 
\item[] [117]	Cloud Security Alliance, Quantum Artificial Intelligence: Exploring the Relationship Between AI and Quantum Computing (Jan. 20, 2025), https://cloudsecurityalliance.org/blog/2025/01/20/quantum-artificial-intelligence-exploring-the-relationship-between-ai-and-quantum-computing. 
\item[] [118]	Swayne, M., AI for Quantum Error Correction: A Comprehensive Guide to Using Artificial Intelligence to Improve Quantum Error Correction, Quantum Insider (Jan. 6, 2025), https://thequantuminsider.com/2025/01/06/ai-for-quantum-error-correction-a-comprehensive-guide-to-using-artificial-intelligence-to-improve-quantum-error-correction/.
\item[] [119]	Wang, Z.; Tang, H., Artificial Intelligence for Quantum Error Correction, A Comprehensive Review, arXiv 2412.20380v1 (Dec. 29, 2024), https://arxiv.org/pdf/2412.20380v1. 
\item[] [120]	Swayne, M., AI Power for Quantum Errors: Google Develops AlphaQubit to Identify, Correct Quantum Errors, Quantum Insider (Nov. 20, 2024), https://thequantuminsider.com/2024/11/20/ai-power-for-quantum-errors-google-develops-alphaqubit-to-identify-correct-quantum-errors/. 
\item[] [121]	Microsoft Quantum, Roadmap to Fault Tolerant Quantum Computation Using Topological Qubit Arrays (Apr. 8, 2025), https://arxiv.org/pdf/2502.12252.
\item[] [122]	Bausch, J.; Senior, A.; Heras, F.; Edlich, T.; Davies, A.; Newman, M. et al., Learning High-Accuracy Error Decoding for Quantum Processors. Nature 635, 834 to 840 (Nov. 20, 2024), https://www.nature.com/articles/s41586-024-08148-8.
\item[] [123]	Moore, G.E., Cramming More Components onto Integrated Circuits, Electronics 38(8), 114 to 117 (1965).
\item[] [124]	Jurvetson, S., Rose’s Law for Quantum Computers (Oct. 4, 2012). https://www.flickr.com/photos/jurvetson/8054771535.
\item[] [125]	Ivezic, M., Post-Quantum Cryptography (PQC) Meets Quantum AI (QAI), PostQuantum, (Sep. 24, 2024), https://postquantum.com/post-quantum/pqc-quantum-ai-qai/.
\item[] [126]	Arnault, P.; Arrighi, P.; Herbert, S.; Kasnetsi, E., A Topology of Quantum Algorithms, (Jul. 6, 2024) https://arxiv.org/html/2407.05178v1.
\item[] [127]	Arlt, S.; Duan, H.; Li, F.; Xie, S.; Wu, Y.; Krenn, M., Meta-Designing Quantum Experiments with Language Models. arXiv:2406.02470 (Jun. 4, 2024) https://arxiv.org/abs/2406.02470.
\item[] [128]	Frohnert, F.; Gu, X.; Krenn, M.; van Nieuwenburg, E., Discovering Emergent Connections in Quantum Physics Research via Dynamic Word Embeddings. arXiv:2411.06577v1 (Nov. 10, 2024) https://arxiv.org/abs/2411.06577.
\item[] [129] Mankowitz, D. J., et al., Faster Sorting Algorithms Discovered Using Deep Reinforcement Learning. Nature 618, 257 to 263 (June 7, 2023). https://www.nature.com/articles/s41586-023-06004-9.
\item[] [130]	Jordan, S.; Shutty, N.; Wootters, M.; Zalcman, A.; Schmidhuber, A.; King, R.; Isakov, S.; Babbush, R., Optimization by Decoded Quantum Interferometry, arXiv:2408.08292 (Mar. 12, 2025) https://arxiv.org/abs/2408.08292.
\item[] [131]	Goswami, K.; Veereshi, G.A.; Schmelcher, P.; Mukherjee, R., Solving the Travelling Salesman Problem Using a Single QuBit, (2024) https://arxiv.org/abs/2407.17207.
\item[] [132]	Lordi, N.; Trank-Greene, M.; Kyle, A.; Combes, J., Quantum Permutation Puzzles with Indistinguishable Particles, arXiv:2410.22287v2 (Mar. 14, 2025) https://arxiv.org/abs/2410.22287. 
\item[] [133]	Swayne, M. Quantum Contest Offers 1 Bitcoin for Cracking Encryption with Shor’s Algorithm, Quantum Insider (April 18, 2025), https://thequantuminsider.com/2025/04/18/quantum-contest-offers-1-bitcoin-for-cracking-encryption-with-shors-algorithm.
\item[] [134]	Olaluwe, A., et al. Machine Learning and Side-Channel Attacks on Post-Quantum Cryptography. (Sep 2025), https://eprint.iacr.org/2025/1754.pdf. 
\item[] [135]	Valsaraj, H. et al., When Randomness Isn’t Random: Practical Fault Attack on Post-Quantum Lattice Standards, (Mar. 22, 2026), https://eprint.iacr.org/2025/2009.
\item[] [136]	Lai, Z., et al., You Only Decapsulate Once: Ciphertext-Independent Single-Trace Passive Side-Channel Attacks on HQC (Dec. 1, 2025), https://eprint.iacr.org/2025/2162. 
\item[] [137]	Soulami, S.; Connan, Y.; Duquesne, S., Full Secret Key Recovery of First-order Masked Crystals-Kyber implementation using multiple distinct chosen-ciphertexts, (Mar. 19, 2026), https://eprint.iacr.org/2026/528. 
\item[] [138]	Bashiri, K.; Geuenich, J.; Mittmann, J., Exploiting noisy single-bit leakage in ML-DSA, (Mar. 24, 2026), https://eprint.iacr.org/2026/580. 
\item[] [139]	Wang, H., et al., A Practical Neighborhood Search Attack on Oracle MLWE, (Apr. 11, 2026), https://eprint.iacr.org/2026/177.
\item[] [140]	Alfarano, A., et al., Improving ML Attacks on LWE with Data Repetition and Stepwise Regression, (Mar. 30, 2026), https://eprint.iacr.org/2026/612. 
\item[] [141]	Pichollet, A.; Schrottenloher, A., Quantum Truncated Differential Attacks using Convolutions, (Feb. 18, 2026), https://eprint.iacr.org/2026/305. 
\item[] [142]	May, A.; Sá Diogo, G., Multi-Instance Security Degradation of Code-Based KEMs, (Mar. 15, 2026), https://eprint.iacr.org/2026/517. 
\item[] [143]	Guo, Q.; Nabokov, D.; Johansson, T., Unlocking the True Potential of Decryption Failure Oracles: A Hybrid Adaptive-LDPC Attack on ML-KEM Using Imperfect Oracles, (Jan. 20, 2026), https://eprint.iacr.org/2026/070. 
\item[] [144]	Pulles, L.; Vie, P., Accelerating the Primal Hybrid Attack against Sparse LWE using GPUs, (Oct. 30, 2025), https://eprint.iacr.org/2025/1990.
\item[] [145]	Agrawal, S.; Bagchi, A.; Kumar, R., Attacks on Sparse LWE and Sparse LPN with new Sample-Time tradeoffs, (Mar. 31, 2026), https://eprint.iacr.org/2026/614. 
\item[] [146]	Hou, J.; Jiang, H., Careful with the Ring: Enhanced Hybrid Decoding Attacks against Module/Ring-LWE, (Mar. 17, 2026), https://eprint.iacr.org/2026/366. 
\item[] [147]	Li, Y.; Zheng, Z., Unified Dual Attack Analyses: Covariance-Based Score Distribution Prediction for LWE (May 27, 2026). https://eprint.iacr.org/2026/1048.pdf. 
\item[] [148] IBM Quantum, IBM Lays Out Clear Path to Fault-Tolerant Quantum Computing. IBM Quantum Blog (June 10, 2025). https://www.ibm.com/quantum/blog/large-scale-ftqc.
\item[] [149] Google Quantum AI and Collaborators, Quantum Error Correction Below the Surface Code Threshold. Nature 638, 920 to 926 (2025). https://doi.org/10.1038/s41586-024-08449-y.
\item[] [150]	LMArena (formerly LMSYS), Arena LLM Leaderboard, https://arena.ai/leaderboard/ (accessed June 15, 2026).
\item[] [151]	NIST, Advanced Encryption Standard (AES), FIPS PUB 197 (2001; updated 2023) https://csrc.nist.gov/pubs/fips/197/final.
\item[] [152]	Bonnetain, X.; Naya-Plasencia, M.; Schrottenloher, A., Quantum Security Analysis of AES. IACR Transactions on Symmetric Cryptology 2019(2), pp. 55 to 93 (2019). https://tosc.iacr.org/index.php/ToSC/article/view/8314.
\item[] [153] Amodei, D., Machines of Loving Grace. darioamodei.com (October 2024). https://darioamodei.com/essay/machines-of-loving-grace.
\item[] [154] Amodei, D., The Adolescence of Technology. darioamodei.com (January 2026). https://darioamodei.com/essay/the-adolescence-of-technology.
\item[] [155] Anthropic, Response to the OSTP Request for Information on an AI Action Plan. submission to the United States Office of Science and Technology Policy (March 2025). https://assets.anthropic.com/m/4e20a4ab6512e217/original/Anthropic-Response-to-OSTP-RFI-March-2025-Final-Submission-v3.pdf.
\item[] [156]	Legg, S., interviewed on the Dwarkesh Podcast: 2028 AGI, Superhuman Alignment, New Architectures (Oct. 26, 2023). https://www.dwarkesh.com/p/shane-legg.
\item[] [157]	CNBC, Human-level AI will be here in five to 10 years, Google DeepMind CEO says (Mar. 17, 2025). https://www.cnbc.com/2025/03/17/human-level-ai-will-be-here-in-5-to-10-years-deepmind-ceo-says.html.
\item[] [158]	Suleyman, M., interview with R. Khalaf, Mustafa Suleyman Sets Out Microsoft AI’s Goal of ‘Humanist Superintelligence’, Financial Times (Feb. 2026). https://www.ft.com/video/2c428045-bf4f-45bd-ada2-8ba53983cd81.
\item[] [159]	Aschenbrenner, L., Situational Awareness: The Decade Ahead (June 2024). https://situational-awareness.ai/.
\item[] [160]	LeCun, Y., interview, Meta’s AI Chief Yann LeCun on AGI, Open-Source, and AI Risk. TIME (Feb. 13, 2024). https://time.com/6694432/yann-lecun-meta-ai-interview/.
\item[] [161]	Kwiatkowski, K.; Kampanakis, P.; Westerbaan, B.; Stebila, D., Post-Quantum Hybrid ECDHE-MLKEM Key Agreement for TLSv1.3, IETF draft-ietf-tls-ecdhe-mlkem (2026). https://datatracker.ietf.org/doc/draft-ietf-tls-ecdhe-mlkem/.
\item[] [162]	IETF, ML-KEM Post-Quantum Key Agreement for TLS 1.3 (June 24, 2026), https://datatracker.ietf.org/doc/draft-ietf-tls-mlkem/. 
\item[] [163]	NIST, SP 800-140C, Approved Security Functions https://csrc.nist.gov/projects/cryptographic-module-validation-program/sp-800-140-series-supplemental-information/sp800-140c.
\item[] [164] NSA. Announcing the Commercial National Security Algorithm Suite 2.0; U/OO/194427-22; NSA: Fort Meade, MD, USA, September 2022. https://media.defense.gov/2022/Sep/07/2003071834/-1/-1/0/CSA\_CNSA\_2.0\_ALGORITHMS\_.PDF.
\item[] [165]	Couzens, B., Quantum \& Post-Quantum Cryptography 2026 (July 4, 2026), https://www.sitg-consulting.com/post/quantum-post-quantum-cryptography-2026. 
\item[] [166]	Hyoung, Y.L., et al., A Layered Risk Scoring Model for TLS Connections Against Quantum Threats. (June 8, 2026), https://eprint.iacr.org/2026/1174.pdf.
\item[] [167]	Noetzold, D.; Barbosa, J.; De Paz, J.; Leithardt, V., “A Modular Risk Assessment Module for Adaptive Cryptographic Selection in Q-OPSEC,” (June 2026). https://eprint.iacr.org/2026/1341.pdf. 
\item[] [168]	Cremers, C.; Nakarmi, A.; Peltonen, A.; Ronen, E., “3PaaS: Privacy-Preserving Post-Compromise Security as a Service,” (July 2026). https://eprint.iacr.org/2026/1353.pdf. 
\item[] [169]	Seyedi, Z; Rass, S.; Ahmad, S.; Rahmati, F., “Adaptive Quantum-Resistant Hybrid Encryption Framework for Secure IoMT Edge Data Sharing,” (July 2026). https://eprint.iacr.org/2026/1354.pdf. 
\item[] [170]	Choi, C., “Independent Labs Crack Google’s Secret Cryptography Work: Using crowdsourcing and AI swarms, one startup surpassed Google’s results in 72 hours,” IEEE Spectrum (July 7, 2026). https://spectrum.ieee.org/google-quantum-cryptography-zero-knowledge. 
\item[] [171]	O’Donnell, J., “PsiQuantum has a plan to make a massive quantum computer out of light,” MIT Technology Review, (July 14, 2026). https://www.technologyreview.com/2026/07/14/1140356/psiquantum-plan-massive-quantum-computer-out-of-light/. 
\item[] [172]	Butler, A., Prosperity at Risk: The Quantum Computer Threat to the US Financial System, Hudson Institute (April 2023), https://www.hudson.org/technology/prosperity-risk-quantum-computer-threat-us-financial-system.
\item[] [173]	IBM Research, “Cryptography Abstraction Layer,” https://research.ibm.com/blog/cryptography-abstraction-layer.
\item[] [174]	Rameshan, N. and Messmer, G., “Cryptographic Agility for Applications: An Assessment Framework and Principled API Design,” Eurocrypt 2026 workshop paper (May 10, 2026). https://research.ibm.com/publications/cryptographic-agility-for-applications-an-assessment-framework-and-principled-api-design.
\item[] [175]	Executive Order 13526, Classified National Security Information, §3.3 (automatic declassification at 25 years, with exemptions to 50 and 75 years) (Dec. 29, 2009), https://www.archives.gov/isoo/policy-documents/cnsi-eo.html.
\item[] [176]	Swidler \& Berlin v. United States, 524 U.S. 399 (1998) (attorney-client privilege survives the client’s death), https://supreme.justia.com/cases/federal/us/524/399/.
\item[] [177]	HIPAA documentation-retention requirement, 45 CFR § 164.316(b)(2)(i) (six years), with patient medical-record retention set by state law (three to twenty-plus years), https://www.law.cornell.edu/cfr/text/45/164.316.
\item[] [178]	SEC Rule 17a-4, 17 CFR § 240.17a-4 (broker-dealer record retention, six years); PCI DSS v4.0, Requirement 3 (cardholder-data retention and disposal), https://www.law.cornell.edu/cfr/text/17/240.17a-4.
\item[] [179]	Campbell, R., Enterprise Migration to Post-Quantum Cryptography: Timeline Analysis and Strategic Frameworks, Computers 15(1):9 (MDPI, 2026) (operational-technology and embedded-system lifecycles; enterprise migration durations), https://www.mdpi.com/2073-431X/15/1/9.
\item[] [180]	Häner, T., et al. (IonQ), Computing 256-bit elliptic curve discrete logarithms in 26 days on a fault-tolerant trapped-ion quantum computer with 20,000 qubits, ePrint 2026/1916 (Sept. 7, 2026); arXiv:2609.05625. https://eprint.iacr.org/2026/1916.
\item[] [181]	Jo, S.; Lee, G. J., Quantum Discrete Logarithms on Elliptic Curves with 5n/2 + o(n) Logical Qubits and Õ(n²) Toffoli Gates, ePrint 2026/2014 (Sept. 14, 2026). https://eprint.iacr.org/2026/2014.
\item[] [182]	Dallaire-Demers, P.-L.; Doyle, W.; Foo, T. (Pauli Group), Brace for impact: ECDLP challenges for quantum cryptanalysis, arXiv:2508.14011 (v2, Mar. 25, 2026; v3, Oct. 29, 2026). https://arxiv.org/abs/2508.14011.
\item[] [183]	Anthropic, Discovering cryptographic weaknesses with Claude (July 28, 2026). https://www.anthropic.com/research/discovering-cryptographic-weaknesses.
\item[] [184]	Nasr, M.; Carlini, N., Cryptanalysis of 7-Round AES via the Algebraic Structure of its S-box, Anthropic (July 28, 2026). https://anthropic.com/document/aes\_mobius\_bridge.pdf.
\item[] [185]	Anthropic, Mythos Preview's Chain of Thought in Discovering the AES Möbius Bridge (July 28, 2026). https://anthropic.com/document/aes\_mobius\_bridge\_cot.pdf.
\item[] [186]	Straznickas, Z.; Weis, S. A. (Anthropic), HAWK-n Key Recovery Reduces to SVP in Dimension n/2 + 1, ePrint 2026/1593 (Aug. 3, 2026); announced by S. A. Weis on the NIST pqc-forum (July 28, 2026). https://eprint.iacr.org/2026/1593; https://groups.google.com/a/list.nist.gov/g/pqc-forum/c/2r2u6SbHun4.
\item[] [187]	NIST, Nine Candidates Advance to Third Round of Additional Digital Signatures for PQC, news release (May 14, 2026; updated July 29, 2026 with the HAWK withdrawal). https://www.nist.gov/news-events/news/2026/05/nine-candidates-advance-third-round-additional-digital-signatures-pqc.
\item[] [188]	Fluri, L.; Shafran, A.; Carlini, N.; Jagielski, M.; Nasr, M.; Dunkelman, O.; Ronen, E.; Tramèr, F., CryptanalysisBench: Can LLMs do Cryptanalysis? arXiv:2607.18538 (v2, July 29, 2026). https://arxiv.org/abs/2607.18538.
\item[] [189]	OpenAI, The Quasi-Riemann Hypothesis: A Zero-Free Half-Plane Re(s) > 7/8, openai/math result family 003, preprints/The-Quasi-Riemann-Hypothesis-September-30-2026/paper.pdf (Sept. 30, 2026; released Oct. 6, 2026); withdrawals and revisions recorded in history.md. https://github.com/openai/math.
\item[] [190]	Goldblatt, D., openai-zeta-proof-check: independent re-run of the Lean check for the 7/8 zero-free half-plane on two kernels under the standard axioms (Oct. 6 to 7, 2026; one operator, run with Claude Code). https://github.com/davegoldblatt/openai-zeta-proof-check.
\item[] [191]	Kakaes, K., AI Proved the Quasi-Riemann Hypothesis. What Happens Now? Quanta Magazine (Oct. 19, 2026). https://www.quantamagazine.org/ai-proved-the-quasi-riemann-hypothesis-what-happens-now-20261019/.
\item[] [192]	Visser, M., Primes between 8th powers? arXiv:2610.20058 [math.NT] (Oct. 14, 2026). https://arxiv.org/abs/2610.20058.
\item[] [193]	Anthropic (Claude, under J. Leder's direction), formal-math: a Lean 4 proof that θ(p\_c) = 0 for bond percolation on Z\textasciicircum{}d, d ≥ 2, via Kozma and Nitzan's Conjecture 3 (Aug. 28, 2026). https://github.com/anthropics/formal-math/blob/795efb86f191735c5481675763537cfb4ff37e55/percolation/README.md.
\item[] [194]	Duminil-Copin, H., Care for a little more AI? Proofs and Prompts (Aug. 30, 2026). https://proofsandprompts.com/2026/08/30/care-for-a-little-more-ai/.
\item[] [195]	Lu, E., Factoring RSA-260, Cognition (Sept. 9, 2026). https://cognition.com/blog/factoring-rsa-260.
\item[] [196]	Weis, S. A., RSA-896 (Sept. 2026). https://saweis.net/posts/rsa-896.html.
\item[] [197]	Zhao, Y.; He, D.; Cheng, W.; Liu, Y.; Ming, J.; Zhou, Y., When Module Lattice Leaks: Horizontal Fusion Attacks on ML-DSA Implementation, ACM CCS 2026; ePrint 2026/1904 (Sept. 2026). https://eprint.iacr.org/2026/1904.
\item[] [198]	ElGhamrawy, M.; Eisenbarth, T.; Hermelink, J.; Rabich, A.; Sieck, F.; Streit, S.; Thietke, J.; Zhang, Z., Keep Track of Your Errors: Solving ILWE and Improving Attacks on ML-DSA, ePrint 2026/2091 (Sept. 2026). https://eprint.iacr.org/2026/2091.
\item[] [199]	Zhou, Y.; Wang, W.; Sun, Y.; Yu, Y., Every Signing Leaks: Breaking Falcon via Floating-Point Conversion Leakage, ePrint 2026/2124 (Sept. 2026). https://eprint.iacr.org/2026/2124.
\item[] [200]	Jahandideh, V.; Batina, L., Breaking First-Order Masked Kyber with Belief Propagation Alone, ePrint 2026/2448 (Oct. 2026). https://eprint.iacr.org/2026/2448.
\item[] [201]	Das, B. S., One Unverified Coordinate Is Enough: Key Recovery from Faulty Fujisaki-Okamoto Rejection Checks in ML-KEM, ePrint 2026/2239 (Sept. 27, 2026). https://eprint.iacr.org/2026/2239.
\item[] [202]	Qu, H.; Tian, C.; Wang, G.; Xu, G., Provable dual attack on LWE via lattice projection, ePrint 2026/2117 (Sept. 2026). https://eprint.iacr.org/2026/2117.
\item[] [203]	Li, J.; Wang, L.-P.; Wang, H., A Provable Correctness Analysis of the MATZOV-Style FFT Dual Attack on ML-KEM, ePrint 2026/2166 (Sept. 2026). https://eprint.iacr.org/2026/2166.
\item[] [204]	Li, J.; Wang, L.-P.; Wang, H., Provable FFT-Accelerated Dual Attack on LWE Using the General Dual Lattice, ePrint 2026/2186 (Sept. 2026). https://eprint.iacr.org/2026/2186.
\item[] [205]	Hashimoto, R.; Shikata, J.; Takayasu, A.; Wang, Y., AH-BKZ: A Lattice Reduction Algorithm with Asynchronous Hybrid Processing, ICICS 2026, LNCS 17214, pp. 3 to 23; ePrint 2026/1876. https://eprint.iacr.org/2026/1876.
\item[] [206]	Cid, C.; Elkouss, D.; Goulão, M., Superposition Key-Recovery Attacks on Dilithium and Fiat-Shamir Signature Schemes, ePrint 2026/2258 (Sept. 2026). https://eprint.iacr.org/2026/2258.
\item[] [207]	Kuninets, A.; Bakharev, A., Poisoned Cryptography: Solving Direct-Sum LIP in Half the Dimension, ePrint 2026/2281 (Sept. 30, 2026). https://eprint.iacr.org/2026/2281.
\item[] [208]	Lin, H.; Vasudevan, P. N., Provably Subexponential Algorithms for Bounded-Error LWE with Few Samples, ePrint 2026/2426 (Oct. 2026). https://eprint.iacr.org/2026/2426.
\item[] [209]	Metaculus, When will the first weakly general AI system be devised, tested, and publicly announced? (question 3479), community forecast read Nov. 2, 2026 (28 Aug. 2027, interquartile 20 Sept. 2026 to 17 Oct. 2029, about 1,800 forecasters; question open, resolution under review). https://www.metaculus.com/questions/3479/.
\item[] [210]	Metaculus, When will the first general AI system be devised, tested, and publicly announced? (question 5121), community forecast read Nov. 2, 2026 (18 Sept. 2030, interquartile 22 Mar. 2028 to June 2036, about 1,900 forecasters). https://www.metaculus.com/questions/5121/.
\item[] [211]	Lifland, E.; Kokotajlo, D.; Halstead, B., Q2.5 2026 Timelines Update: Uplift and Revenue, AI Futures Project (Aug. 16, 2026). https://blog.aifutures.org/p/q25-2026-timelines-update-uplift.
\item[] [212]	Campbell, R., Runtime Cryptographic Evidence to Bounded Assurance Verdicts: Deterministic Conformance and Policy Appraisal for SHA-256 and AES-256-GCM, Computers 15(10):659 (Sept. 29, 2026). https://doi.org/10.3390/computers15100659.
\item[] [213]	Canetti, R.; Ha, J.; Kaptchuk, G., Analyzing Cryptography in Context: A Cryptography-Native Approach to Threat Modeling, USENIX Security 2026; ePrint 2026/1507. https://eprint.iacr.org/2026/1507.
\item[] [214]	Song, M.-H.; Eum, S.-W.; Lee, S.-W.; Kim, H.-G.; Hyoung, Y.-L.; Seo, H., A Post-Quantum Cryptography Recommendation System with TLS Validation for Heterogeneous Environments, ePrint 2026/1938 (Sept. 2026). https://eprint.iacr.org/2026/1938.
\item[] [215]	MITRE for the Post-Quantum Cryptography Coalition, Post-Quantum Cryptography (PQC) Capability Development Report, MITRE Case 26-2150 (Sept. 2026; evidence collected July 2026; peer review Aug. to Sept. 2026; released Oct. 15, 2026). https://pqcc.org/wp-content/uploads/2026/10/PQC-Capability-Development-Report.pdf (archived Oct. 16, 2026 at web.archive.org).
\item[] [216]	Bindel, N.; Hale, B.; Connolly, D.; Driscoll, F., Hybrid Signature Spectrums, RFC 9955, IETF (July 2026). https://www.rfc-editor.org/rfc/rfc9955.
\item[] [217]	Shea, L.; Haller, M.; Suhl, A.; Heninger, N.; Thomé, E., Forging 1024-bit RSA signatures in nearly SNFS time, ePrint 2026/2131 (Sept. 20, 2026). https://eprint.iacr.org/2026/2131.
\item[] [218]	OpenSSL Project, OpenSSL TLS 1.3 server may choose unexpected key agreement group (CVE-2026-2673), security advisory (Mar. 13, 2026; fixed in 3.6.2 and 3.5.6, Apr. 7, 2026). https://openssl-library.org/news/secadv/20260313.txt.
\item[] [219]	wolfSSL, Release notes 5.9.0 (Mar. 18, 2026), 5.9.2 (June 23, 2026) and 5.9.4 (Sept. 25, 2026) (CVE-2026-3503, CVE-2026-6330, CVE-2026-10097). https://github.com/wolfssl/wolfssl/releases.
\item[] [220]	Quantinuum, 48 error-corrected logical qubits at 2:1 encoding on Helios (Nov. 5, 2025; Dasu, S., et al., Computing with many encoded logical qubits beyond break-even, arXiv:2602.22211, Feb. 25, 2026), and the Helix architecture (Berthusen, N., et al., Experimental validation of a compact fault-tolerant architecture for trapped ions, arXiv:2609.03194, Sept. 2, 2026). https://arxiv.org/abs/2602.22211; https://arxiv.org/abs/2609.03194.
\item[] [221]	Bluvstein, D.; Geim, A. A.; Li, S. H.; et al. (Harvard, MIT and QuEra Computing), A fault-tolerant neutral-atom architecture for universal quantum computation, Nature 649, 39 to 46 (online Nov. 10, 2025); arXiv:2506.20661. https://www.nature.com/articles/s41586-025-09848-5.
\item[] [222]	Ye, M.; Maksymov, A.; Delfosse, N. (IonQ), Real-time decoder for a MegaQuOp quantum computer using a single CPU, arXiv:2608.25027 (v1, Aug. 25, 2026; v2, Sept. 3, 2026). https://arxiv.org/abs/2608.25027.
\item[] [223]	Regev, O., Quantum computation and lattice problems, SIAM Journal on Computing 33(3), 738 to 760 (2004). arXiv:cs/0304005.
\item[] [224]	Kuperberg, G., A subexponential-time quantum algorithm for the dihedral hidden subgroup problem, SIAM Journal on Computing 35(1), 170 to 188 (2005). arXiv:quant-ph/0302112.
\item[] [225]	Brakerski, Z.; Kirshanova, E.; Stehlé, D.; Wen, W., Learning With Errors and Extrapolated Dihedral Cosets, PKC 2018, LNCS 10770, pp. 702 to 727. arXiv:1710.08223.
\item[] [226]	Bai, S.; Jangir, H.; Kirshanova, E.; Ngo, T.; Youmans, W., A Quasi-polynomial Time Algorithm for the Extrapolated Dihedral Coset Problem over Power-of-Two Moduli, CRYPTO 2025; ePrint 2025/1046. https://eprint.iacr.org/2025/1046.
\item[] [227]	Wen, W.; Zheng, J., Module Learning With Errors and Structured Extrapolated Dihedral Cosets, CRYPTO 2026, LNCS 16802, pp. 263 to 294; ePrint 2026/155. https://eprint.iacr.org/2026/155.
\item[] [228]	Simon, D. R. (Amazon Web Services), A Polynomial-Time Quantum Algorithm for the Dihedral Coset Problem, ePrint 2026/1591 (received Aug. 3, 2026; seventh revision Nov. 3, 2026). https://eprint.iacr.org/2026/1591.
\item[] [229]	Gupte, A.; Ragavan, S.; Zhandry, M., The ePrint:2026/1591 Quantum Algorithm Does Not Solve DCP, ePrint 2026/1693 (Aug. 15, 2026; rev. Sept. 1, 2026), with a Lean 4 formalization at https://github.com/sragavan99/lean-ePrint-2026-1591-refutation. https://eprint.iacr.org/2026/1693.
\item[] [230]	Gupte, A.; Ragavan, S.; Zhandry, M., A Fourier-Label Information-Loss Barrier for Dihedral Coset Algorithms, arXiv:2609.40062 (Sept. 30, 2026). https://arxiv.org/abs/2609.40062.
\item[] [231]	Guo, Y.; Yang, S., Rigorous Statements and Proofs of the Lemmas in Simon's Algorithm for the Dihedral Coset Problem and Their Underlying Hypothesis, ePrint 2026/1714 (Aug. 17, 2026); arXiv:2608.16598. https://eprint.iacr.org/2026/1714.
\item[] [232]	Pappalardo, A.; Goudarzi, D.; Hertz, V.; Jacinto, H.; Kroell, L.; Schrottenloher, A.; Shanahan, P.; Gouzien, É.; de La Vieuville, R.; Londe, V. (Alice \& Bob), Shor's algorithm compiled on cat qubits: 22 593 qubits for RSA-2048 in 3.14 days, ePrint 2026/2665 (Oct. 28, 2026). https://eprint.iacr.org/2026/2665.
\item[] [233]	Xiong, Z.; Wang, M., Faster Polynomial Factorization over Finite Fields via Batched Modular Composition, ePrint 2026/2625 (Oct. 24, 2026). https://eprint.iacr.org/2026/2625.
\item[] [234]	Mamun, S. M., Evidence-Tiered Horizons: An Anatomical Review of Grover et al.'s CRQC+AI Scenario Instrument for TLS 1.3 and Ten Competing Threat Models, Zenodo (Aug. 27, 2026). https://doi.org/10.5281/zenodo.22990284.
\item[] [235]	Gershteyn, I. M.; Alber, J. A., Quantum Horizon: An evaluation of quantum computing as a threat to Bitcoin and Ethereum, arXiv:2606.14484 (June 12, 2026). https://arxiv.org/abs/2606.14484.
\item[] [236]	NIST, Submission Requirements and Evaluation Criteria for the Post-Quantum Cryptography Standardization Process, Section 4.A.5 (MAXDEPTH) (Dec. 2016); the same document as [49]. https://csrc.nist.gov/CSRC/media/Projects/Post-Quantum-Cryptography/documents/call-for-proposals-final-dec-2016.pdf.
\item[] [237]	Grassl, M.; Langenberg, B.; Roetteler, M.; Steinwandt, R., Applying Grover's Algorithm to AES: Quantum Resource Estimates, PQCrypto 2016, LNCS 9606, pp. 29 to 43; arXiv:1512.04965.
\item[] [238]	Jaques, S.; Naehrig, M.; Roetteler, M.; Virdia, F., Implementing Grover Oracles for Quantum Key Search on AES and LowMC, EUROCRYPT 2020, LNCS 12106, pp. 280 to 310; ePrint 2019/1146. https://eprint.iacr.org/2019/1146.
\item[] [239]	Lavasani, A.; Reed, A. P.; Dreiling, J. M.; et al. (Quantinuum), Quantum error correction well below the color code threshold, arXiv:2610.24077 (Oct. 16, 2026). https://arxiv.org/abs/2610.24077.
\item[] [240]	Rosenfeld, E.; Gidney, C.; Roberts, G.; et al. (Google Quantum AI), Magic state cultivation on a superconducting quantum processor, Science, eaee7961 (Oct. 15, 2026). https://doi.org/10.1126/science.aee7961.
\item[] [241]	Bao, Z.; Guo, J.; List, E.; Wang, H., The Cross-ratio Property and Its Use for Cryptanalysis of Round-reduced AES, ePrint 2026/1549 (July 29, 2026). https://eprint.iacr.org/2026/1549.
\item[] Disclaimer/Publisher’s Note: The statements, opinions and data contained in all publications are solely those of the individual author(s) and contributor(s) and not of the publisher and/or the editor(s). The publisher and/or the editor(s) disclaim responsibility for any injury to people or property resulting from any ideas, methods, instructions or products referred to in the content.\end{itemize}
\end{document}